%% file: paper.tex
\documentclass[english,twocolumn,showpacs,aps,prd,a4paper]{revtex4-1}

\usepackage[hyperindex,breaklinks,colorlinks]{hyperref}

\usepackage{atlasphysics}
\usepackage{units}
\usepackage{subfigure}
\usepackage{xspace}
\usepackage{placeins}
\usepackage{booktabs}
\usepackage{ifthen}
\usepackage{amsmath}
\usepackage{multirow}
\usepackage{hhline}
\usepackage{graphicx}
\graphicspath{{figures/}}
\usepackage{rotating} 

\newboolean{blind}

\usepackage{preprintcover}
\PreprintCoverPaperTitle{Search for supersymmetry in events with four or more leptons in $\ensuremath{\sqrt{s}}=8\,$TeV $pp$ collisions with the ATLAS detector}
\PreprintIdNumber{CERN-PH-EP-2014-074}
\PreprintCoverAbstract{
Results from a search for supersymmetry in events with four or more leptons including electrons, muons and taus are presented.
The analysis uses a data sample corresponding to 20.3~\mbox{fb$^{-1}$} of proton-proton collisions delivered by the Large Hadron Collider at $\ensuremath{\sqrt{s}}=8\,$TeV and recorded by the ATLAS detector.
Signal regions are designed to target supersymmetric scenarios that can be either enriched in or depleted of events involving the production of a \ensuremath{Z} boson.
No significant deviations are observed in data from standard model predictions and results are used to set upper limits on the event yields from processes beyond the standard model.
Exclusion limits at the 95\% confidence level on the masses of relevant supersymmetric particles are obtained.
In $R$-parity-violating simplified models with decays of the lightest supersymmetric particle to electrons and muons, limits of 1350 and 750~GeV are placed on gluino and chargino masses, respectively.
In $R$-parity-conserving simplified models with heavy neutralinos decaying to a massless lightest supersymmetric particle, heavy neutralino masses up to 620~GeV are excluded.
Limits are also placed on other supersymmetric scenarios.
}

\PreprintJournalName{Phys. Rev. D}

\begin{document}

\vspace{5mm}
\title{Search for supersymmetry in events with four or more leptons in $\rts=8\tev$ $pp$ collisions with the ATLAS detector}

\author{The ATLAS Collaboration}

\date{September 4, 2014}

\begin{abstract}
Results from a search for supersymmetry in events with four or more leptons including electrons, muons and taus are presented.
The analysis uses a data sample corresponding to 20.3~\ifb{} of proton-proton collisions delivered by the Large Hadron Collider at $\rts=8\tev$ and recorded by the ATLAS detector.
Signal regions are designed to target supersymmetric scenarios that can be either enriched in or depleted of events involving the production of a \Zboson{} boson.
No significant deviations are observed in data from standard model predictions and results are used to set upper limits on the event yields from processes beyond the standard model.
Exclusion limits at the 95\% confidence level on the masses of relevant supersymmetric particles are obtained.
In $R$-parity-violating simplified models with decays of the lightest supersymmetric particle to electrons and muons, limits of $1350$ and $750\gev$ are placed on gluino and chargino masses, respectively.
In $R$-parity-conserving simplified models with heavy neutralinos decaying to a massless lightest supersymmetric particle, heavy neutralino masses up to $620\gev$ are excluded.
Limits are also placed on other supersymmetric scenarios.
\end{abstract}


\maketitle

\section{Introduction}
\label{sec:intro}

Supersymmetry (SUSY)~\cite{Miyazawa:1966,Ramond:1971gb,Golfand:1971iw,Neveu:1971rx,Neveu:1971iv,Gervais:1971ji,Volkov:1973ix,Wess:1973kz,Wess:1974tw} 
is a space-time symmetry that postulates the existence
of new SUSY particles, or sparticles, with spin ($S$) differing by
one half-unit with respect to their standard model
(SM) partners. In supersymmetric extensions of the SM, each SM fermion (boson) is associated with a SUSY boson (fermion), having the same quantum numbers as its partner
except for $S$.
The scalar superpartners of the SM fermions are called sfermions (comprising the sleptons, $\slepton$, the sneutrinos, $\snu$, and the squarks, \squark), while the gluons have fermionic superpartners 
called gluinos ($\gluino$). 
The SUSY partners of the Higgs and electroweak (EW) gauge bosons, known as higgsinos, winos and the bino,
mix to form the mass eigenstates known as charginos ($\tilde{\chi}_l^\pm$,~$l=1,2$) and neutralinos ($\tilde{\chi}_m^0$,~$m=1,...,4$). 

In generic SUSY models with minimal particle content, the superpotential includes terms that violate conservation of lepton ($L$) and baryon ($B$) number~\cite{Weinberg:1981wj,Sakai:1981pk}:

\begin{equation}\label{eq:rpvpotential}
\frac{1}{2}\rpvlambda L_iL_j\bar{E_k} +
\rpvlambdaP L_iQ_j\bar{D_k} + 
\frac{1}{2}\rpvlambdaPP \bar{U_i}\bar{D_j}\bar{D_k} +
\kappa_iL_iH_2,
\end{equation}

\noindent
where $L_i$ and $Q_i$ indicate the lepton and quark SU(2)-doublet superfields, respectively, while $\bar{E_i}$, $\bar{U_i}$ and $\bar{D_i}$ 
are the corresponding singlet superfields. The indices $i$, $j$ and $k$ refer to quark and lepton generations. The Higgs SU(2)-doublet superfield 
$H_2$ is the Higgs field that couples to up-type quarks. The 
\rpvlambda, \rpvlambdaP{} and \rpvlambdaPP{} parameters are new Yukawa couplings, while the $\kappa_i$ parameters have dimensions of mass and vanish at the unification scale.

In the absence of a protective symmetry, $L$- and $B$-violating terms
may allow for proton decay at a rate that is in conflict with
the tight experimental constraints on the proton's lifetime~\cite{Ahmed:2003sy}.
This difficulty can be avoided by imposing the conservation 
of $R$-parity~\cite{Fayet:1976et,Fayet:1977yc,Farrar:1978xj,Fayet:1979sa,Dimopoulos:1981zb}, defined as $P_R=(-1)^{3(B-L)+2S}$. 
However, experimental bounds on proton decay can also be evaded in $R$-parity-violating (RPV) scenarios,
as long as the Lagrangian conserves either $L$ or $B$.

In $R$-parity-conserving (RPC) models,
the lightest SUSY particle (LSP) is stable and 
leptons can originate from unstable weakly interacting sparticles decaying into the LSP.
In RPV models, the LSP is unstable and decays to SM particles, 
including charged leptons and neutrinos when at least one of the $\rpvlambda$ parameters is nonzero.
Therefore, both the RPC and RPV SUSY scenarios can result in signatures with large lepton multiplicities and substantial missing transverse momentum, which can be utilized to suppress SM background processes effectively.
In this paper, it is assumed that the LSP
is either the lightest neutralino ($\ninoone$) or the neutral and weakly interacting superpartner of the graviton, the gravitino ($\tilde{G}$).

A search for new physics is presented in final states with at least four isolated leptons, including electrons, muons and $\tau$ leptons (taus).
Electrons and muons are collectively referred to as ``light leptons,'' which include those from leptonic tau decays, while taus refer to hadronically decaying taus in the rest of this paper.
Final states with two, three or at least four light leptons are considered, requiring at least two, one and zero taus, respectively. 
Events are further classified according to the presence or absence of a \Zboson{} boson candidate.  
In final states with four light leptons the backgrounds with four prompt leptons (\ZZ{} and \ttbarZ) dominate; 
these are estimated using Monte Carlo (MC) simulations.
On the other hand, in final states with taus the main background arises from events where light-flavor jets are misidentified as taus, and these are estimated with a data-driven method.

The analysis uses $\lumi$ of proton-proton collision data recorded in 2012 with the ATLAS detector at the Large Hadron Collider (LHC) at 
a center-of-mass energy of $\sqrt{s}\,$$=\,8\tev$.
Results are  interpreted in terms of model-independent limits on the event yields from new physics processes leading to the given signature,  
as well as in a variety of specific SUSY scenarios. 
These scenarios include RPV and RPC simplified models, which describe the interactions of a minimal set of particles, 
 as well as models with general gauge-mediated SUSY breaking (GGM)~\cite{ggm2008,Buican:2008ws},  
which is a generalization of gauge-mediated SUSY breaking theories (GMSB)~\cite{Dine:1981gu,AlvarezGaume:1981wy,Nappi:1982hm,Dine:1993yw,Dine:1994vc,Dine:1995ag}  
where the parametrization does not depend on the details of the SUSY breaking mechanism. 

This analysis updates and extends results presented previously by ATLAS~\cite{ATLAS:2012kr}.
Results from similar searches interpreted in RPV models have been reported 
by other experiments~\cite{Chatrchyan:2013xsw,Chatrchyan:2012mea,Abazov:2006nw,Heister:2002jc,Abdallah:2003xc,Achard:2001ek,Abbiendi:2003rn},
while previous ATLAS searches requiring photons in the final state have constrained closely related GGM models with different neutralino compositions~\cite{Aad:2012zza,Aad:2012jva}.

\section{New physics scenarios \label{sec:scenarios}}

Lepton-rich signatures are expected in a variety of new physics scenarios. 
The SUSY models used for the interpretation of results from this analysis are described briefly below.


\subsection{RPV simplified models}
\label{sec:RPVmodels}

\noindent

\begin{figure}[tb]
\centering
\subfigure[ ~Chargino NLSP]{\includegraphics[width=0.23\textwidth]{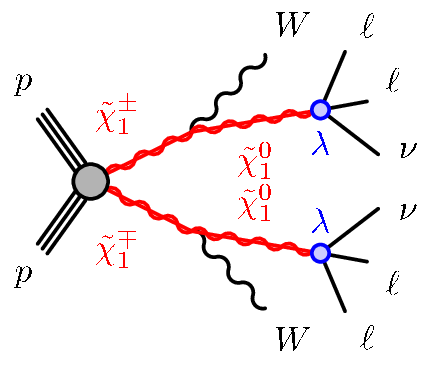}}
\subfigure[ ~R(L)-slepton NLSP]{\includegraphics[width=0.23\textwidth]{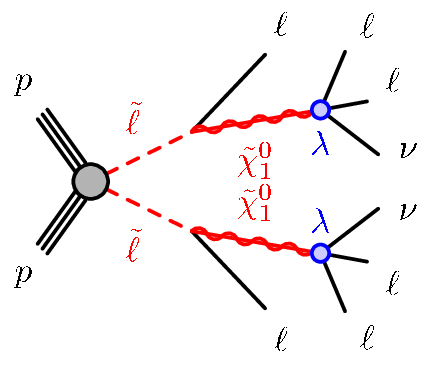}}

\subfigure[ ~Sneutrino NLSP]{\includegraphics[width=0.23\textwidth]{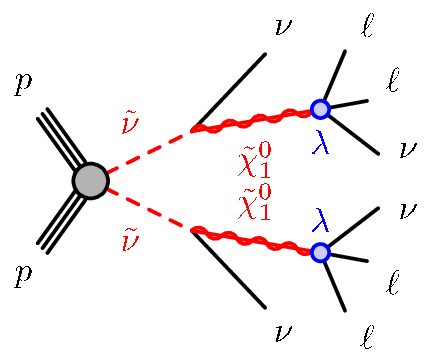}}
\subfigure[ ~Gluino NLSP]{\includegraphics[width=0.23\textwidth]{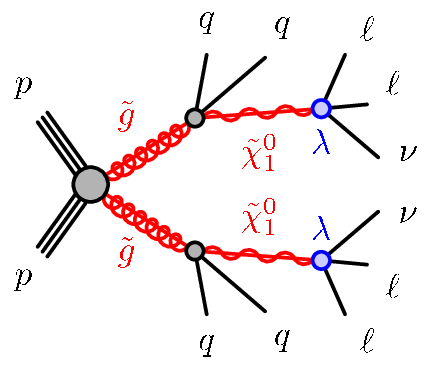}}
\caption{Representative diagrams for the RPV simplified models considered in this analysis.
\label{fig:RPVgraphs} }
\end{figure}

In the RPV simplified models used in this analysis, a binolike \ninoone{} is assumed to decay into two charged leptons and a neutrino via the \rpvlambda{} term in \Eqref{eq:rpvpotential}.
The observed final-state signature is driven by this decay, but the cross section and, to a lesser extent, the signal acceptance depend on the sparticle production mechanism.
Four event topologies are tested, resulting from different choices for the next-to-lightest SUSY particles (NLSPs):
a chargino (\chinoonepm) NLSP;
slepton NLSPs, referring to mass-degenerate \sel, \smu{} and \stau{} sleptons;
sneutrino NLSPs, referring to mass-degenerate \snue, \snumu{} and \snutau{} sneutrinos;
and a gluino NLSP, the latter being a benchmark for how the experimental reach may increase when strong production is introduced.
In the slepton case, both the left-handed and right-handed sleptons (L-sleptons and R-sleptons, respectively) are considered, as the different production cross sections for the two cases substantially affect the analysis sensitivity.
The assumed decays of each NLSP choice are described in \tabref{tab:RPVmodeldecay} and illustrated in \figref{fig:RPVgraphs}.
All SUSY particles are generated on shell, and forced to decay at the primary vertex.
The masses of the NLSP and LSP are varied; other sparticles are decoupled by assigning them a fixed mass of $4.5\tev$.
Direct pair production of $\ninoone\ninoone$ is not considered, as the production cross section is found to be negligible in most cases.

\begin{table}[ht]
\centering
 \caption{Sparticle decays in the SUSY RPV simplified models used in this analysis.
The neutralino LSP is assumed to decay to two charged leptons and a neutrino.
For the chargino model, the $\Wpm$ from the 
\chinoonepm{} decay may be virtual.
 \label{tab:RPVmodeldecay}}
\small{
 \begin{tabular}{l c l }
\hline
  RPV Model NLSP& & Decay   \\ 
\hline\hline
  Chargino	& & $\chinoonepm \rightarrow W^{\pm(*)} ~\ninoone$   \\ 
  L-slepton 	& & $\sleptonL \rightarrow \ell ~\ninoone$    \\ 
        	& & $\stauL \rightarrow \tau ~\ninoone$   \\ 
  R-slepton 	& & $\sleptonR \rightarrow \ell ~\ninoone$    \\ 
        	& & $\stauR \rightarrow \tau ~\ninoone$   \\ 
  Sneutrino 	& & $\snul \rightarrow \nu_{\ell} ~\ninoone$    \\ 
        	& & $\snutau \rightarrow \nu_{\tau} ~\ninoone$  \\
  Gluino 	& & $\gluino \rightarrow q \qbar ~\ninoone$     \\ 
                & & $q \in u,d,s,c$ \\
\hline\hline
 \end{tabular}
}
\end{table}

The NLSP mass ranges explored are as follows: 500--$1700\gev$ for the gluino model, 200--$1000\gev$ for the chargino model, and 75--$600\gev$ for the slepton and sneutrino models.
In each case, the choice of lower bound is guided by the limits from the previous searches at the Large Electron Positron collider (LEP) and the Tevatron; the production cross sections at those values lie between 0.4~pb (chargino and R-slepton models) and 4.5~pb (gluino model).
The upper bound is high enough that the production cross section is 0.1~fb or smaller in all cases.
For a fixed value of $m_{\mathrm{NLSP}}$, $m_{\mathrm{LSP}}$ is allowed to vary between $10$ and $m_{\mathrm{NLSP}}-10\gev$.
These lower and upper limits are designed to allow enough phase space for prompt decays of the LSP to SM particles and of the NLSP to the LSP, respectively.

\subsection{RPC simplified models \label{sec:rpc}}

Simplified models with $R$-parity conservation assume the pair production of degenerate higgsinolike \ninotwo{} and \ninothree.
These decay to a binolike \ninoone{} LSP via a cascade, resulting also in the production of charged leptons.

\begin{figure}[tb]
\centering
\subfigure[ ~R-slepton RPC]{\includegraphics[width=0.23\textwidth]{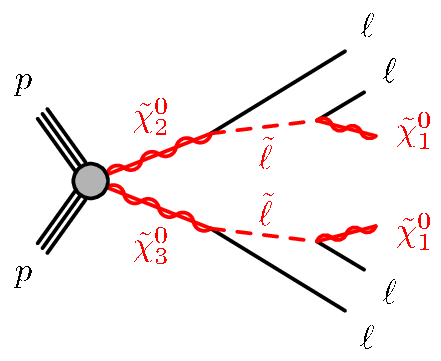}}
\subfigure[ ~Stau RPC]{\includegraphics[width=0.23\textwidth]{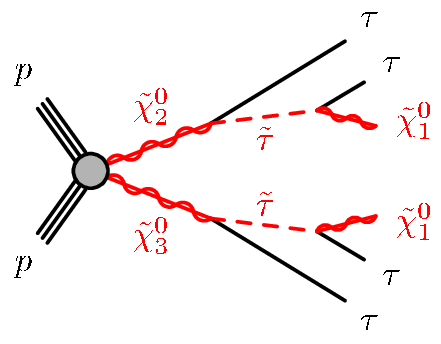}}
\subfigure[ ~\Zboson{} RPC]{\includegraphics[width=0.23\textwidth]{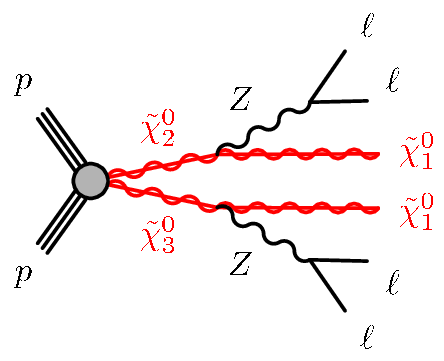}}
\caption{Representative diagrams for the RPC simplified models considered in this analysis.
\label{fig:N2N3graphs} }
\end{figure}

Three decay chains for the \ninotwo{} and \ninothree{} are considered (see also \tabref{tab:RPCmodeldecay} and \figref{fig:N2N3graphs}):
a light-lepton-rich ``R-slepton RPC'' scenario, with intermediate right-handed smuons and selectrons;
a tau-rich ``stau RPC'' scenario, with intermediate right-handed staus; and a lepton-rich ``\Zboson{} RPC'' scenario, 
with intermediate \Zboson{} bosons. 
The choice of right-handed sleptons in the decay chain ensures a high four-lepton yield, while suppressing the leptonic branching fraction of any associated chargino, thus enhancing the rate of four-lepton events with respect to events with lower lepton multiplicities.
In more realistic models, mixing occurs among the four neutralino states, leading to a small wino component.
This component ensures equal branching ratios to selectrons and smuons, as assumed in the R-slepton model.
The simplified model assumes the same neutralino branching fraction to both sleptons.

Masses between $100$ and $700\gev$ are considered for the \ninotwo{} and \ninothree, with production cross sections varying from approximately 1.7~pb to 0.2~fb over this range.
In the R-slepton model, the LSP mass is also varied, from $0$ up to $m_{\ninotwothree}-20\gev$, while in the stau and \Zboson{} models only a massless LSP is considered.
Where relevant, the masses of intermediate sparticles (sleptons and staus) in the decay chains are assumed to be the average of the \ninotwothree{} and \ninoone{} masses; all other sparticles are decoupled.

\begin{table}[tb]
\centering
 \caption{Sparticle decays in the SUSY RPC simplified models used in this analysis.
For \Zboson{} boson decays, the gauge boson may be virtual.
\label{tab:RPCmodeldecay}}
\small{
 \begin{tabular}{l c l }
\hline
  RPC Model	& & Decay   \\ 
\hline\hline
  R-slepton 	& & $\ninotwothree \rightarrow \ell^{\pm} \sleptonR^{\mp} \rightarrow \ell^{+}\ell^{-} \ninoone$  \\ 
  Stau  	& & $\ninotwothree \rightarrow \tau^{\mp} \stauone^{\pm} \rightarrow \tau^{\mp} \tau^{\pm} \ninoone$  \\ 
  \Zboson{}	& & $\ninotwothree \rightarrow  \Zboson^{(*)} \ninoone \rightarrow \ell^{\pm} \ell^{\mp} \ninoone$   \\ 
\hline\hline
 \end{tabular}
}
\end{table}

\subsection{RPC GGM SUSY models} \label{sec:RPCGGM}

In all GGM scenarios the gravitino $\gravino$ is the LSP and, 
unlike GMSB SUSY models, the colored sparticles are not required to be heavier than the electroweak sparticles,  
which allows for an enhanced discovery potential at the LHC~\cite{ggm2008,rudermanGGM}. 
The GGM parametrization uses the following principal variables: 
the bino mass $M_{1}$, the wino mass $M_{2}$, the gluino mass $M_3$, the higgsino mass parameter $\mu$, the ratio of the SUSY Higgs vacuum expectation values $\tan\beta$, and the proper decay length of the NLSP, $c\tau_{\mathrm{NLSP}}$.

\begin{figure}[tb]
\centering
\subfigure[~Weak production GGM]{\includegraphics[width=0.23\textwidth]{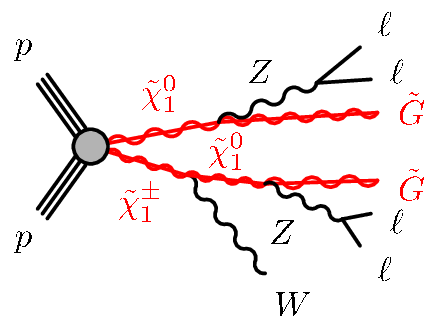}}
\subfigure[~Strong production GGM]{\includegraphics[width=0.23\textwidth]{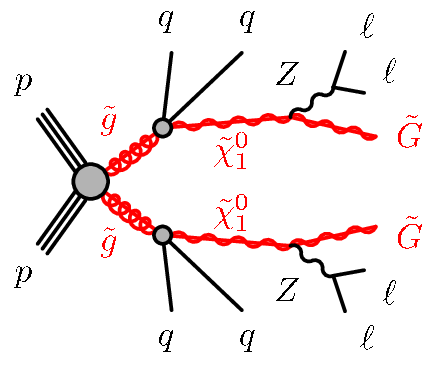}}
\caption{Representative diagrams of relevant processes for GGM models considered in this analysis.
\label{fig:GGMgraphs} }
\end{figure}

Two GGM scenarios are considered for this analysis, one with $\tan\beta\,$$=\,$1.5 and the other with $\tan\beta\,$$=\,$30.
For both it is assumed that
$M_1 = M_2 =1\tev$ and $c\tau_{\mathrm{NLSP}}\,$$<\,$0.1$\,$mm, while $\mu$ and $m_{\gluino}=M_{3}$ are varied between set values.  
As a result, both sets of models have higgsinolike $\ninoone$, $\ninotwo$ and $\chinoonepm$ co-NLSPs. 
In the $\tan\beta\,$$=\,$1.5 models, the neutralino NLSPs decay nearly exclusively (branching ratio $\sim$97\%) to a \Zboson{} boson plus a gravitino ($\ninoone \rightarrow \Zboson \gravino$), 
while in the $\tan\beta\,$$=\,$30 models the NLSP can also decay to a Higgs boson plus a gravitino 
($\ninoone \rightarrow h \gravino$), with an assumed Higgs boson mass of $125\gev$ and Higgs boson branching ratios set to those of the SM.
The branching ratio of NLSP decays to a Higgs boson ranges widely, from 0\% for 
$\mu=100\gev$ to $\sim40\%$ for $\mu=500\gev$. 
Gluino masses of up to $1.2\tev$ are considered, and the requirement $200\gev < \mu < m_{\gluino}-10\gev$ is also made, where the lower limit excludes models with nonprompt sparticle decays.
Production of strongly interacting sparticle pairs dominates across the bulk of the GGM parameter space, 
but as the gluino mass increases, production of weakly interacting sparticles becomes more important. 
Representative diagrams for the relevant processes are shown in \figref{fig:GGMgraphs}.
The total SUSY production cross section in both models varies from 1.2--1.9~pb for $m_{\gluino}=600\gev$ to 3.1~fb for the highest masses considered.
However, for $\mu=200\gev$ the cross section never falls below 0.6~pb, due to contributions from \ninoone, \chinoonepm{} and \ninotwo{} production.

\section{The ATLAS detector \label{sec:atlasdet}}

The ATLAS detector~\cite{atlas-det} is a multipurpose particle physics detector 
with forward-backward symmetric cylindrical geometry~\footnote{
  ATLAS uses a right-handed coordinate
  system with its origin at the nominal interaction point (IP) in the
  center of the detector and the $z$-axis along the beam pipe. The
  $x$-axis points from the IP to the center of the LHC ring, and the
  $y$-axis points upward. Cylindrical coordinates $(r,\phi)$ are used
  in the transverse plane, $\phi$ being the azimuthal angle around the
  beam pipe. The pseudorapidity is defined in terms of the polar angle
  $\theta$ as $\eta=-\ln\tan(\theta/2)$.
}. 
The inner tracking detector (ID) consists of a silicon pixel detector, a silicon microstrip detector, and a transition radiation tracker (TRT), 
and covers pseudorapidities of $\abseta\,$$<\,$2.5. The ID is surrounded by a thin superconducting solenoid providing a 2$\,$T axial magnetic field. 
A high-granularity lead/liquid-argon (LAr) sampling calorimeter measures the energy and the position of electromagnetic showers within $\abseta\,$$<\,$3.2. 
LAr sampling calorimeters are also used to measure hadronic showers in the end-cap (1.5$\,$$<\,$$\abseta\,$$<\,$3.2) and forward (3.1$\,$$<\,$$\abseta\,$$<\,$4.9) regions, 
while an iron/scintillator tile calorimeter measures hadronic showers in the central region ($\abseta\,$$<\,$1.7). 
The muon spectrometer (MS) surrounds the calorimeters and consists of three large superconducting air-core toroid magnets, each with eight coils, 
a system of precision tracking chambers ($\abseta\,$$<\,$2.7), and fast trigger chambers ($\abseta\,$$<\,$2.4). A three-level trigger system~\cite{Aad:2012xs}  selects events to be recorded for offline analysis.

\section{Monte Carlo simulations\label{sec:MCsamples}}

MC simulations are used to aid in the description of SM backgrounds and to model the SUSY signals.
Details of the MC generation are listed in \tabref{tab:MCsamples}.
When the parton shower is generated with HERWIG-6.520~\cite{herwig}, the underlying event is simulated by JIMMY-4.31~\cite{Butterworth:1996zw}.
All samples are processed using the full ATLAS detector simulation~\cite{simulation} based on GEANT4~\cite{Agostinelli:2002hh}, except for the \tWZ, \tZ{} and $\Wboson/\Zboson H(\to\mu\mu)$ samples, which are instead simulated with a parametrization of the performance of the ATLAS electromagnetic and hadronic calorimeters and with GEANT4 for other detector components~\cite{atlfastII}.
The effect of multiple proton-proton interactions in the same or nearby bunch crossings (pileup) is taken into account in all MC simulations,
and the distribution of the number of interactions per bunch crossing in the MC simulation is reweighted to that observed in the data.
Specific notes on some of the generated processes follow.

The \ZZ{} and \WZ{} diboson processes are simulated using POWHEG~\cite{powheg,powheg3,powheg4,powheg1}, including off shell photon contributions and internal conversion events where two leptons are produced from photon radiation in the final state.
The $gg\to\ZZ$ process is simulated separately, but does not include the $\ZZ\rightarrow 4\tau$ process, which is estimated to be negligible in the signal regions used in this analysis.
Triboson processes are also generated, including those with six electroweak vertices and a \VV+2-jet final state, where \Vboson{} is a \Wboson{} or \Zboson{} boson, as indicated in \tabref{tab:MCsamples}.
Five mechanisms are considered for SM Higgs boson production ($m_H=125\gev$ assumed) which can give rise to four or more leptons in the final state:
 gluon fusion (\ggF); vector-boson fusion (\VBF); associated production with a \Wboson{} (\Wboson$H$) or \Zboson{} boson (\Zboson$H$); 
 and associated production with a \ttbar{} pair (\ttH).
Top quark samples are generated assuming a top quark mass of $172.5\gev$.

SUSY signal cross sections are calculated to next-to-leading order (NLO) in the strong coupling constant 
using PROSPINO2~\cite{Beenakker:1996ch}.
The inclusion of the resummation of soft gluon emission at next-to-leading-logarithmic (NLL)
accuracy (NLO+NLL)~\cite{Beenakker:1996ch,Kulesza:2008jb,Kulesza:2009kq,Beenakker:2009ha,Beenakker:2011fu} 
is performed in the case of strong sparticle pair production.
For neutralino, chargino, slepton and sneutrino production, the NLO cross sections used are in agreement with the NLO+NLL calculation within $\sim$2\%~\cite{Fuks:2012qx,Fuks:2013vua,Fuks:2013lya}.
The nominal cross section and its uncertainty are taken from an envelope of cross section predictions using 
different parton density function (PDF) sets and factorization and renormalization scales, as described in Ref.~\cite{Kramer:2012bx}.
For all models, additional MC samples are generated to test how the event acceptance varies with modified initial- and final-state radiation (ISR/FSR), and renormalization and factorization scales.
MadGraph is used to generate these additional samples for the RPV and RPC simplified models, while PYTHIA-6.426~\cite{pythia} is used for the GGM models.

 \begin{table*}[ht]
 \centering
 \caption{The MC-simulated samples used in this paper. The generators and the parton shower they are interfaced to, cross section predictions used for yield normalization, tunes used for the underlying event (UE) and PDF sets are shown.
Where two PDF sets are given, the second refers to the generator used for fragmentation and hadronization.
Samples preceeded by (S) are used for systematic studies only, and ``HF'' refers to heavy-flavor jet production.
Cross sections are calculated at leading-order (LO), NLO, next-to-next-to-LO (NNLO) and next-to-next-to-leading-logarithm (NNLL) QCD precision.
Certain samples include NLO EW corrections in the calculation.
See text for further details of the event generation and simulation.
   \label{tab:MCsamples} }
 \scriptsize{
   \begin{tabular}{ ccccc }
\hline
 Process & Generator & Cross section  & UE tune   & PDF set  \\
  & + fragmentation/hadronization & calculation & & \\
\hline\hline
 {\bf Dibosons} & & & & \\
  \multirow{2}{*}{\WW, \WZ, \ZZ} & POWHEG-BOX-1.0~\cite{powheg,powheg3,powheg4,powheg1} & NLO  & \multirow{2}{*}{AU2~\cite{Pythia8tunes}} & \multirow{2}{*}{CT10~\cite{pdf:ct10}} \\
  & + PYTHIA-8.165~\cite{Sjostrand:2007gs} & with MCFM-6.2~\cite{mcfm1,mcfm2}  & & \\
      (S) \ZZ & aMC@NLO-4.03~\cite{Frederix:2011ss} & MCFM-6.2~\cite{mcfm1,mcfm2} & AUET2B~\cite{mc11ctunes} & CT10 \\
          \ZZ{} via gluon fusion & gg2ZZ~\cite{Kauer:2012hd} + HERWIG-6.520~\cite{herwig} & NLO & AUET2B & CT10/CTEQ6L1 \\
\hline
 {\bf Tribosons} & & & & \\
         \WWW, \WWZ, \ZZZ & MadGraph-5.0~\cite{madgraph} + PYTHIA-6.426~\cite{pythia} & NLO~\cite{Campanario:2008yg} & AUET2B & CTEQ6L1~\cite{CTEQ6L1} \\
         \VV + 2 jets & SHERPA-1.4.0~\cite{SherpaMC} & LO & SHERPA default & CT10 \\
\hline
 {\bf Higgs} & & & & \\
       via gluon fusion & POWHEG-BOX-1.0~\cite{Alioli:2008tz} + PYTHIA-8.165  & NNLL QCD, NLO EW~\cite{Dittmaier:2012vm} & AU2 & CT10  \\
       via vector boson fusion & POWHEG-BOX-1.0~\cite{Nason:2009ai} + PYTHIA-8.165 & NNLO QCD, NLO EW~\cite{Dittmaier:2012vm} & AU2 & CT10  \\
 associated \Wboson/\Zboson & PYTHIA-8.165 & NNLO QCD, NLO EW~\cite{Dittmaier:2012vm} & AU2 & CTEQ6L1 \\
       associated \ttbar & PYTHIA-8.165 & NLO~\cite{Dittmaier:2012vm} & AU2 & CTEQ6L1 \\
\hline
 {\bf Top+Boson} & & & & \\
           \ttbarW, \ttbarZ & ALPGEN-2.14~\cite{alpgen} + HERWIG-6.520 & NLO~\cite{ttZ, ttW} & AUET2B & CTEQ6L1 \\
           (S) \ttbarZ & MadGraph-5.0 + PYTHIA-6.426 & NLO~\cite{ttZ} & AUET2B & CTEQ6L1 \\
           \ttbarWW, \tZ, \tWZ & MadGraph-5.0 + PYTHIA-6.426 &  LO & AUET2B &  CTEQ6L1 \\
\hline
 \boldmath $\ttbar$ & POWHEG-BOX-1.0~\cite{Frixione:2007nw} + PYTHIA-6.426 & NNLO+NNLL~\cite{Cacciari:2011hy,Baernreuther:2012ws,Czakon:2012zr,Czakon:2012pz,Czakon:2013goa,Czakon:2011xx} & Perugia~2011C~\cite{Skands:2010ak} & CT10/CTEQ6L1  \\
\hline
 {\bf Single top} & & & & \\
         $t$-channel & AcerMC-38~\cite{AcerMC:2002dd} & NNLO+NNLL~\cite{xs:singletop1}  & AUET2B & CTEQ6L1 \\ 
         $s$-channel, \Wt &  MC@NLO-4.03~\cite{mcatnlo} & NNLO+NNLL~\cite{xs:singletop2,xs:singletop3} & AUET2B & CT10 \\ 
\hline
 {\bf \boldmath \Wjets, \Zjets} & & & & \\
 $M_{\ell\ell}\,$$>\,40\gev$ ($30\gev$ HF) & ALPGEN-2.14 + PYTHIA-6.426 & with DYNNLO-1.1~\cite{Catani:2009sm} & Perugia~2011C & CTEQ6L1 \\
 $10\gev\,$$<\,$$M_{\ell\ell}\,$$<\,40\gev$ & ALPGEN-2.14 + HERWIG-6.520 & with MSTW2008 NNLO~\cite{Martin:2009iq} & AUET2B & CTEQ6L1 \\
\hline
 {\bf Multijet} & PYTHIA-8.165 & LO & AU2 & CTEQ6L1 \\
\hline
 {\bf SUSY signal} & & & & \\
 RPV simplified models & HERWIG++~2.5.2~\cite{Bahr:2008pv} & See text & UE-EE-3~\cite{ueee3} & CTEQ6L1 \\
 RPC simplified models & MadGraph-5.0 + PYTHIA-6.426 & NLO, see text & AUET2B & CTEQ6L1 \\
 GGM & PYTHIA-6.426 & NLO, see text & AUET2B & CTEQ6L1 \\
\hline
     \end{tabular} 
 }       
 \end{table*}

\section{Event reconstruction and preselection \label{sec:presel}}
 
\begin{table}[ht]
\centering
 \caption{Offline $\pt$ and $\et$ thresholds used in this analysis for different trigger channels.
For dilepton triggers, the two numbers refer to the leading and subleading triggered lepton, respectively.
\label{tab:triggerThresholds}}
\small{
 \begin{tabular}{c c}
\hline
~Trigger channel~ & ~\pt{} or \et{} threshold [\gev]~ \\
\hline\hline
~Single isolated $e/\mu$~ & 25 \\
\hline
\multirow{2}{*}{Double $e$} & 14, 14 \\
                            & 25, 10 \\
\hline
\multirow{2}{*}{Double $\mu$} & 14, 14 \\
                              & 18, 10 \\
\hline
\multirow{2}{*}{$e+\mu$}      & 14($e$), 10($\mu$) \\
                              & 18($\mu$), 10($e$) \\
\hline\hline
 \end{tabular}
}
\end{table}

For all physics channels considered in this analysis, including those with one or more taus in the final state, 
events are required to pass at least one of a selection of single isolated or double electron/muon triggers.
Double lepton triggers have asymmetric or symmetric transverse momentum and energy ($\pt$ and \et) thresholds, depending on the lepton flavors involved.
Thresholds on the \pt{} or \et{} of reconstructed leptons matching the triggering objects are chosen to ensure that the trigger efficiency is high and independent of the lepton \pt{} or \et; these thresholds are listed in \tabref{tab:triggerThresholds}.
Triggering is restricted  to $\abseta\,$$<\,$2.4 and $\abseta\,$$<\,$2.47 for muons and electrons, respectively.
The overall trigger efficiency for SUSY signal events varies between approximately 80\% for events with two muons and two taus, and more than 99\% for events with four light leptons.

After applying standard data-quality requirements, events are analyzed if the primary vertex
has five or more tracks with $\pt\,$$>\,$400$\mev$ 
associated with it.
The vertex with the highest scalar sum of the squared transverse momenta of associated tracks is taken to be the primary vertex of the event.

Candidate electrons must satisfy the ``medium'' 
identification criteria, following Ref.~\cite{Aad:2014fxa} and modified for 2012 operating conditions, and have $\abseta\,$$<\,$2.47 and $\et\,$$>\,10\gev$, 
where \et{} and \abseta{} are determined from the calibrated clustered energy deposits in the electromagnetic calorimeter and the matched 
ID track, respectively. 
Muon candidates are reconstructed 
by combining tracks in the ID and the MS~\cite{ATLAS-CONF-2013-088}, and have $\abseta\,$$<\,$2.5 and \hbox{$\pt\,$$>\,10\gev$}.
The quality of the ID track associated with a muon is ensured by imposing
requirements described in Ref.~\cite{Aad:2014zya}.

Jets are reconstructed with the anti-$k_t$ algorithm~\cite{Cacciari:2008gp} 
with a radius parameter of $R\,$$=\,$0.4 using three-dimensional calorimeter energy clusters~\cite{topoclusters} as input.
The clusters are calibrated using ``local cluster weighting'' calibration, where the energy deposits 
arising from electromagnetic and hadronic showers are independently calibrated~\cite{JESuncert1}.
The final jet energy calibration corrects the calorimeter response to the true particle-level jet energy~\cite{JESuncert1,Aad:2012vm}.
The correction factors are obtained from simulation and are refined and validated using data.
An additional correction subtracts the expected contamination from pileup, calculated as a product of the jet area and the average energy density of the event~\cite{Cacciari:2008gn}.
Events containing jets failing to satisfy the quality criteria described in Ref.~\cite{JESuncert1} are 
rejected to suppress events with large calorimeter noise or noncollision backgrounds. 
Jets are required to have $\pt\,$$>\,20\gev$ and $\abseta\,$$<\,$4.5.

Jets are identified as containing a $b$-quark (``$b$-tagged'')
using a multivariate technique based on quantities such as the impact parameters of the tracks associated with a reconstructed secondary vertex. 
For this analysis, the $b$-tagging algorithm~\cite{Aad:2012qf}
is configured to achieve an efficiency of 80\% 
for correctly identifying $b$-quark jets in a simulated sample of \ttbar{} events.

Tau candidates are reconstructed using calorimeter ``seed'' jets with $\pT\,$$>\,10\gev$ and $\abseta\,$$<\,$2.47.
The tau reconstruction algorithm uses the cluster shapes in the electromagnetic and hadronic calorimeters as well as tracks within a cone of size $\Delta R\equiv \sqrt{(\Delta\phi)^2+(\Delta\eta)^2}=0.2$ of the seed jet. 
The tau energy scale is set using an $\eta$- and $\pT$-dependent calibration~\cite{tauperf}.
In this analysis, one- or three-prong tau decays are selected if they have unit charge,  $\pt\,$$>\,20\gev$, and $\abseta\,$$<\,$2.47.

To remove overlaps and resolve ambiguities between particle objects, a procedure is applied based on geometrical proximity 
using the variable $\Delta R$.
Objects are removed at each step in the procedure before moving on to the next.
If two candidate electrons are identified within $\Delta R\,$$=\,$0.05 
of each other, the lower energy electron is discarded. 
If a candidate electron and a candidate jet are within $\Delta R\,$$=\,$0.2 of each other, the jet is discarded. 
All leptons are required to be separated by more than $\Delta R\,$$=\,$0.4 from the closest remaining jet.
In the rare occurrence when a candidate electron overlaps with a candidate muon
within $\Delta R\,$$=\,$0.01, both particles are discarded since it usually means that they were reconstructed using the same track.
Similarly, if two muons are separated by less than $\Delta R\,$$=\,$0.05 then they are unlikely to be well reconstructed, and both are removed. 
Candidate taus are required to be separated by more than $\Delta R\,$$=\,$0.2 from the closest electron or muon; otherwise the tau is discarded. 

Candidate objects that are not removed by the above procedure are classified as ``baseline.'' 
``Signal'' objects are baseline objects that also satisfy additional criteria described in the following.

Signal light leptons are required to originate from the primary vertex, with a closest approach in the transverse plane of less than five (three) standard deviations and a longitudinal distance $z_0$ satisfying $|z_{0}\sin{\theta}|\,$$<\,$0.4$\,$(1.0)$\,$mm for electrons (muons)~\cite{Note1}.
Signal electrons must also satisfy the ``tight'' criteria defined in Ref.~\cite{Aad:2014fxa}, which includes requirements placed on the ratio of calorimetric energy to track momentum, and the number of high-threshold hits in the TRT.
Signal light leptons are required to be isolated from hadronic activity in the event.
Track isolation is calculated as the scalar sum of transverse momenta of tracks with $\pt>400\mev$ ($1\gev$) within a cone of radius $\Delta R\,$$=\,$0.3 around each baseline electron (muon), excluding the track of the lepton itself.
Calorimeter isolation is calculated, for electrons only, by summing the transverse energies of topological clusters within a radius of $\Delta R\,$$=\,$0.3 around the electron, and it is corrected for the effects of pileup.
In order to maintain sensitivity to some RPV scenarios with highly boosted particles, contributions to the lepton isolation from tracks or clusters of other electron and muon candidates that satisfy all signal criteria, except the isolation requirements, are removed.
The track isolation must be less than 16\% (12\%) of the electron's $\et$ (muon's $\pt$), and the calorimeter isolation for electrons must be less than 18\% of the electron's $\et$.

Signal jets are baseline jets with $\abseta<2.5$. 
Additionally, in order to suppress jets from a different interaction in the same beam bunch crossing, 
a jet with $\pt<50\gev$ is discarded if more than half of the \pt-weighted sum of its tracks does not come from the tracks which are associated with the primary vertex.
 
Signal taus must satisfy the ``medium'' identification criteria of a boosted decision tree~\cite{tauBDT} algorithm, based on various track and cluster variables 
for particle discrimination. 
Tau objects arising from misidentified electrons are discarded using a ``loose'' electron veto based on TRT and calorimeter information.
A muon veto is also applied.
If a signal tau 
and a jet are within $\Delta R\,$$=\,$0.2 of each other, the tau is kept while the jet is discarded.

The missing transverse momentum vector, $\ptmissvec$, and its magnitude, $\MET$, are calculated from the transverse momenta of 
calibrated electrons, muons, photons and jets, as well as all the topological clusters with $|\eta |<4.9$
 not associated with such objects~\cite{topoclusters,metPerf}.
Hadronically decaying taus are calibrated as jets in the \MET, which is found not to adversely affect sensitivity to SUSY events.

All particle selections are applied identically to data and to the MC events.
To account for minor differences between data and MC simulation in the electron,
muon and tau reconstruction and identification efficiencies, $\pt$- and $\eta$-dependent
scale factors derived from data in dedicated regions are applied to signal leptons.
Although $b$-tagging is not used to discriminate SUSY events from the SM background, 
it is used to compare the MC simulation of leptons arising from heavy-flavor jets to data. 
For this measurement, the $b$-tagging efficiency and mistag rates are themselves adjusted by scale factors 
derived from \ttbar{} and light-jets data in dedicated regions~\cite{ATLAS-CONF-2014-004,ATLAS-CONF-2012-040,ATLAS-CONF-2012-043}.

\section{Signal regions\label{sec:SRsel}}

\begin{table*}[ht]
\centering
 \caption{The selection requirements for the signal regions, where $\ell=e,\mu$ and ``SFOS'' indicates two same-flavor opposite-sign light leptons. The invariant mass of the candidate \Zboson{} boson in the event selection can be constructed using two or more of the light leptons present in the event: all possible lepton combinations are indicated for each signal region.
 \label{tab:SRdef}}
\small{
 \begin{tabular}{l c c c c c c }
\hline\hline
    		& ~~~N($\ell$)~~~ & ~~~N($\tau$)~~~ & \Zboson-veto & $\MET$~[\gev] & & $\meff$~[\gev]   \\ 
\hline
SR0noZa  & $\geq$4 & $\geq$0 & SFOS, SFOS+$\ell$, SFOS+SFOS & $>$50 & & -- \\
SR1noZa    & $=$3 & $\geq$1  & SFOS, SFOS+$\ell$ & $>$50 & & -- \\
SR2noZa    & $=$2 & $\geq$2  & SFOS & $>$75 & & -- \\
\hline
SR0noZb     & $\geq$4 & $\geq$0 & SFOS, SFOS+$\ell$, SFOS+SFOS & $>$75 & or & $>$600 \\
SR1noZb    & $=$3 & $\geq$1     & SFOS, SFOS+$\ell$ & $>$100 & or & $>$400 \\
SR2noZb    & $=$2 & $\geq$2     & SFOS & $>$100 & or & $>$600 \\
\hline\hline
    		& N($\ell$) & N($\tau$) & \Zboson-requirement & $\MET$~[\gev] & &   \\ 
\hline
SR0Z     & $\geq$4 & $\geq$0 & SFOS & $>$75   &  & --  \\
SR1Z     & $=$3    & $\geq$1 & SFOS & $>$100  &  & --  \\
SR2Z     & $=$2    & $\geq$2 & SFOS & $>$75   &  & --  \\
\hline\hline
 \end{tabular}
}
\end{table*}

Nine signal regions (SRs) are defined in order to give good sensitivity to the SUSY signal models considered.
The SRs require at least four leptons, and are classified depending on the number of light 
leptons required. The number of light leptons can be equal to two, three or at least four, with the corresponding number of taus 
in the same regions required to be at least two, one or zero, respectively.
Events with five or more leptons are not vetoed, to retain potential signals with higher lepton multiplicities.

The SRs are further subdivided between those vetoing against the presence of a 
\Zboson{} boson (``noZ'' regions) and those requiring the presence of one (``Z'' regions).
The noZ regions target signals from RPV and RPC simplified models, while the Z regions target the GGM and \Zboson~RPC models.
The noZ regions are further divided into ``noZa'' regions, designed to target the RPC \ninotwo\ninothree{} decays via
an $\MET$ selection, and ``noZb'' regions, optimized for RPV decays and implementing 
a combination of selections on $\MET$ and $\meff$, the latter defined as the scalar sum of the \MET, the $\pt$ of signal leptons and the $\pt$ 
of signal jets with $\pt>40\gev$.
The definitions of the different SRs are given in \tabref{tab:SRdef} and discussed in more detail below.

In four-lepton events with at least two light leptons, the dominant SM backgrounds are rich in 
\Zboson{} bosons, such as those from \ZZ{} and \Zjets{} processes. 
These can be suppressed by means of a ``\Zboson-veto,'' which rejects events where light-lepton 
combinations yield invariant mass values in the 81.2--$101.2\gev$ interval.
For events with only two light leptons, the invariant mass combination is 
unambiguously constructed
from the only possible choice, when it exists, of two same-flavor opposite-sign light leptons in the event (called an SFOS pair). 
When more than two light leptons are present, all possible SFOS pairs are considered.
To suppress radiative \Zboson{} boson decays, combinations of an SFOS pair with an additional light lepton (SFOS+$\ell$) and with a second SFOS pair (SFOS+SFOS)  are also taken into account.

For events that pass the \Zboson-veto,  
two classes of signal regions are defined: SR$x$noZa and SR$x$noZb, where $x=0,1,2$ is the minimum number of taus required. In SR$x$noZa regions, 
a relatively soft requirement on $\MET$ ($>$50--$75\gev$) provides effective rejection of 
SM backgrounds to \ninotwo\ninothree{} signals, while in SR$x$noZb regions, 
in order to improve sensitivity to signal, events are accepted if they satisfy either a moderate requirement on $\MET$ ($>$75--$100\gev$) or have a relatively large $\meff$ ($>$400--$600\gev$).

Three signal regions (SR$x$Z, where $x=0,1,2$ is the minimum number of taus required) are defined aimed at the GGM and \Zboson~RPC scenarios, all requiring the 
presence of an SFOS light-lepton pair with invariant mass in the 81.2--$101.2\gev$ mass interval.  
No attempt is made to recover radiative \Zboson{} boson decays in these regions. 
In the Z regions, an \MET{} selection is applied ($>$75--$100\gev$), to remove SM background contributions from \Zboson+$X$ events.

\section{Determination of the standard model background\label{sec:SMbkgd}}

Several SM processes can mimic a four-lepton signal. Backgrounds 
can be classified into ``irreducible'' processes (with at least four prompt leptons) 
and ``reducible'' processes (with fewer than four prompt leptons). 
``Nonprompt leptons'' include leptons originating from semileptonic decays in heavy-flavor jets or photon conversions as well as misidentified light-flavor jets.
Background events with fewer than two prompt leptons are found to be negligible using MC simulation and are not considered. 
The irreducible component of the background (\ZZ, \WWZ, \ZZZ, \tWZ, \ttbarZ/\WW, and Higgs boson decays) is estimated from simulation, 
while the relevant reducible background (\WWW, \WZ, \ttbarW; \Zjets, \ttbar, \Wt, \WW) is estimated from data using the
``weighting method.''

In the weighting method, the number of reducible background events in a given region is estimated from data using 
MC-based probabilities for a nonprompt lepton to pass or fail the signal lepton selection.
Leptons are first classified as ``loose'' or ``tight,'' 
based on isolation criteria and reconstruction quality.
Loose leptons are baseline leptons that fail any of the other requirements imposed on signal leptons.
Tight leptons coincide with signal leptons as defined previously.
The ratio $F=f/{\bar{ f}}$ for nonprompt leptons defines the ``fake ratio,'' where $f$ ($\bar f$) is the probability that a nonprompt lepton is misidentified as a tight (loose) lepton. 

For each SR, two control regions (CRs) are used for the extraction of the data-driven background predictions.
The CR definitions only differ from that of their associated SR in the quality of the required leptons: 
CR1 requires exactly three tight leptons and at least one loose lepton; 
while CR2 requires exactly two tight leptons and at least two loose leptons.

The number $N^\textrm{SR}_\textrm{red}$ of background events 
with one or two nonprompt leptons from reducible sources in each SR can then be determined from the number of events $N^\textrm{CR1}$ and $N^\textrm{CR2}$ in 
regions CR1 and CR2, respectively:

\begin{align}
\label{eq:fakes}
N^\textrm{SR}_\textrm{red} & = [N^\textrm{CR1}_\textrm{data}-N^\textrm{CR1}_\textrm{irr}]\times F \\
           & \quad -[N^\textrm{CR2}_\textrm{data}-N^\textrm{CR2}_\textrm{irr}]\times F_1\times F_2\notag,
\end{align}

\noindent
where $F$ is the uniquely defined fake ratio in CR1,
while 
$F_1$ and $F_2$ are the two fake ratios that can be constructed using the two loose leptons in CR2.
The number of irreducible background events in CR1 and CR2, $N^\textrm{CR1}_\textrm{irr}$ and $N^\textrm{CR2}_\textrm{irr}$, 
are subtracted from the corresponding number of events  seen in data, $N^\textrm{CR1}_\textrm{data}$ and $N^\textrm{CR2}_\textrm{data}$,
and the resulting quantities are subtracted from one another 
so that events with two nonprompt leptons are not double-counted. 

Fake ratios are calculated from MC simulation, separately for light-flavor jets, heavy-flavor jets (including charm) and photon conversions (electrons and taus only).
For taus, light jets are separated further into quark- and gluon-jet categories.
These categories are referred to as ``fake types.''
The fake ratios additionally depend on the lepton kinematics and the hard process producing the nonprompt lepton.
The hard processes considered are the following: \ttbar; \Zgamma{} production in association with jets; \WZ{} production; 
\ttbarZ{} production where one top quark decays hadronically; and \ZZ{} production where one lepton is either out of the acceptance or not reconstructed.
For all lepton flavors, the dependence of the fake ratio on the lepton \pt{} is taken into account.
In addition, electron fake ratios are parametrized in \abseta, while tau fake ratios include the dependence on \abseta{} and the number of associated tracks (one or three).

To account correctly for the relative abundances of fake types and production processes, a weighted average $F_\textrm{SR}$ of fake ratios 
is computed in each SR, as

\begin{equation}
\label{eq:fr}
F_\textrm{SR}=\sum_{i, j} \left(  R_\textrm{SR}^{ij} \times s^{i} \times F^{ij} \right).
\end{equation}

\noindent
The factor $R_\textrm{SR}^{ij}$ is a ``process fraction'' that depends on the process and fake type, 
which in each SR gives the fraction of nonprompt leptons of fake type $i$ originating from process category $j$,
while $F^{ij}$ is the corresponding fake ratio, and the scale factor $s^{i}$ is a correction that depends on the fake type,
as explained below.

The process fractions are obtained from four-lepton MC events, appropriately
taking into account the four-lepton yields and how the \met{} and \meff{} selection efficiency depends on the process and fake type in the SR where the 
process fraction is calculated.
Systematic uncertainties arising from the modeling of process fractions are estimated by varying the nonprompt lepton abundances for each fake type and process by a factor of two.

Scale factors are applied to the fake ratios to account for possible differences between data and simulation.
These are assumed to be independent of the physical process, and are determined from data
in dedicated regions enriched in objects of a given fake type.

For nonprompt light leptons from heavy-flavor jets, the scale factor is measured in 
a $b\bar{b}$-dominated control sample, which selects events with only one 
$b$-tagged jet containing a muon, and an additional baseline light lepton. 
The scale factors are found to be $0.69\pm0.05$ and $0.84\pm 0.11$ for 
electrons and muons respectively, where both the statistical and systematic uncertainties are included.
The systematic uncertainty, for these and other measured scale factors, arises from uncertainties in the subtraction of the background from the selected region and variation of the selection criteria used to define the region.
For taus, the heavy-flavor scale factor cannot be reliably measured using data. Instead, it is assumed to vary within the same range as for other measured scale factors, and a value of 1.0$\pm$0.2 is used.

The scale factor for nonprompt taus originating from light-flavor jets is measured separately for one- and three-prong tau decays 
as a function of $\pt$ and $\eta$, in a 
\Wboson+jets-dominated control sample, where events with one muon with 
$\pt \,$$>\,25\gev$ and one baseline tau are selected, and events with $b$-tagged jets 
are vetoed to suppress heavy-flavor contributions.
The scale factors are close to unity (0.89--1.06, with uncertainties between 0.03 and 0.06) in the lowest \pt{} bin (20--$30\gev$), and decrease to between 0.5 and 0.6 at high \pt{} [$\mathcal{O}(100\gev)$].

For electron candidates originating from photon conversions, the scale factor is determined in a sample of photons 
from final-state radiation of \Zboson{} boson decays to muon pairs.
The scale factor is found to be $1.11\pm0.07$, 
where both the statistical and systematic uncertainties are included. 
For taus, a scale factor from photon conversions of 1.0$\pm$0.2 is applied, as in the case of the heavy-flavor correction.

For the processes considered, the most common fake types are misidentified light-flavor jets in the case of taus, 
while for light leptons the fake types are typically dominated by nonprompt leptons in heavy-flavor jets. 
The fake ratios have in general a significant dependence on the lepton \pt. 
The \pt-averaged fake ratios  are in the range 0.01--0.18 (0.09--0.24) for electrons (muons) 
and 0.02--0.15 (0.004--0.04) for one-prong (three-prong) tau decays.

\section{Background Model Validation \label{sec:validation}}

Before data is inspected in the SRs, the adequacy of the reducible background model is tested 
by verifying agreement between data and SM background expectations. 
Six validation regions (VRs) are introduced for this purpose, defined by the selections listed in \tabref{tab:VRnames}.
They use the same selection criteria as for the corresponding SRs, except that either one or both of \MET{} and \meff{} must lie below some predefined value,
to ensure that SRs and VRs do not overlap and that signal contamination in the VRs is minimal.
In VRs applying a \Zboson-veto, it is required that \MET$\,<\,50\gev$ and \meff$\,<\,400\gev$,
while in VRs with a \Zboson{} boson requirement only \met$\,<\,50\gev$ is applied.
The reducible background, which is significant in the one- and two-tau signal regions, has a similar composition in the SRs and the corresponding VRs.
On the other hand, the irreducible background can be substantially different between SRs and VRs, due to processes with genuine \met{} (especially \ttbarZ), which are significant in the SRs but negligible in the VRs.
Therefore the VRs are primarily used to validate the model for the reducible background estimation, as well as to test the \ZZ{} MC simulation.
It was verified that contamination in the VRs from the considered SUSY models is not significant.

\begin{table*}[ht]
  \centering
    \caption{Summary of the selection requirements that define the six validation regions used in the analysis. 
\label{tab:VRnames}}

    \small{
 \begin{tabular}{l c c c c c }
\hline\hline
             & ~~~N($\ell$)~~~ & ~~~N($\tau$)~~~ & \Zboson{}-veto & ~~~$\MET$~[\gev]~~~ &  ~~~$\meff$~[\gev]~~~   \\ 
\hline
VR0noZ  & $\geq$4  & $\geq$0 & SFOS, SFOS+$\ell$, SFOS+SFOS   & $<$50 & $<$400\\
VR1noZ  & $=$3     & $\geq$1 & SFOS, SFOS+$\ell$              & $<$50 & $<$400 \\
VR2noZ  & $=$2     & $\geq$2 & SFOS                           & $<$50 & $<$400 \\
\hline\hline
                & N($\ell$) & N($\tau$) & \Zboson{}-requirement & $\MET$~[\gev] &    \\ 
\hline
VR0Z      & $\geq$4  & $\geq$0 & SFOS & $<$50 & -- \\
VR1Z      & $=$3     & $\geq$1 & SFOS & $<$50 & -- \\
VR2Z      & $=$2     & $\geq$2 & SFOS & $<$50 & -- \\
\hline\hline
 \end{tabular}
    }
\end{table*}

The background model adopted in the VRs is the same as in the SRs, with
the irreducible background obtained from MC simulation and the reducible background estimated using the weighting method.
The irreducible background in the VRs is dominated by \ZZ,
\Zjets{} and \WZ{} processes, depending on tau multiplicity.  
Observed and expected event yields in each VR are shown in \tabref{tab:VRyields}, together with the corresponding CL$_b$ value~\cite{CLs}.
Perfect agreement between expected and observed yields corresponds to a CL$_b$ value of 0.5, while values approaching 0 or 1 indicate poor agreement.
Good agreement between data and SM background predictions 
is observed in all regions, within statistical and systematic uncertainties (which are discussed in \Secref{sec:systematics}).

\begin{table*}[ht]
  \centering
    \caption{Observed and expected event yields in the six validation regions. Both the statistical and systematic uncertainties are included, also taking into 
account correlations between irreducible and reducible backgrounds. The CL$_b$ value is also quoted for each region.
    \label{tab:VRyields}}
\renewcommand\arraystretch{1.3}
\footnotesize
    \begin{tabular}{ c   ccccccc  rr }
\hline
    & \ZZ & \tWZ & \ttbarZ & \VVV & Higgs & Reducible & $\Sigma $ SM & Data & CL$_b$   \\
\hline\hline
VR0noZ & $3.6\pm0.7$  & $0.017\pm0.010$  & $0.034^{+0.036}_{-0.033}$  & $0.090^{+0.032}_{-0.033}$  & $0.18\pm0.13$  & $0.5^{+0.4}_{-0.5}$  & $4.4\pm0.9$  & $3\phantom{.0}$  & $0.29$  \\
VR1noZ & $1.43\pm0.27$  & $0.010\pm0.006$  & $0.033\pm0.022$  & $0.071\pm0.029$  & $0.28\pm0.19$  & $7.1^{+1.8}_{-1.7}$  & $8.9^{+1.8}_{-1.7}$  & $7\phantom{.0}$  & $0.31$  \\
VR2noZ & $1.53^{+0.18}_{-0.17}$  & $0.007\pm0.004$  & $0.025^{+0.031}_{-0.025}$  & $0.051\pm0.020$  & $0.29\pm0.13$  & $33.2^{+3.3}_{-7.3}$  & $35.1^{+3.4}_{-7.4}$  & $32\phantom{.0}$  & $0.37$  \\
VR0Z & $184^{+20}_{-19}$  & $0.13\pm0.07$  & $1.2\pm0.6$  & $2.13\pm0.33$  & $4.7\pm3.4$  & $0.5^{+3.1}_{-0.5}$  & $193^{+21}_{-19}$  & $216\phantom{.0}$  & $0.81$  \\
VR1Z & $8.8\pm0.9$  & $0.039\pm0.021$  & $0.28\pm0.11$  & $0.19\pm0.08$  & $0.63\pm0.16$  & $21\pm4$  & $31\pm4$  & $32\phantom{.0}$ & $0.55$   \\
VR2Z & $8.2^{+1.0}_{-1.0}$  & $0.0027\pm0.0021$  & $0.09^{+0.12}_{-0.09}$  & $0.069\pm0.013$  & $0.61\pm0.14$  & $90^{+8}_{-22}$  & $99^{+8}_{-22}$  & $101\phantom{.0}$  & $0.54$  \\
\hline
   \end{tabular}
\end{table*}

\begin{figure*}[tb]
\centering
\subfigure[~VR0Z]{\includegraphics[width=0.45\textwidth]{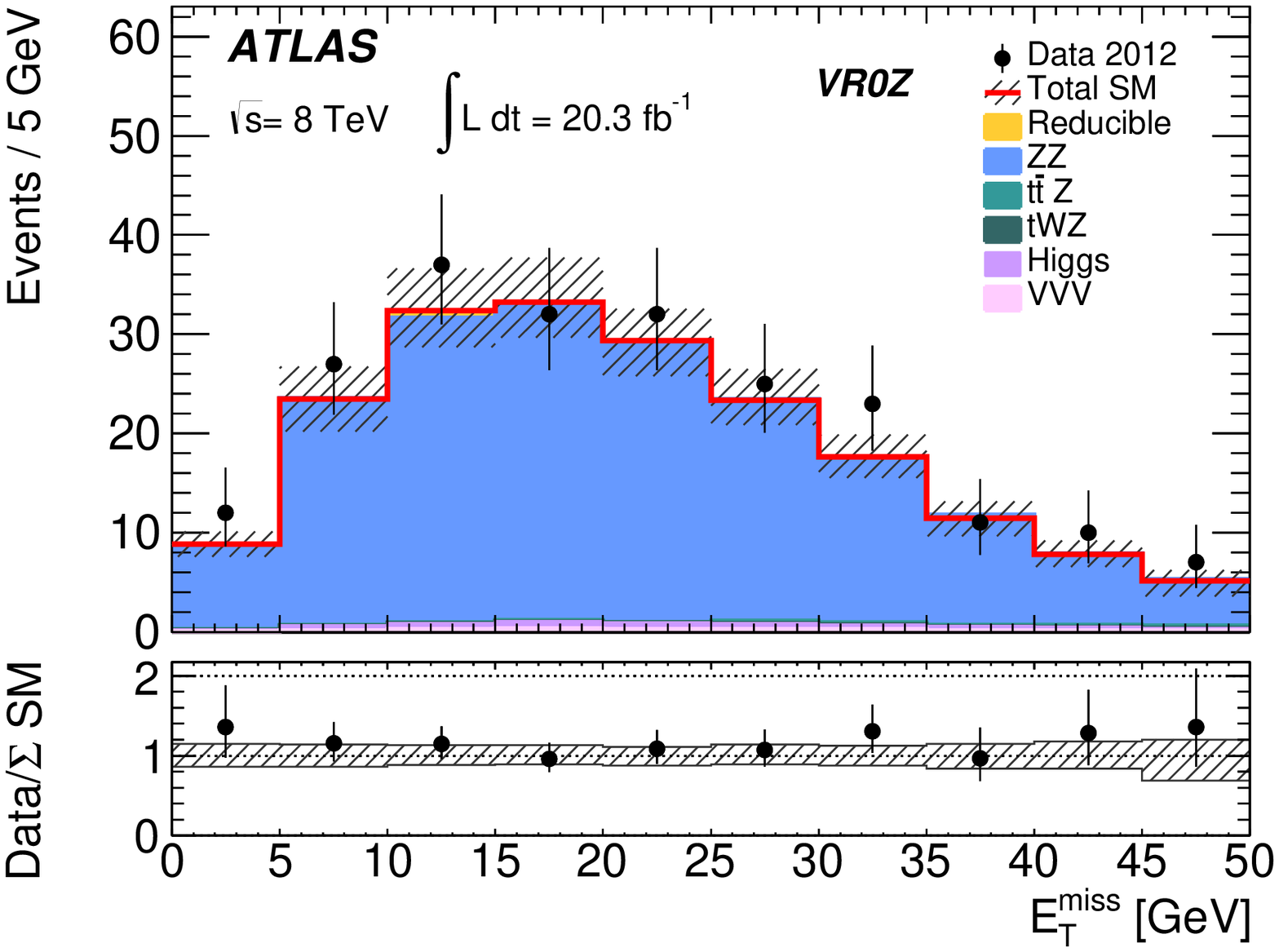}}
\subfigure[~VR0Z]{\includegraphics[width=0.45\textwidth]{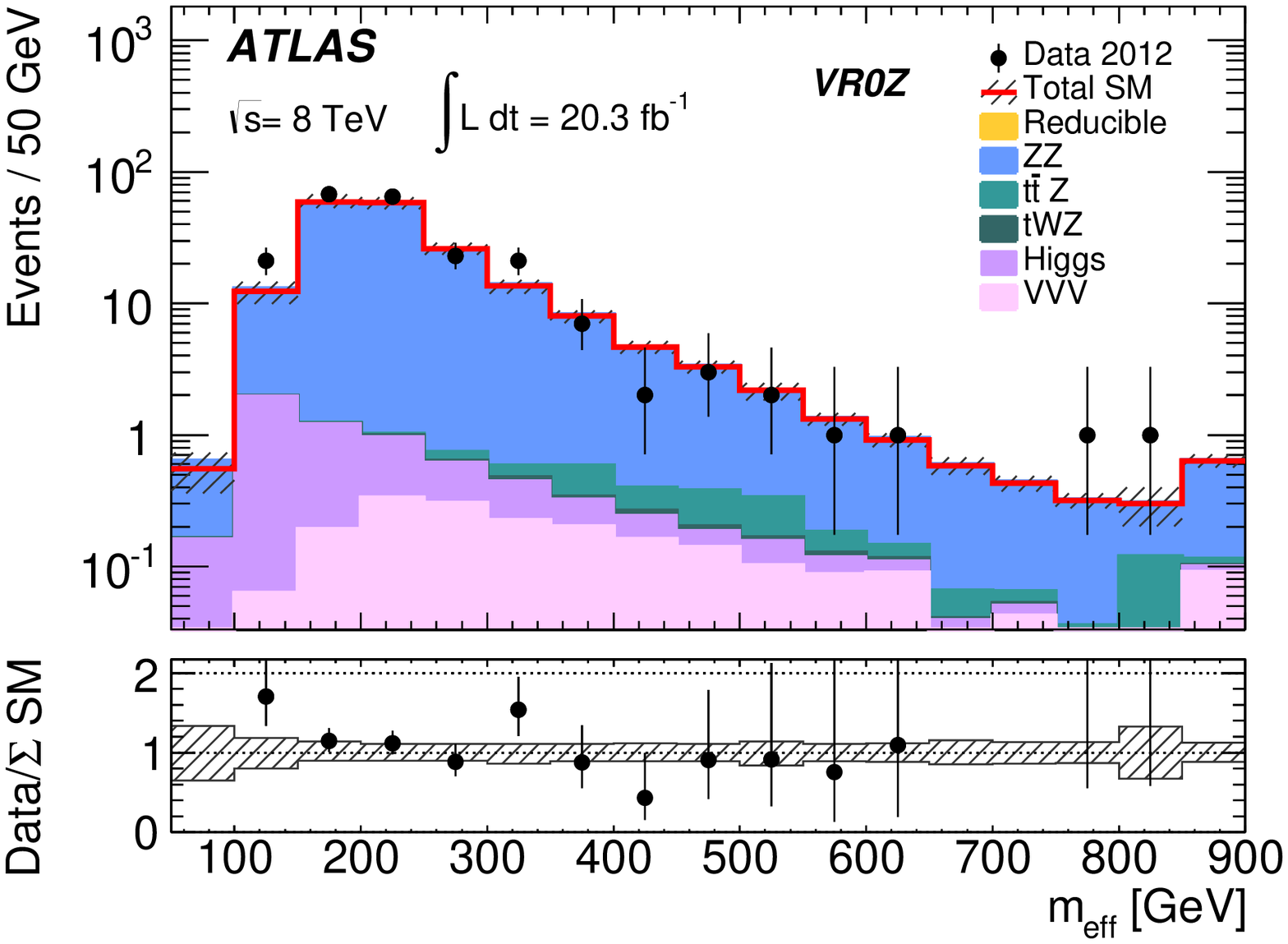}}
\subfigure[~VR2Z]{\includegraphics[width=0.45\textwidth]{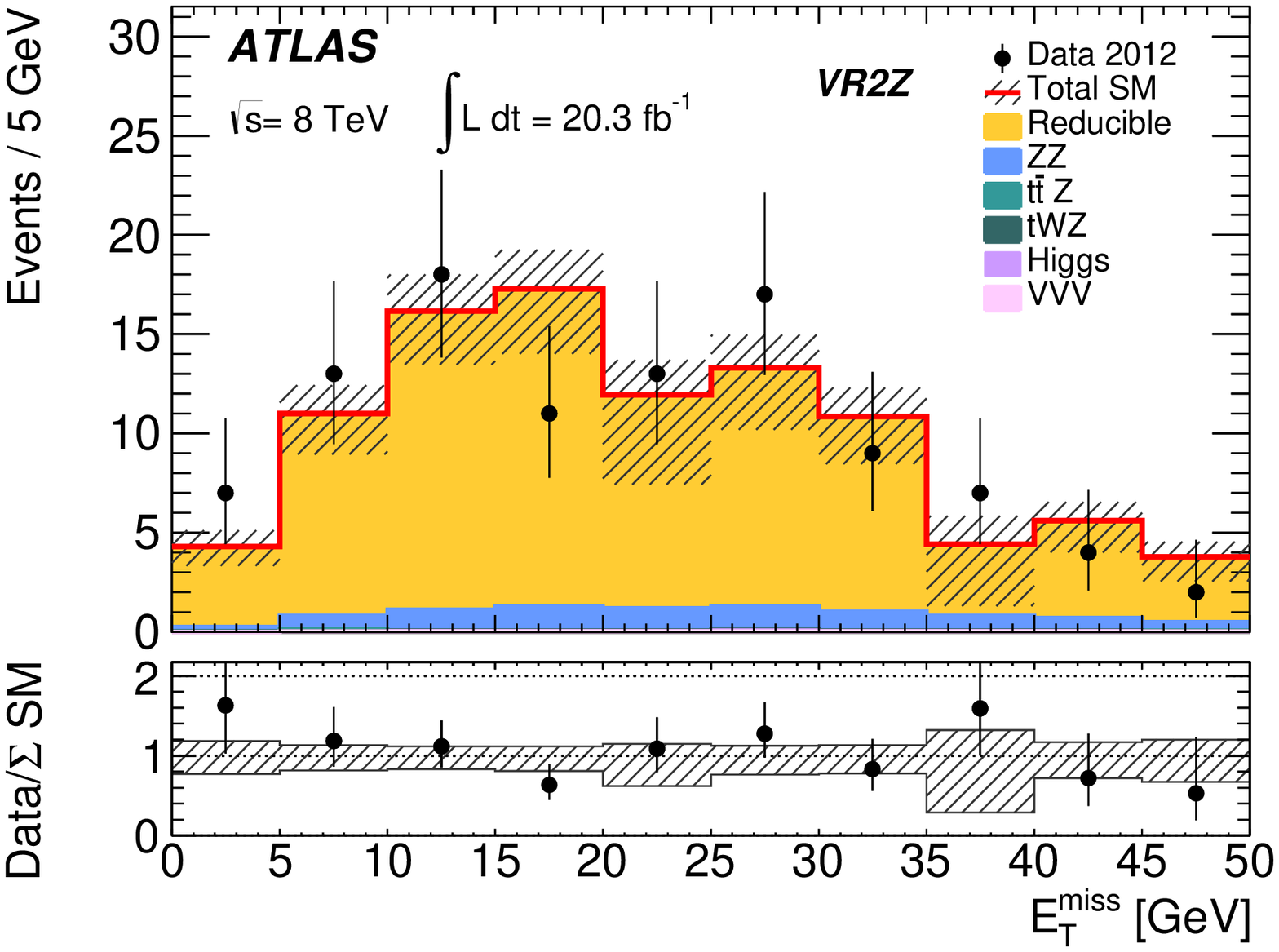}}
\subfigure[~VR2Z]{\includegraphics[width=0.45\textwidth]{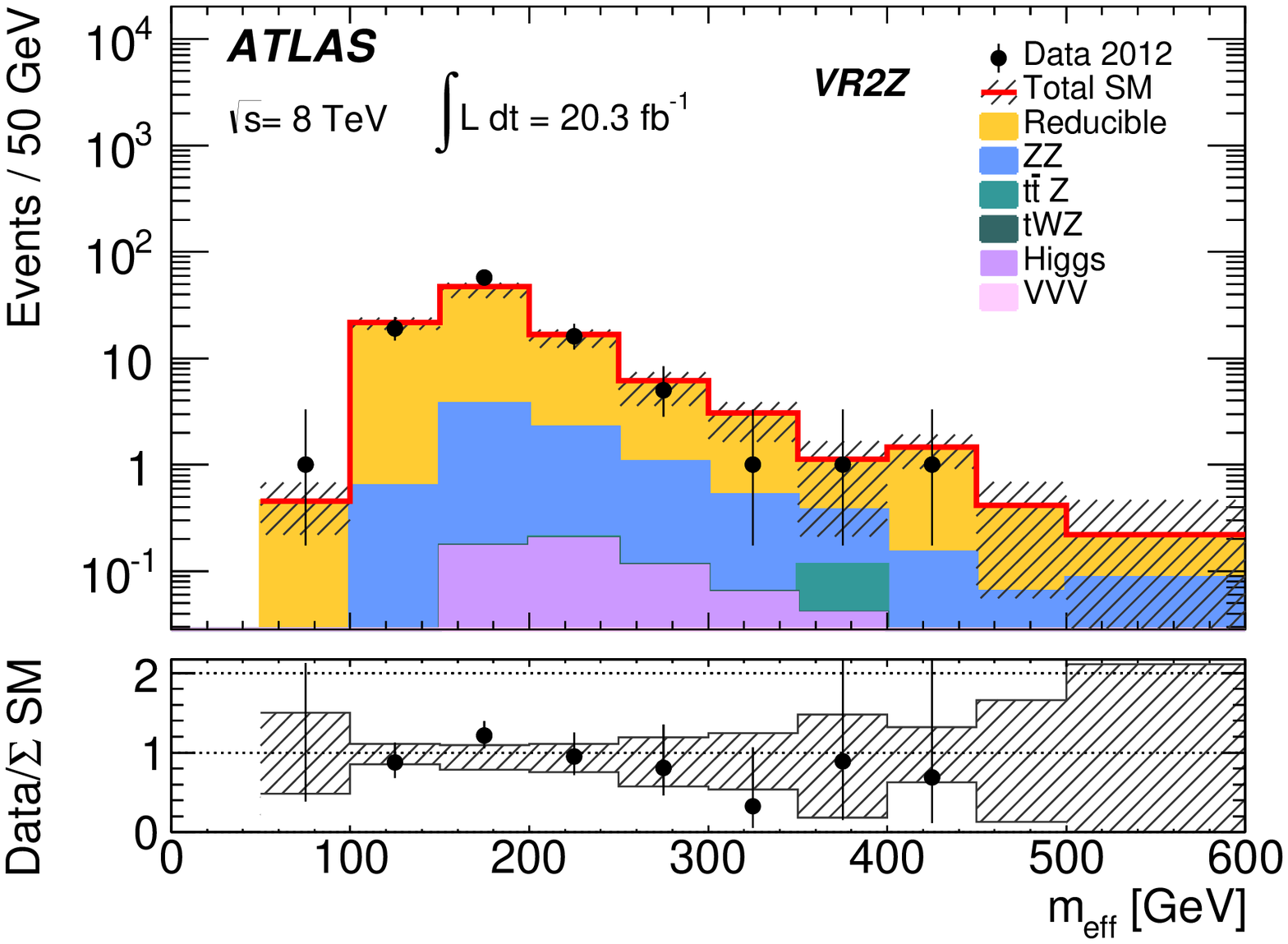}}
\caption{
The  (a),(c) \MET~and (b),(d) \meff~distributions for data and the estimated SM backgrounds, in validation regions VR0Z and VR2Z.
Both the statistical and systematic uncertainties are included in the shaded uncertainty band.
Underneath each plot, the ratio of the observed data to the SM prediction is shown, for comparison with the background uncertainty.
\label{fig:VRmetmeff} }
\end{figure*}

The \MET{} distributions in VR0Z and VR2Z are shown in \figsref{fig:VRmetmeff}(a) and \ref{fig:VRmetmeff}(c), while the \meff{} distributions in the 
same regions are shown in \figsref{fig:VRmetmeff}(b) and \ref{fig:VRmetmeff}(d). 
VR0Z is dominated by irreducible backgrounds, in particular \ZZ{} events, with smaller contributions from Higgs boson and triboson processes,
while VR2Z receives significant contributions from reducible backgrounds, as well as from \ZZ{} events.
In both cases, the shapes of the \MET{} and \meff{} distributions are well described by the background estimate.
Distributions are not shown for other VRs, where event yields are low. 

The \ttbarZ{} process is a significant component of the estimated background in the zero-tau signal regions, but it is small in all validation regions.
The MC simulation of this process was tested in Ref.~\cite{Aad:2014mha} and found to predict the rate of the process well.
Therefore, the MC prediction is used in this analysis, without further correction.

\section{Systematic uncertainties \label{sec:systematics}}

\begin{table}[ht]
\centering
 \caption{Principal experimental and theoretical systematic uncertainties for the irreducible and reducible background estimation.
For experimental uncertainties, the largest value in any SR is quoted.
For theoretical uncertainties, $\sigma$ indicates an uncertainty on the production cross section,
while $A\epsilon$ indicates an uncertainty on the product of acceptance and efficiency.
The uncertainty on the reducible background is indicated as a function of the number of taus
required in the final state.
 \label{tab:SystUncert}}
\small{
 \begin{tabular}{l c | l c }
\hline
  \multicolumn{2}{c|}{Experimental} & \multicolumn{2}{c}{Theoretical} \\ 
\hline\hline
Jet energy scale & 2.4\% & $\sigma$: \ttbarZ/\WW~\cite{ttW,ttZ} & 30\% \\
\multicolumn{2}{l|}{Jet energy resolution} & $A\epsilon$: \ttbarZ &
30--40\% \\
 & 5.5\% & $\sigma$: \ZZ & 5\% \\
$e$ efficiency & 3.5\% & $A\epsilon$: \ZZ & 5--20\% \\
$\tau$ efficiency & 3.3\% & $\sigma$: \VVV/\tWZ & 50\% \\
\MET{} energy scale & 2.7\% & $\sigma A\epsilon$: \VH/\VBF~\cite{Dittmaier:2012vm} & 20\% \\
\MET{} resolution & 2.7\% & $\sigma A\epsilon$: \ggF/\ttH~\cite{Dittmaier:2012vm} & 100\% \\
\cline{3-4} \cline{3-4}
Luminosity & 2.8\% & \multicolumn{2}{c}{Reducible}  \\
\hhline{~~|==}
Trigger & 5\% & $\geq$0$\tau$ SRs & $\sim100\%$ \\
MC sample size & $\lesssim30\%$ & $\geq 1\tau /2 \tau$ SRs & 30--50\% \\
\hline\hline
 \end{tabular}
}
\end{table}

Several sources of systematic uncertainty are considered for the SM background estimates and signal yields.
In the zero-tau signal regions, the background is dominated by the irreducible component, and systematic uncertainties are dominated by theoretical 
uncertainties and by uncertainties stemming from the limited event counts in relevant MC samples.
Moving to higher tau multiplicities, systematic uncertainties on the reducible backgrounds (mainly arising from nonprompt taus) become dominant.
Correlations of systematic
uncertainties between processes and signal/control regions are taken into account when calculating the final uncertainties.
The primary systematic sources, described below, are summarized in \tabref{tab:SystUncert}.

Experimental systematic uncertainties on the jet energy scale (JES) and resolution are determined using \emph{in situ} techniques~\cite{JESuncert1,Aad:2012vm}.
The JES uncertainty includes uncertainties from the quark-gluon composition of the jets, the heavy-flavor fraction and pileup.
Uncertainties on the lepton identification efficiencies, energy scales and resolutions are determined using $\Zboson\to\ell\ell$ events in data, where $\ell=e$, $\mu$ or $\tau$~\cite{Aad:2014fxa,elePerf,ATLAS-CONF-2013-088,tauperf}.
Uncertainties on object momenta are propagated to the \MET{} measurement, and 
additional uncertainties on \MET{} arising from energy deposits not associated with any reconstructed objects are also included.
The uncertainty
on the luminosity is 2.8\%~\cite{Aad:2011dr}.
A 5\% uncertainty is applied to MC samples to cover differences in efficiency observed between the trigger in data and the MC trigger
simulation.

The relative uncertainty on the irreducible background 
is approximately 30--50\% in the noZ signal regions, decreasing to 15--25\% in the Z regions.
It is dominated by theoretical uncertainties in the cross sections and by uncertainties in the MC modeling of the 
irreducible processes.
Theoretical uncertainties in the SM cross sections include PDF 
uncertainties, estimated using variations of appropriate PDF sets, and uncertainties in the QCD modeling, 
estimated by varying the factorization and renormalization scales individually by factors of one half and two. 
Uncertainties on the kinematic acceptance of \MET{} and \meff{} selections arising from the choice of MC generator 
are estimated by comparisons between POWHEG and aMC@NLO for 
\ZZ{} processes, and between ALPGEN and MadGraph for \ttbarZ.
Uncertainties on the acceptance are not considered for the \VVV{} and \tWZ{} processes, which represent a small contribution to the SR yields.
Uncertainties arising from the choice of generator are approximately 5--20\% for \ZZ{} processes, and 30--40\% for \ttbarZ{} in SRs with no taus required, where this background is important.

Uncertainties on the background estimate due to limited statistics of the MC-simulated samples range from a few percent up to 20--30\%.

Relative uncertainties on the reducible backgrounds, as extracted from the weighting method,
are of the order of 100\% in all zero-tau signal regions, and in the range 
of approximately 30--45\% (35--50\%) in regions with at least one (at least two) taus in the final state. 
They are dominated by the systematic uncertainties on the weighting method and statistical uncertainties in the data control regions.
The systematic uncertainties include results of a closure test where the weighting method was applied to MC-simulated events 
and compared with the MC reducible background estimation, as well as uncertainties on the fake ratios.

Systematic uncertainties on the SUSY signal yields from experimental sources typically lie in the 5--20\% range. 
They are usually dominated by the uncertainty on the electron identification and reconstruction efficiency, the electron energy scale, the JES, and the \MET{} energy scale and resolution. 
They include the uncertainties on the signal acceptance, which are typically of the order of a few percent and usually smaller than 10\%. 
The effect of ISR/FSR uncertainties on the signal acceptance is estimated by comparing samples generated with different amounts of ISR/FSR. 
Theoretical uncertainties on cross sections are typically of the order of 10\% but can reach values of approximately 30--40\% for gluino production. 
Uncertainties due to limited statistics of the MC-simulated samples are usually less than 20--30\%.

\section{Results \label{sec:Results}}

The number of events observed in each signal region is reported in \tabref{tab:SRdata}, together with background predictions.
Upper limits at 95\% confidence level (CL) on the number of events originating from beyond-the-SM (BSM) phenomena for each signal region are derived using the CL$_s$ prescription~\cite{CLs} and neglecting any possible signal contamination in the control regions.
These limits are calculated in a profile likelihood fit~\cite{cowan}, where the number of events observed in the signal region is added as an input to the fit, and an additional parameter for the strength of any BSM signal, constrained to be non-negative, is derived from the fit.
All systematic uncertainties and their correlations are taken into account via nuisance parameters in the fit.
By normalizing the limits by the integrated luminosity of the data sample, they can be interpreted as upper limits on the visible BSM cross section, $\sigma_{\mathrm{vis}}$,
defined as the product of acceptance, reconstruction efficiency and production cross section.
The results of both the asymptotic calculations~\cite{cowan} and pseudoexperiments for $\sigma_{\mathrm{vis}}$ are given in \tabref{tab:ModelIndStatInt}.
In addition, the probability ($p_0$) that a background-only experiment is more signal-like than the observation is quoted for each region, as well as the significance of upward fluctuations.
Where the observed number of data events is lower than the background prediction, $p_0$ is truncated at 0.5 and no significance is quoted.
No significant deviation is found from SM expectations in any of the signal regions, within statistical and systematic uncertainties.
The model-independent limits on $\sigma_{\mathrm{vis}}$ all lie below 0.5~fb.

\begin{table*}[ht]
 \centering
  \caption{The number of data events observed in each signal region, together with background predictions in the same regions. 
  Quoted uncertainties include both the statistical and systematic uncertainties, taking into account correlations. 
Where a negative uncertainty reaches down to zero predicted events, it is truncated.
\label{tab:SRdata}}
\footnotesize
\renewcommand\arraystretch{1.3}
\setboolean{blind}{false}
\ifthenelse{\boolean{blind}}{
\begin{tabular}{ c   ccccccc }\hline
}{
\begin{tabular}{ c   ccccccc  c }\hline
}
   & \ZZ & \tWZ & \ttbarZ & \VVV & Higgs & Reducible & $\Sigma $ SM 
\ifthenelse{\boolean{blind}}{}{& Data }
\\
\hline\hline
SR0noZa & $0.29\pm0.08$  & $0.067\pm0.033$  & $0.8\pm0.4$  & $0.19\pm0.09$  & $0.27\pm0.23$  & $0.006^{+0.164}_{-0.006}$  & $1.6\pm0.5$  
\ifthenelse{\boolean{blind}}{}{& $3$ }
\\
SR1noZa & $0.52\pm0.07$  & $0.054\pm0.028$  & $0.21\pm0.08$  & $0.14\pm0.07$  & $0.40\pm0.33$  & $3.3^{+1.3}_{-1.1}$  & $4.6^{+1.3}_{-1.2}$  
\ifthenelse{\boolean{blind}}{}{& $4$ }
\\
SR2noZa & $0.15\pm0.04$  & $0.023\pm0.012$  & $0.13\pm0.10$  & $0.051\pm0.024$  & $0.20\pm0.16$  & $3.4\pm1.2$  & $4.0^{+1.2}_{-1.3}$  
\ifthenelse{\boolean{blind}}{}{& $7$  }
\\
SR0noZb & $0.19\pm0.05$  & $0.049\pm0.024$  & $0.68\pm0.34$  & $0.18\pm0.07$  & $0.22\pm0.20$  & $0.06^{+0.15}_{-0.06}$  & $1.4\pm0.4$
\ifthenelse{\boolean{blind}}{}{& $1$ }
\\
SR1noZb & $0.219^{+0.036}_{-0.035}$  & $0.050\pm0.026$  & $0.17\pm0.07$  & $0.09\pm0.04$  & $0.30\pm0.26$  & $2.1^{+1.0}_{-0.9}$  & $2.9^{+1.0}_{-0.9}$
\ifthenelse{\boolean{blind}}{}{& $1$  }
\\
SR2noZb & $0.112^{+0.025}_{-0.024}$  & $0.016\pm0.009$  & $0.27^{+0.28}_{-0.27}$  & $0.040\pm0.018$  & $0.13\pm0.12$  & $2.5^{+0.9}_{-1.0}$  & $3.0\pm1.0$
\ifthenelse{\boolean{blind}}{}{& $6$  }
\\
\hline
SR0Z & $1.09^{+0.26}_{-0.21}$  & $0.25\pm0.13$  & $2.6\pm1.2$  & $1.0\pm0.5$  & $0.60^{+0.22}_{-0.21}$  & $0.00^{+0.09}_{-0.00}$  & $5.6\pm1.4$
\ifthenelse{\boolean{blind}}{}{& $7$  }
\\
SR1Z & $0.59^{+0.11}_{-0.10}$  & $0.042\pm0.022$  & $0.41\pm0.19$  & $0.22\pm0.11$  & $0.14\pm0.05$  & $1.0\pm0.5$  & $2.5\pm0.6$ 
\ifthenelse{\boolean{blind}}{}{& $3$  }
\\
SR2Z & $0.70^{+0.12}_{-0.11}$  & $0.0018\pm0.0015$  & $0.035\pm0.024$  & $0.039\pm0.014$  & $0.14^{+0.04}_{-0.05}$  & $0.9\pm0.5$  & $1.8\pm0.5$
\ifthenelse{\boolean{blind}}{}{& $1$  }
\\
\hline
\end{tabular}
\end{table*}

\begin{table*}[ht]
\centering
\caption{
Observed and expected $95$\% CL upper limits on the number of signal events ($N_{\mathrm{BSM}}^{\mathrm{obs}}$ and $N_{\mathrm{BSM}}^{\mathrm{exp}}$, respectively), and observed and expected $95$\% CL upper limits
on the visible cross section ($\sigma_{\mathrm{vis}}^{\rm obs}$ and $\sigma_{\mathrm{vis}}^{\rm exp}$, respectively) for
each of the signal regions.
The probability ($p_0$) that a background-only experiment is more signal-like than the observation (truncated at 0.5) and, when $p_0<0.5$, the significance of the difference between the observed data and the expectation expressed as a number of standard deviations ($N_\sigma$) are also given. 
The asymptotic calculation [marked ``(asym.)''] of the results for $\sigma_{\mathrm{vis}}$ is included for comparison with the results using pseudoexperiments.
The number of observed data events and expected background events in each region is also repeated from \tabref{tab:SRdata} for completeness.
\label{tab:ModelIndStatInt}}
\footnotesize
  \renewcommand\arraystretch{1.3}
\begin{tabular}{ c   cc cccccc }
\hline

   & $\Sigma$ SM & Data  & ~$N_{\mathrm{BSM}}^{\mathrm{obs}}$~ & ~$N_{\mathrm{BSM}}^{\mathrm{exp}}$~ & ~$\sigma_{\mathrm{vis}}^{\rm obs}$[fb] (asym.)~ & ~$\sigma_{\mathrm{vis}}^{\rm exp}$[fb] (asym.)~ & ~$p_0$~  & ~$N_\sigma$~ \\
\hline\hline
SR0noZa &$1.6\pm0.5$             & $ 3 $ &  5.9 & $4.4_{-1.0}^{+1.6}$ & 0.29 (0.29) & $0.22_{-0.05}^{+0.08}$ ($0.21^{+0.12}_{-0.07}$) & 0.15 & 1.02\\
SR1noZa &$4.6 ^{+ 1.3} _{- 1.2}$ & $ 4 $ &  5.7 & $5.9_{-1.5}^{+2.5}$ & 0.28 (0.27) & $0.29_{-0.07}^{+0.12}$ ($0.30^{+0.15}_{-0.09}$) & 0.50 & $-$ \\
SR2noZa &$4.0 ^{+ 1.2} _{- 1.3}$ & $ 7 $ &  9.2 & $6.1_{-1.4}^{+2.5}$ & 0.45 (0.45) & $0.30_{-0.07}^{+0.12}$ ($0.31^{+0.15}_{-0.09}$) & 0.13 & 1.14\\
SR0noZb &$1.4\pm0.4$             & $ 1 $ &  3.7 & $3.9\pm1.4$         & 0.18 (0.17) & $0.19\pm0.07$          ($0.19^{+0.11}_{-0.07}$) & 0.50 & $-$ \\
SR1noZb &$2.9 ^{+ 1.0} _{- 0.9}$ & $ 1 $ &  3.5 & $4.7_{-1.2}^{+1.9}$ & 0.17 (0.17) & $0.23_{-0.06}^{+0.09}$ ($0.24^{+0.13}_{-0.08}$) & 0.50 & $-$\\
SR2noZb &$3.0\pm1.0$             & $ 6 $ &  8.7 & $5.6_{-1.3}^{+2.3}$ & 0.43 (0.43) & $0.28_{-0.06}^{+0.11}$ ($0.28^{+0.14}_{-0.09}$) & 0.10 & 1.30\\
\hline
SR0Z &$5.6\pm1.4$                & $ 7 $ &  8.1 & $6.7_{-1.6}^{+2.7}$ & 0.40 (0.40) & $0.33_{-0.08}^{+0.13}$ ($0.34^{+0.16}_{-0.10}$) & 0.29 & 0.55\\
SR1Z &$2.5\pm0.6$                & $ 3 $ &  5.3 & $4.7_{-1.1}^{+1.9}$ & 0.26 (0.26) & $0.23_{-0.05}^{+0.09}$ ($0.23^{+0.13}_{-0.08}$) & 0.34 & 0.40\\
SR2Z &$1.8\pm0.5$                & $ 1 $ &  3.5 & $4.1_{-0.8}^{+1.7}$ & 0.17 (0.17) & $0.20_{-0.04}^{+0.08}$ ($0.21^{+0.12}_{-0.07}$) & 0.50 & $-$\\

\hline
    \end{tabular}
   \end{table*}

\begin{figure*}[htp]
 \begin{center}
  \subfigure[~SR0noZa]{   \includegraphics[width=0.45\textwidth]{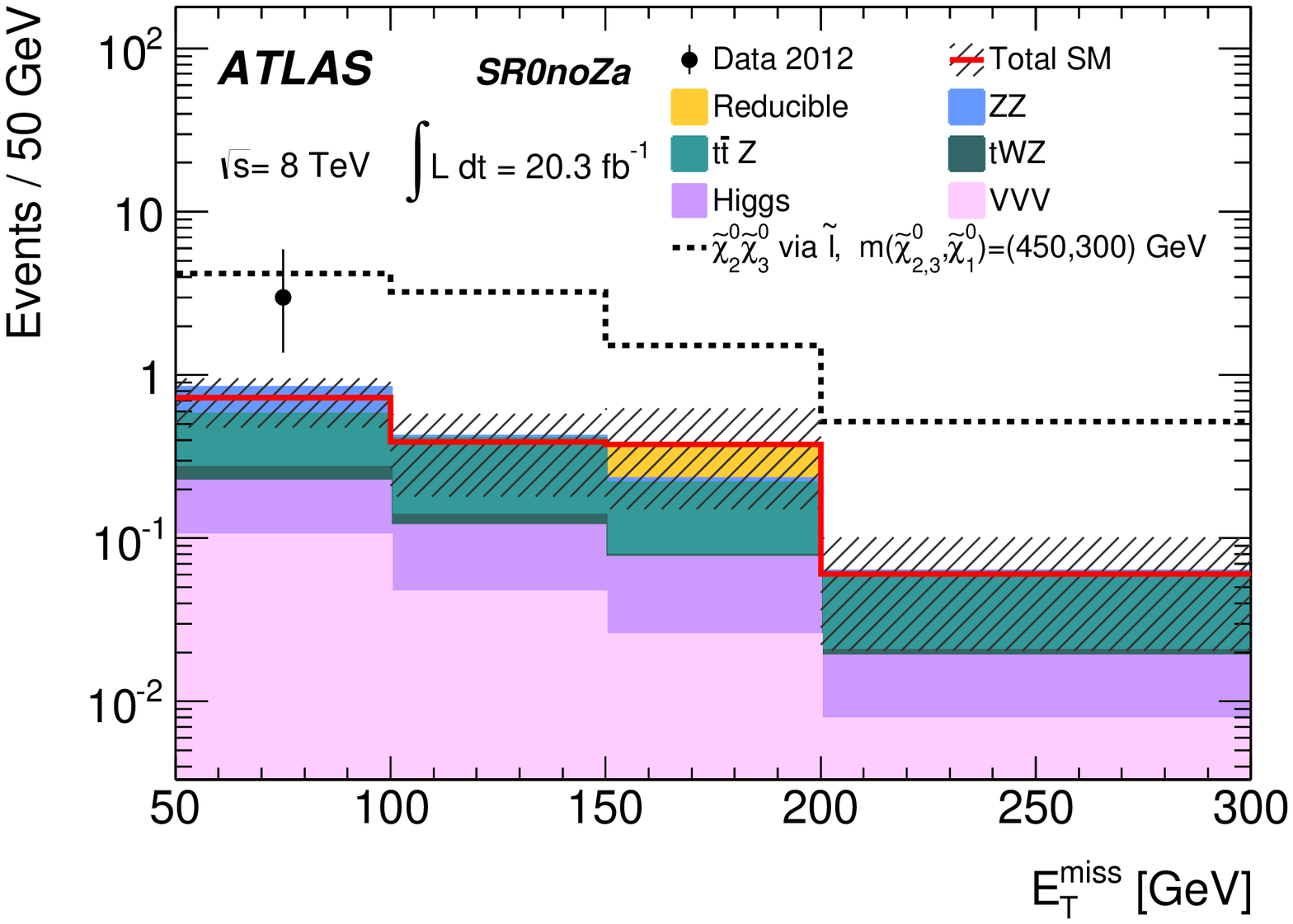}  } %
  \subfigure[~SR0noZa]{   \includegraphics[width=0.45\textwidth]{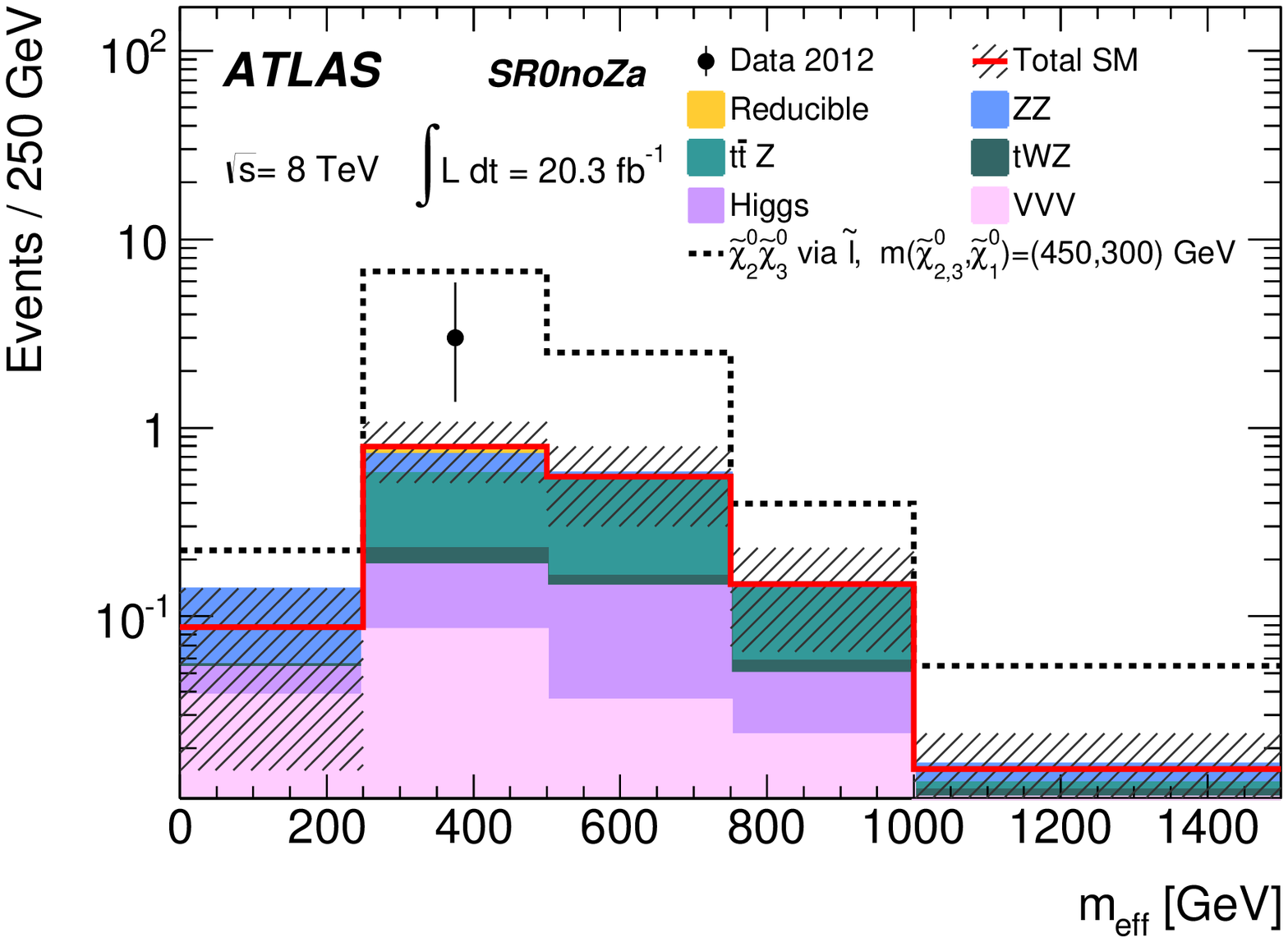}  } 
  \subfigure[~SR1noZa]{   \includegraphics[width=0.45\textwidth]{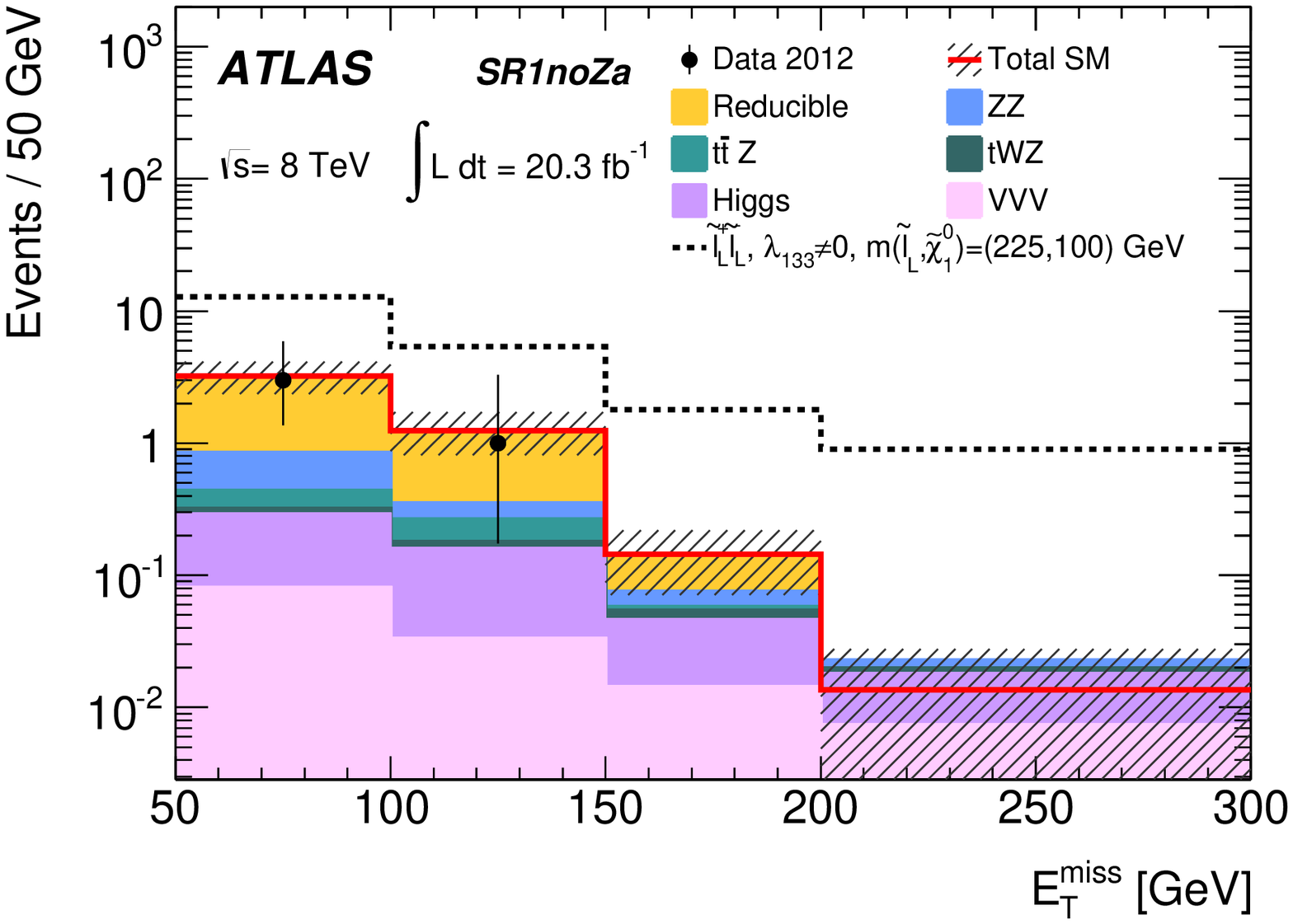}  } %
  \subfigure[~SR1noZa]{   \includegraphics[width=0.45\textwidth]{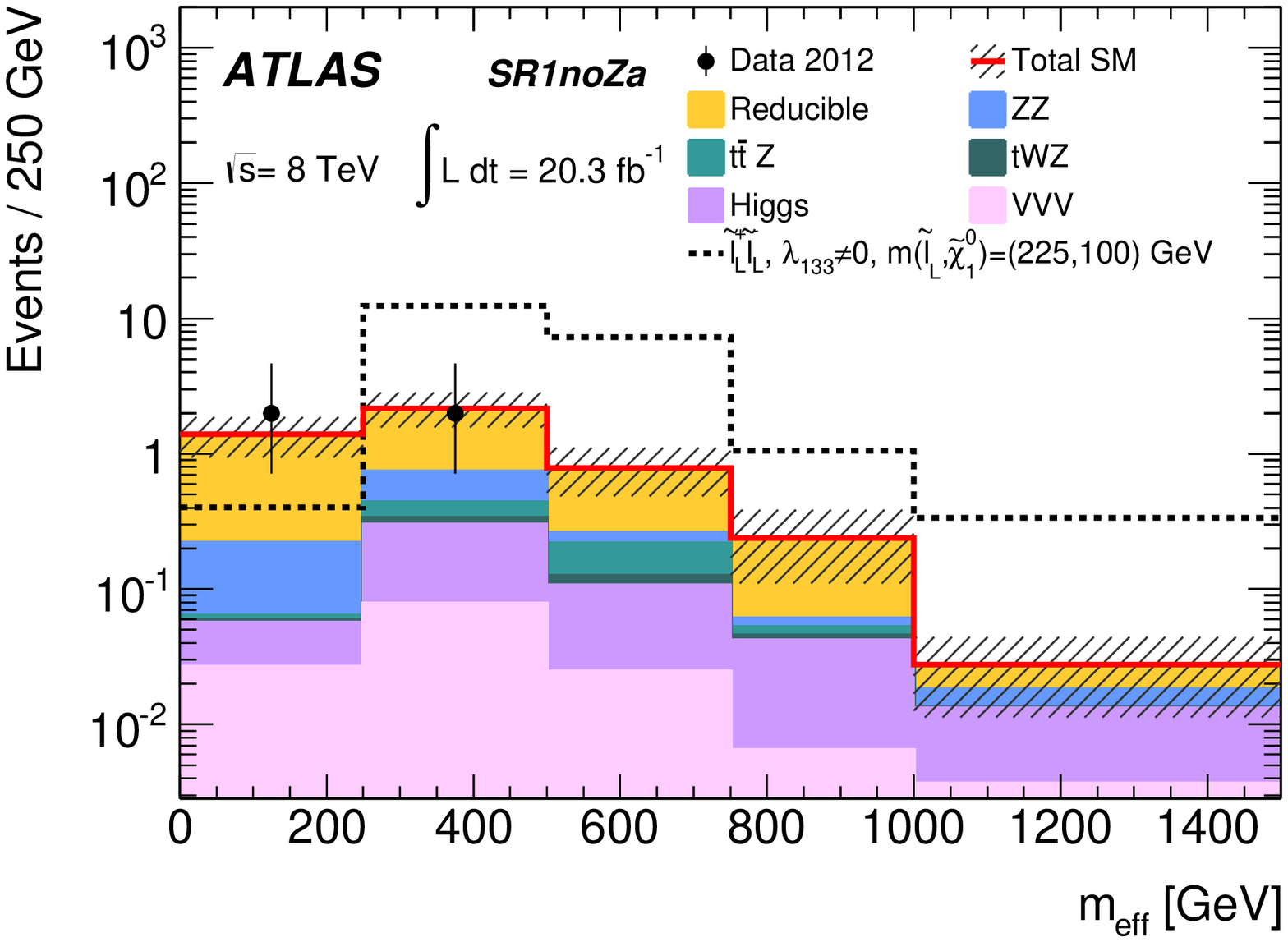}  } 
  \subfigure[~SR2noZa]{   \includegraphics[width=0.45\textwidth]{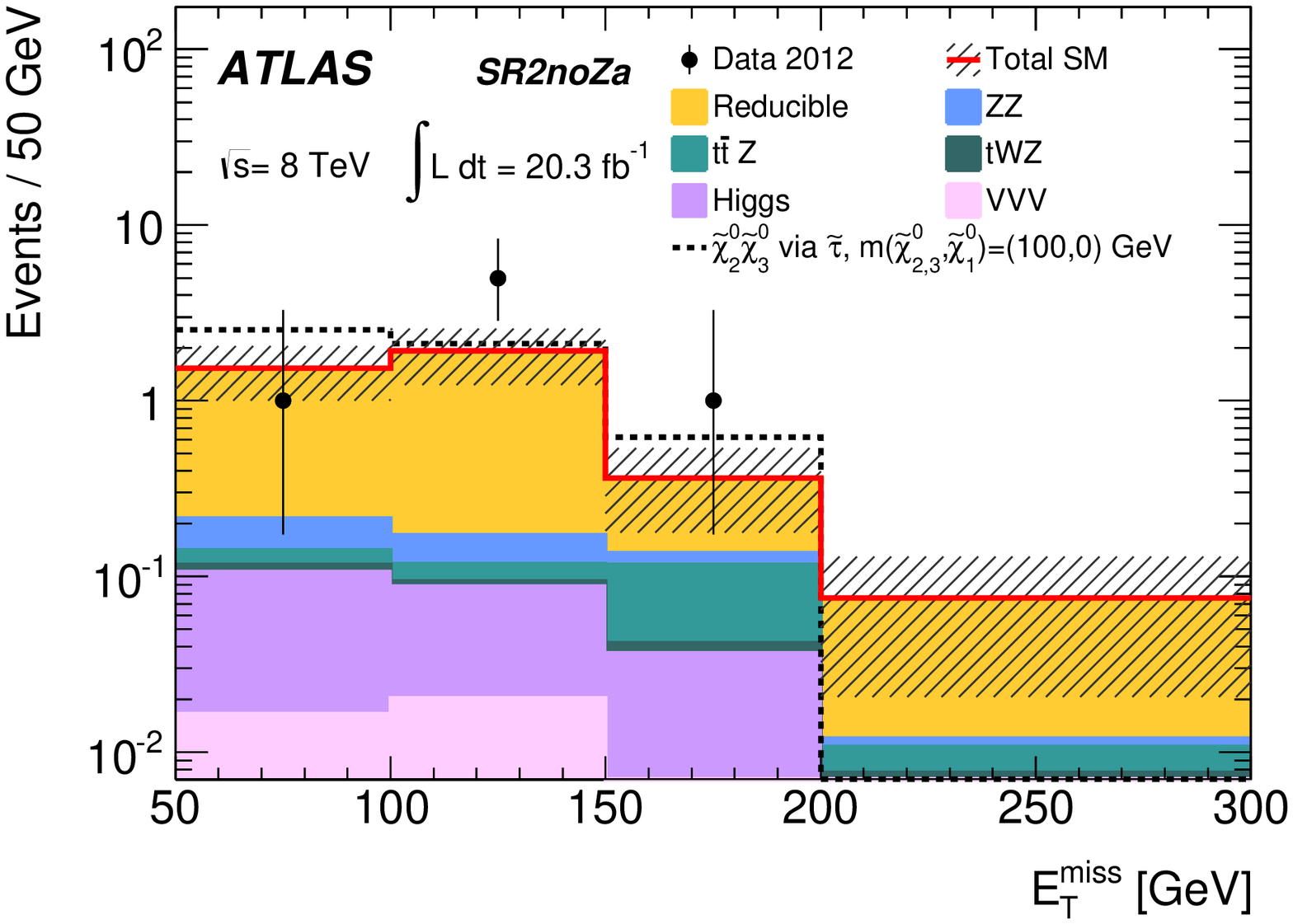}  } %
  \subfigure[~SR2noZa]{   \includegraphics[width=0.45\textwidth]{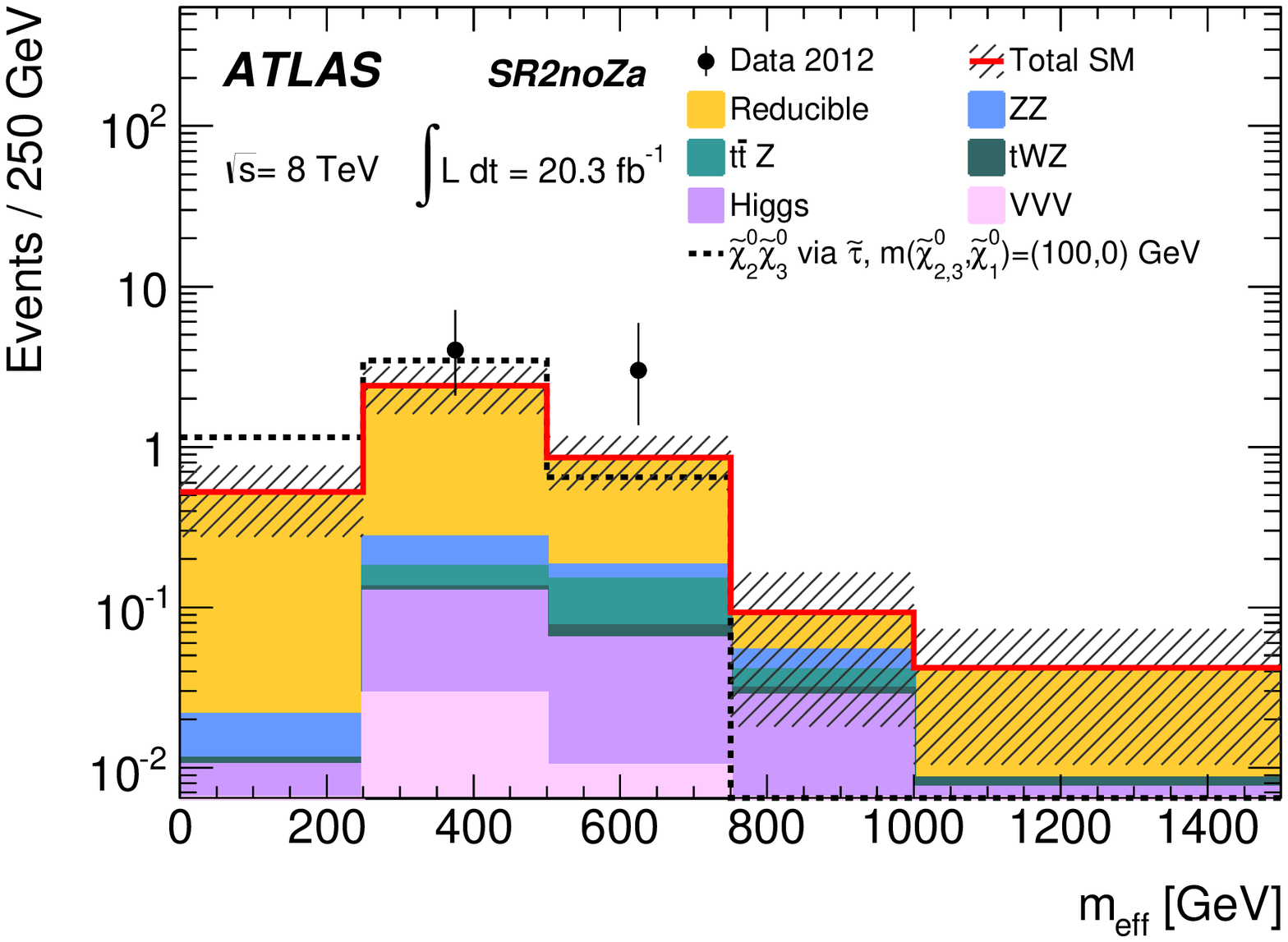}  } 
  \caption{
  The $\MET$ and $\meff$ distributions for data and the estimated SM backgrounds, in signal regions (a)--(b) SR0noZa, (c)--(d) SR1noZa, and (e)--(f) SR2noZa.
  The irreducible background is estimated from MC simulation while the reducible background is estimated from data using the weighting method. Both the statistical and systematic uncertainties are included in the shaded bands.
In each panel the distribution for a relevant SUSY signal model is also shown, where the numbers in parentheses indicate ($m_{\ninotwothree}$, $m_{\ninoone}$) for (a)--(b) and (e)--(f), or ($m_{\mathrm{NLSP}}$, $m_{\mathrm{LSP}}$) for (c)--(d), where all masses are in \gev.
 \label{fig:METMeff1}}
 \end{center}
\end{figure*}

\begin{figure*}[htp]
 \begin{center}
  \subfigure[~SR0noZb]{   \includegraphics[width=0.45\textwidth]{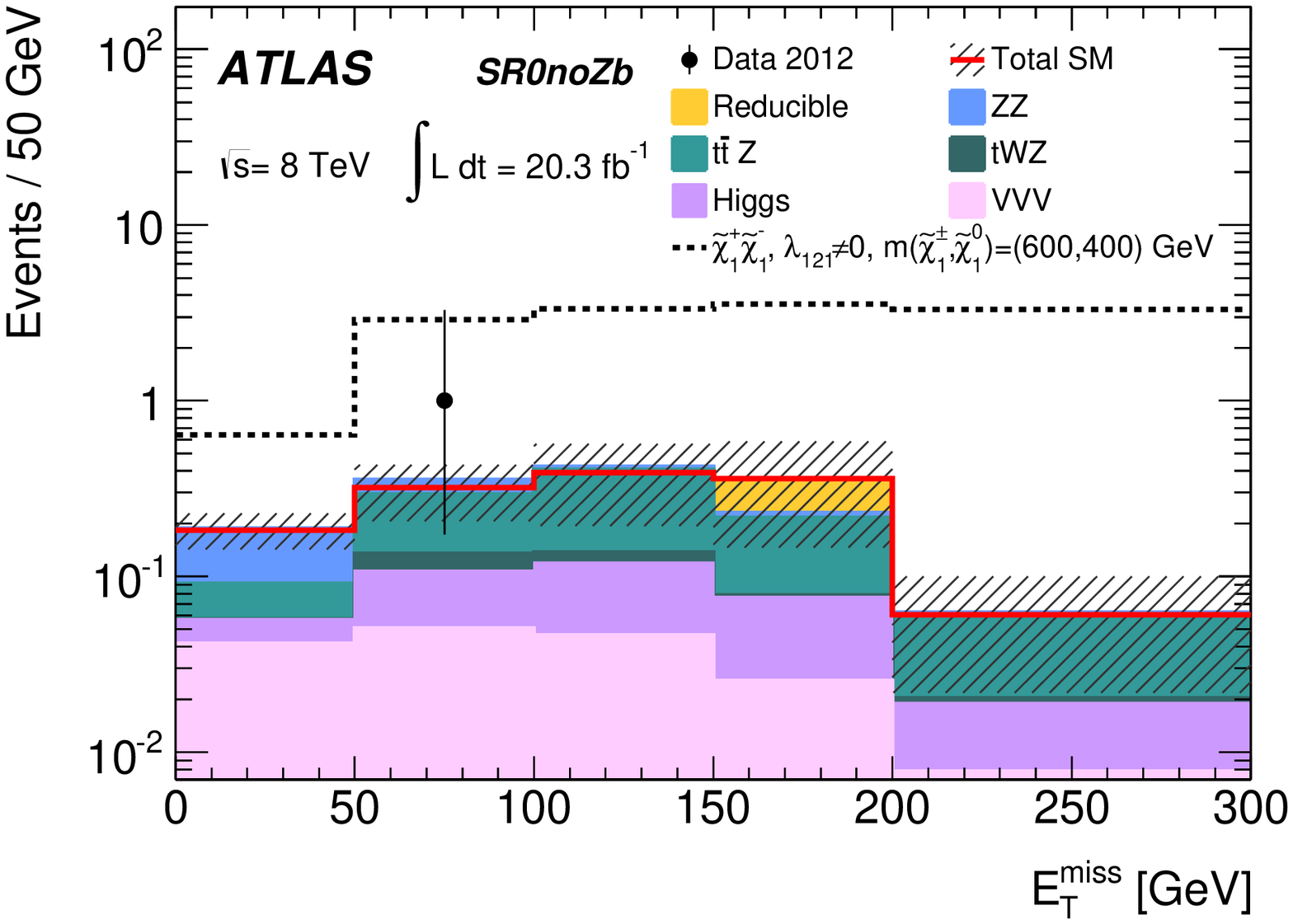}  } 
  \subfigure[~SR0noZb]{   \includegraphics[width=0.45\textwidth]{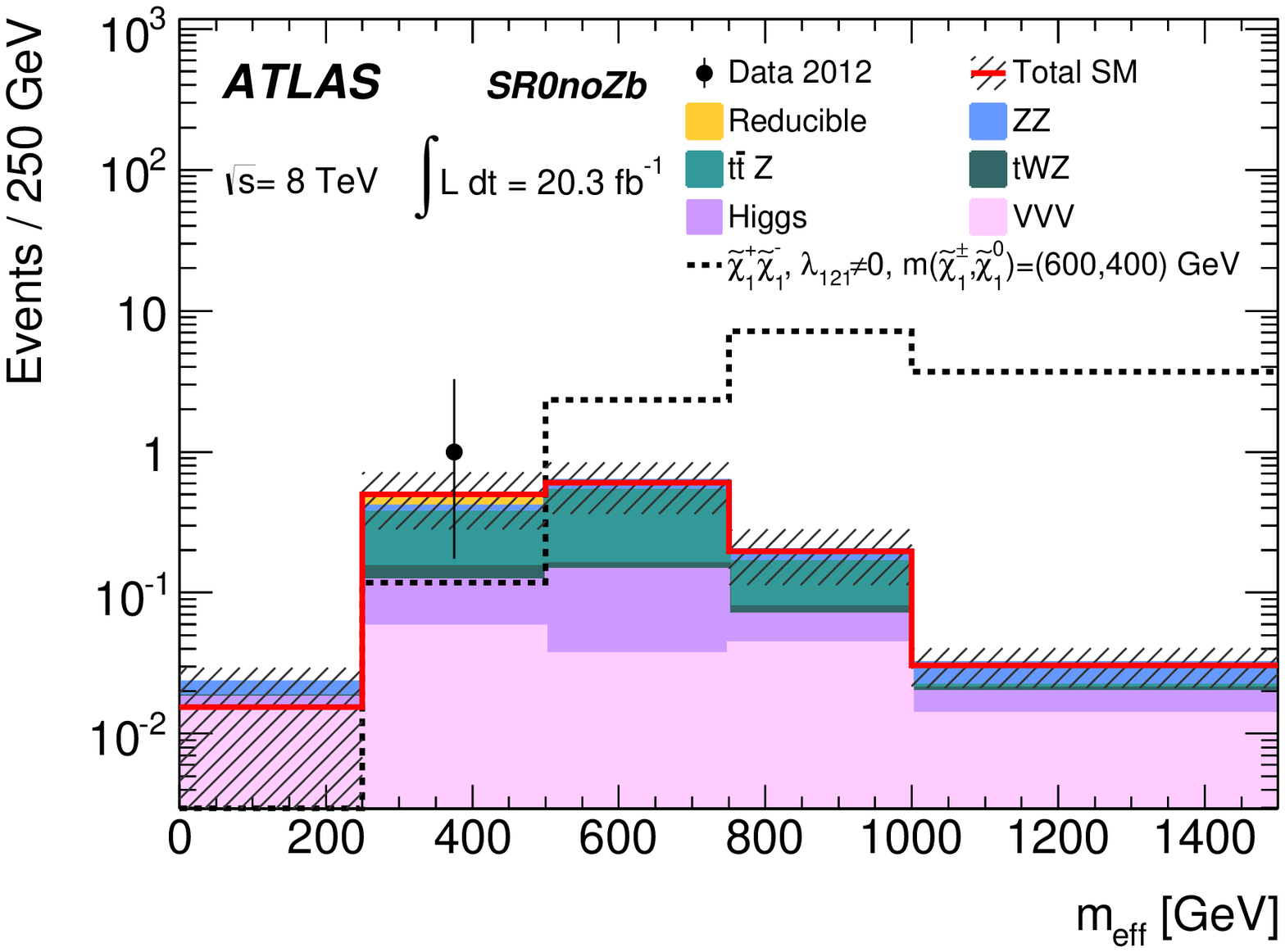}  } 
  \subfigure[~SR1noZb]{   \includegraphics[width=0.45\textwidth]{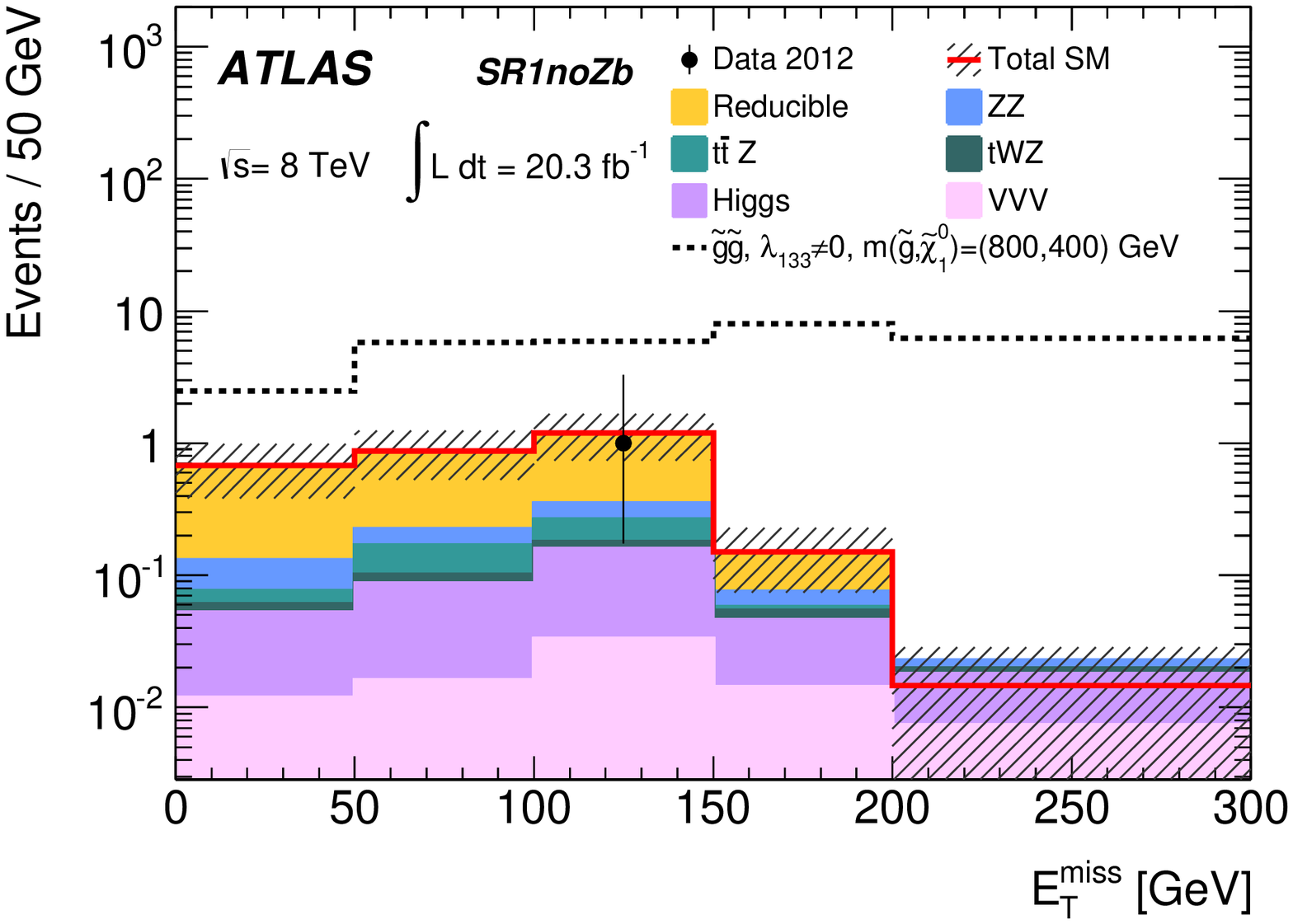}  } 
  \subfigure[~SR1noZb]{   \includegraphics[width=0.45\textwidth]{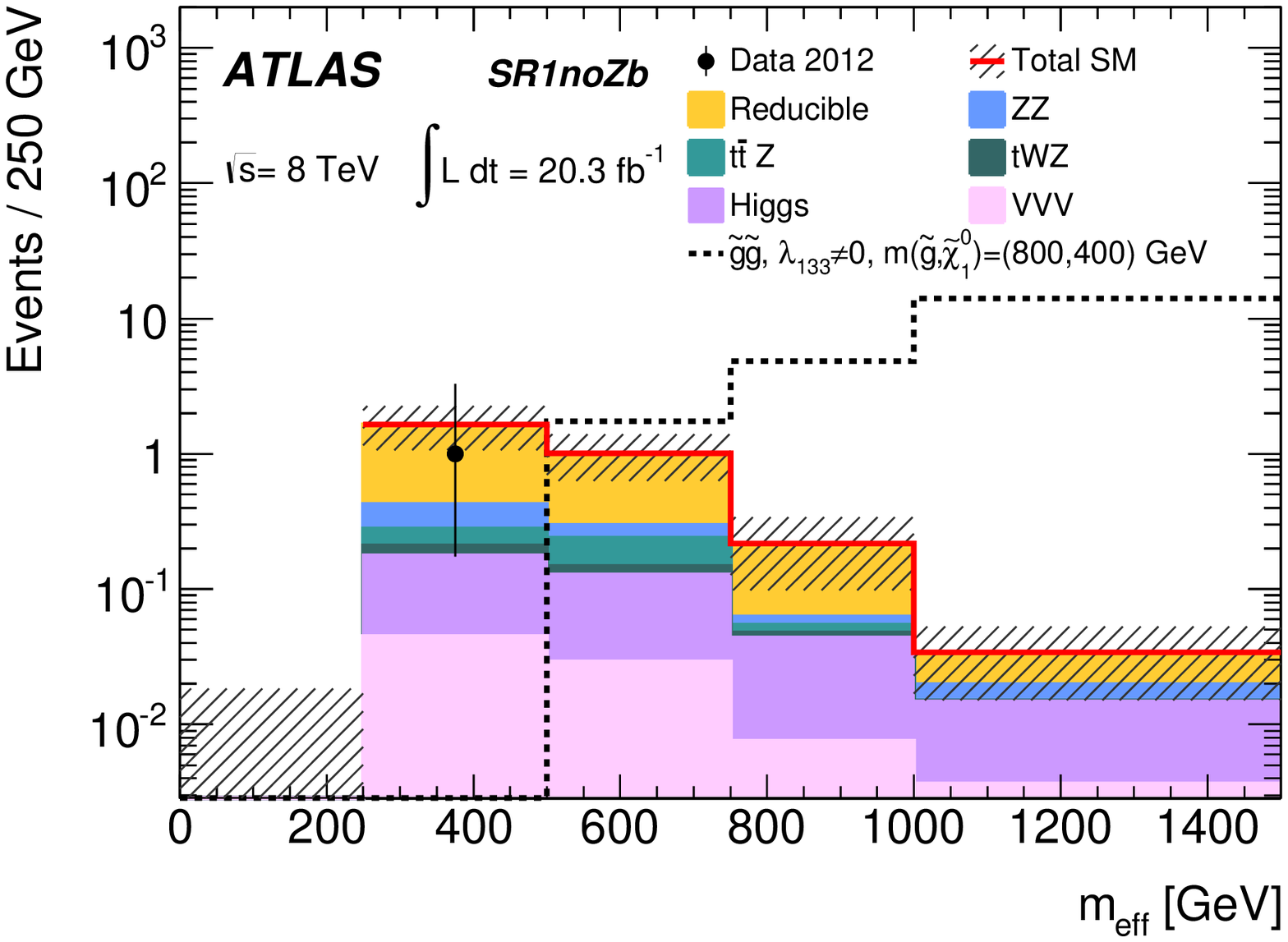}  } 
  \subfigure[~SR2noZb]{   \includegraphics[width=0.45\textwidth]{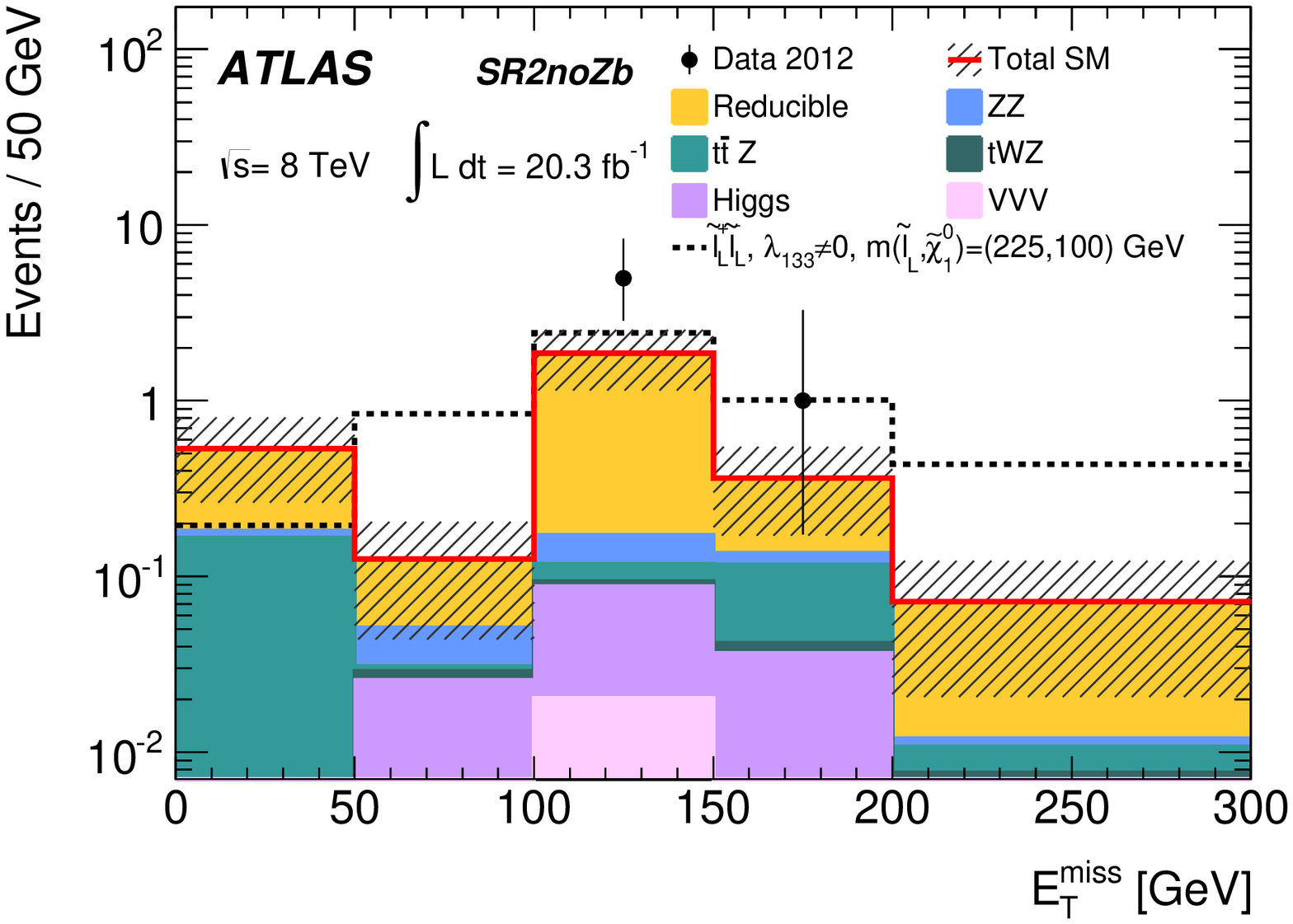}  } 
  \subfigure[~SR2noZb]{   \includegraphics[width=0.45\textwidth]{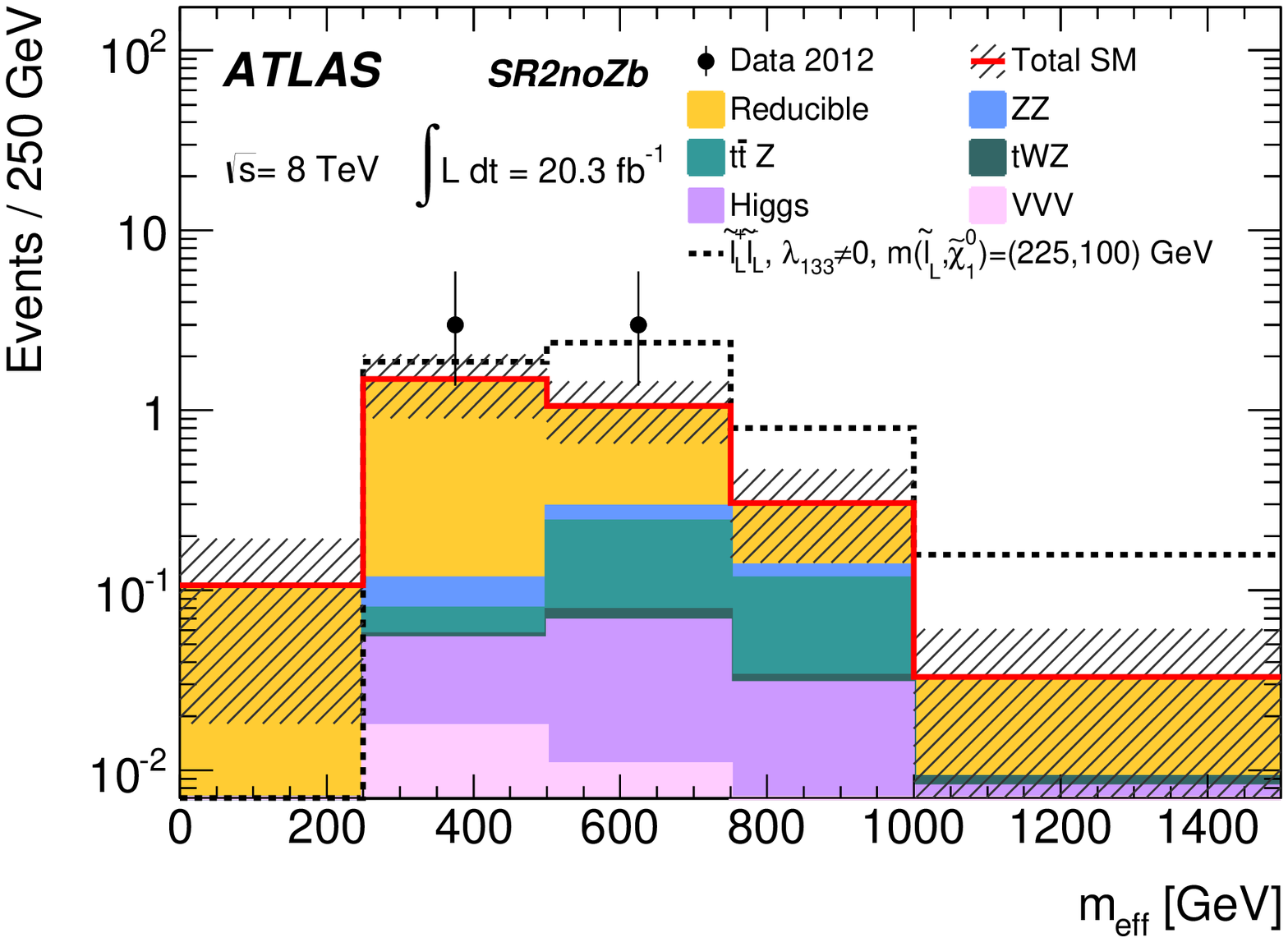}  } 
  \caption{
  The $\MET$ and $\meff$ distributions for data and the estimated SM backgrounds, in signal regions (a)--(b) SR0noZb, (c)--(d) SR1noZb, and (e)--(f) SR2noZb.
  The irreducible background is estimated from MC simulation while the reducible background is estimated from data using the weighting method. Both the statistical and systematic uncertainties are included in the shaded bands.
In each panel the distribution for a relevant SUSY signal model is also shown, where the numbers in parentheses indicate ($m_{\mathrm{NLSP}}$, $m_{\mathrm{LSP}}$) in \gev.
 \label{fig:METMeff2}}
 \end{center}
\end{figure*}

\begin{figure*}[htp]
 \begin{center}
  \subfigure[~SR0Z]{   \includegraphics[width=0.45\textwidth]{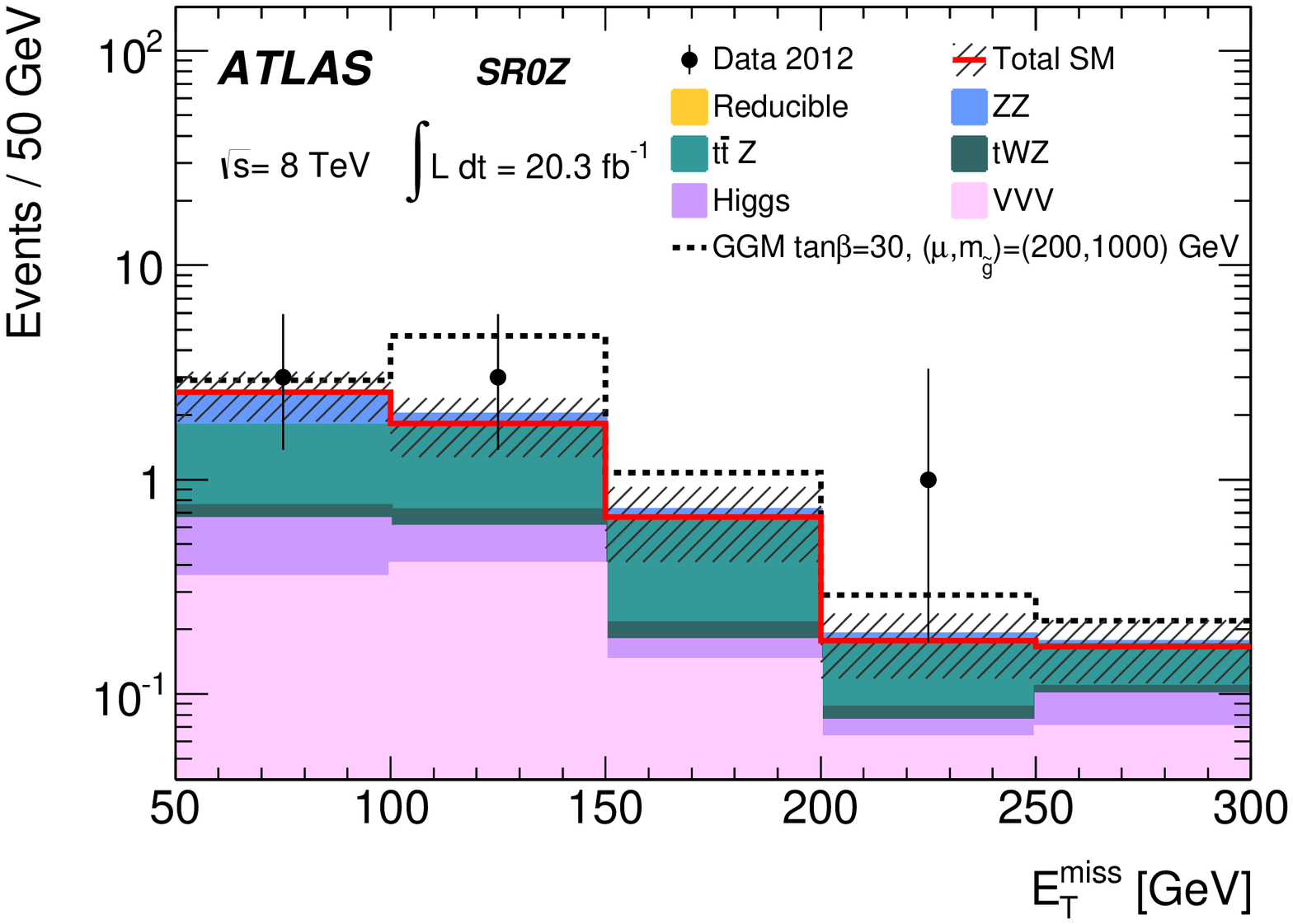}  } 
  \subfigure[~SR0Z]{   \includegraphics[width=0.45\textwidth]{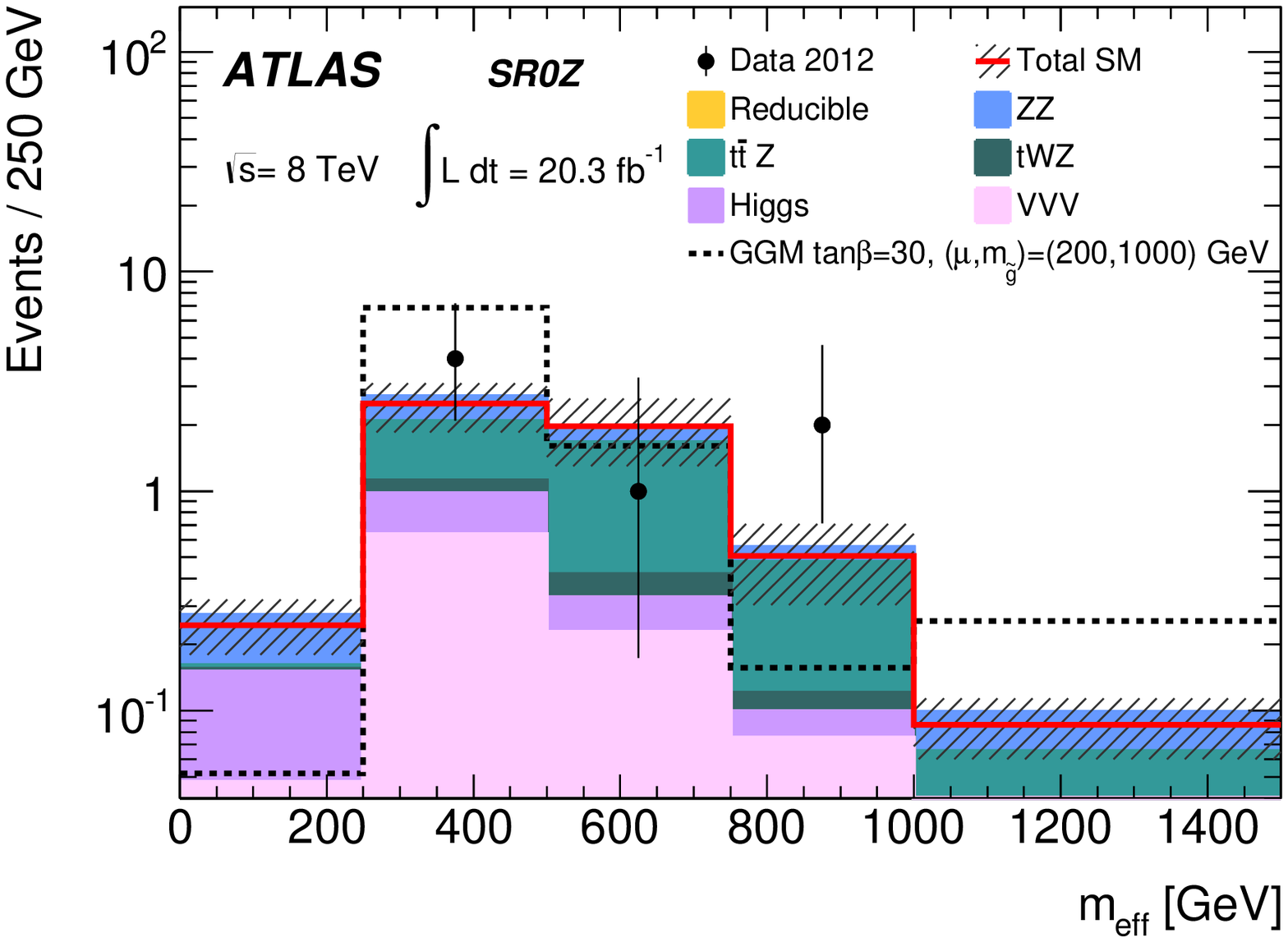}  } 
  \subfigure[~SR1Z]{   \includegraphics[width=0.45\textwidth]{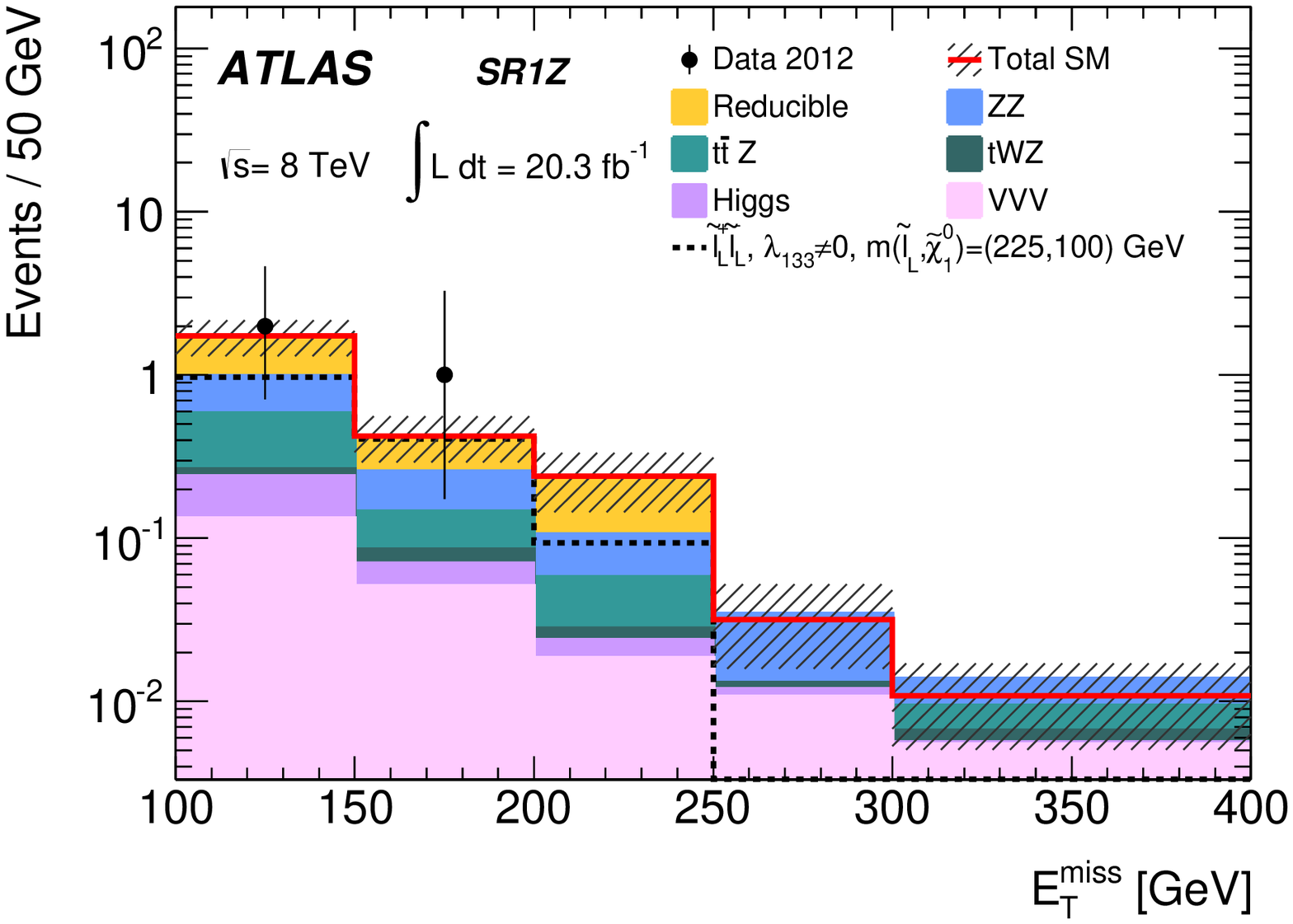}  } 
  \subfigure[~SR1Z]{   \includegraphics[width=0.45\textwidth]{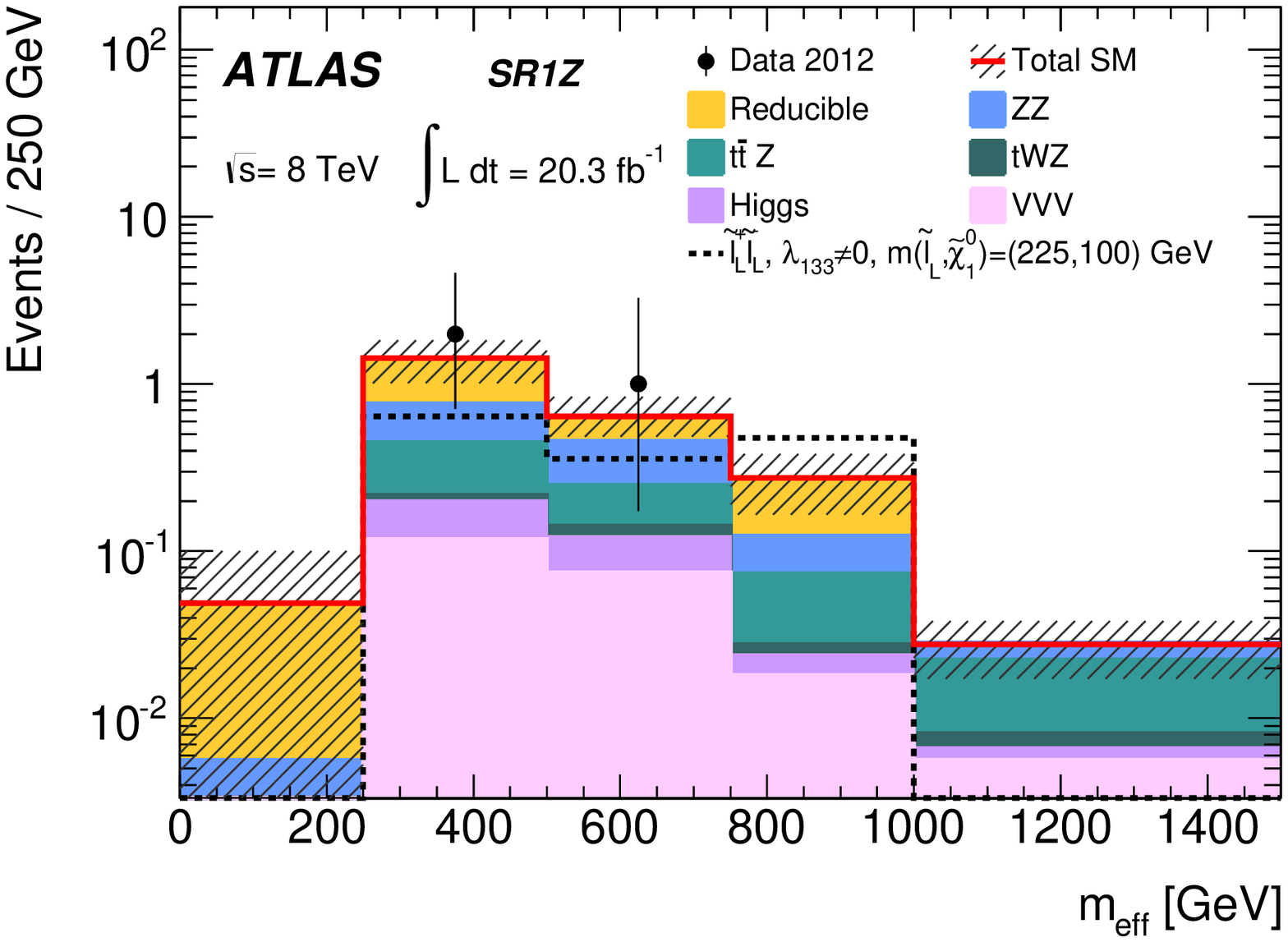}  } 
  \subfigure[~SR2Z]{   \includegraphics[width=0.45\textwidth]{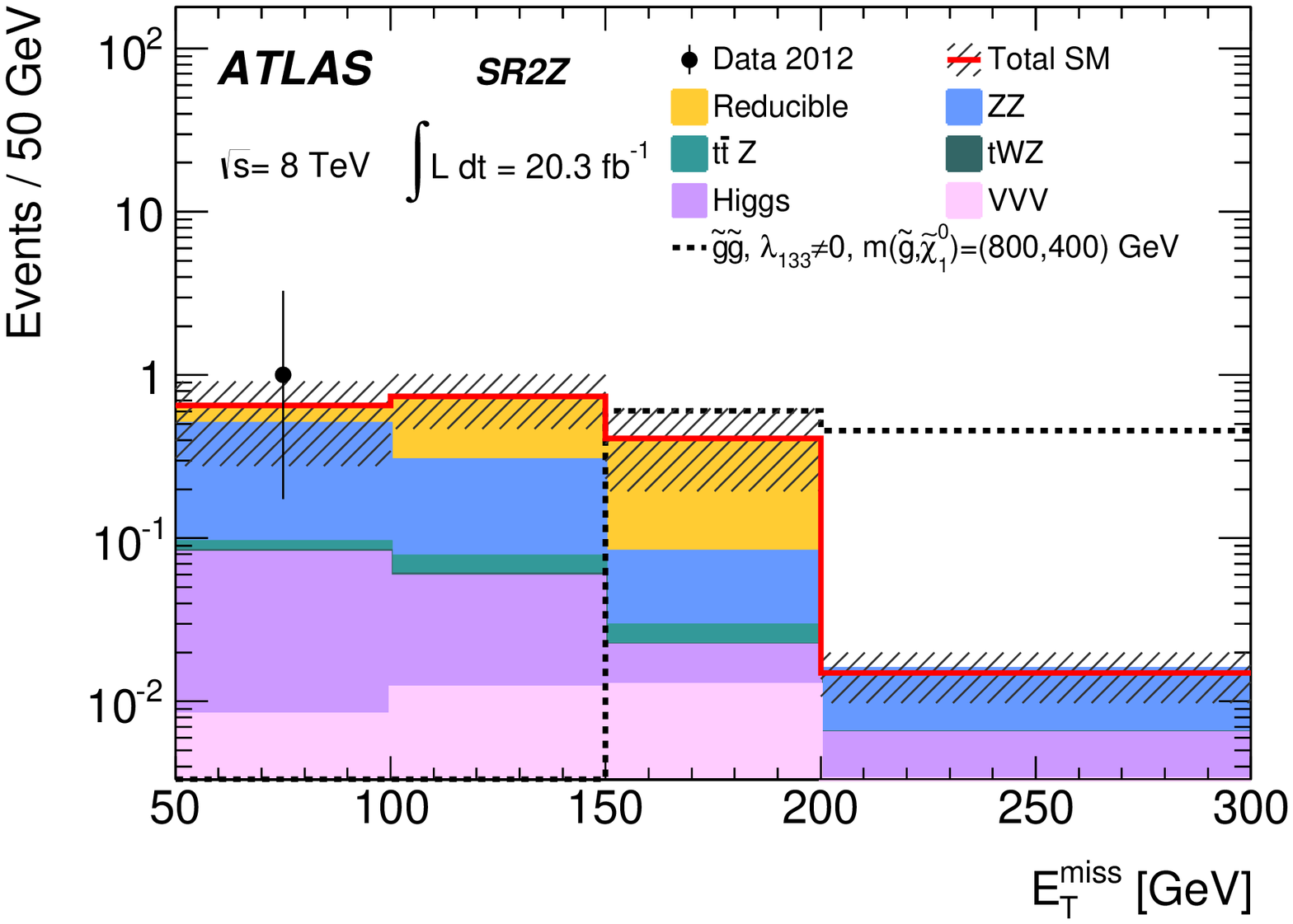}  } 
  \subfigure[~SR2Z]{   \includegraphics[width=0.45\textwidth]{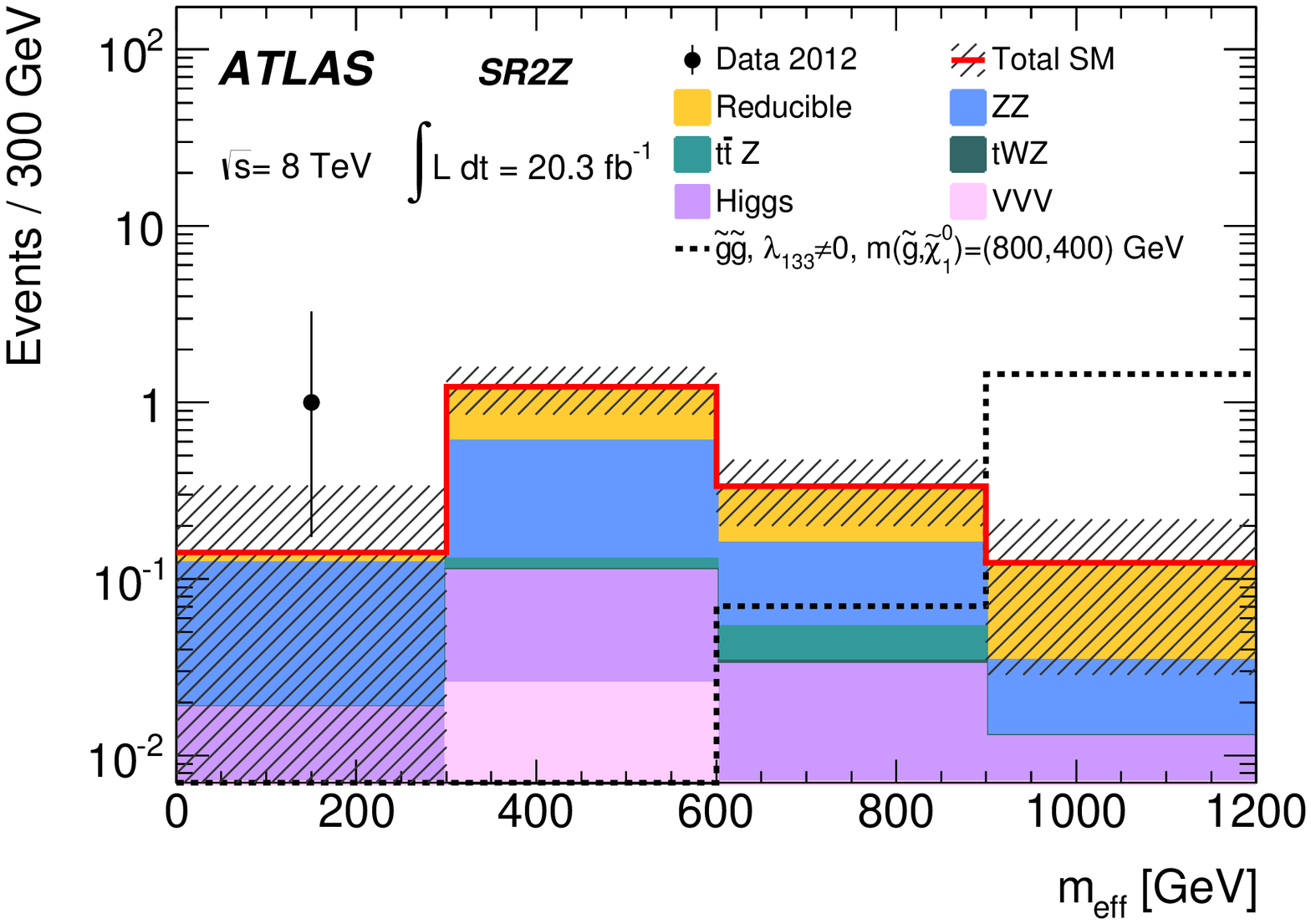}  } 
  \caption{
  The $\MET$ and $\meff$ distributions for data and the estimated SM backgrounds, in signal regions (a)--(b) SR0Z, (c)--(d) SR1Z and (e)--(f) SR2Z.
  The irreducible background is estimated from MC simulation while the reducible background is estimated from data using the weighting method.
 Both the statistical and systematic uncertainties are included in the shaded bands.
In each panel the distribution for a relevant SUSY signal model is also shown, where the numbers in parentheses indicate ($\mu$, $m_{\gluino}$) for (a)--(b), or ($m_{\mathrm{NLSP}}$, $m_{\mathrm{LSP}}$) for (c)--(f), where all masses are in \gev.
 \label{fig:METMeff3}}
 \end{center}
\end{figure*}

The $\MET$ and $\meff$ distributions in all signal regions are shown in \figsref{fig:METMeff1}--\ref{fig:METMeff3}. 
For each signal region, a SUSY signal model is superimposed on the SM background prediction, for illustration.
RPC simplified models are chosen to illustrate SR0noZa and SR2noZa (R-slepton and stau models, respectively), for which these regions are designed.
Similarly, the GGM model with $\tan\beta=30$ illustrates the sensitivity of SR0Z to SUSY.
A variety of RPV simplified models with different experimental signatures are used to illustrate the sensitivity of the remaining signal regions.
Good agreement is again seen between SM background expectations and data, within uncertainties.

\section{Interpretations in new physics scenarios \label{sec:interpretations}}

The results of this analysis are interpreted in RPV simplified models, for various assumed \rpvlambda{} parameters, 
as well as in the RPC simplified models and in RPC GGM models, all presented in \Secref{sec:scenarios}.
As more than one signal region may be sensitive to any particular scenario, a statistical combination of different signal regions is performed to extract the limits.
Section~\ref{sec:SRsel} defines three pairs of overlapping signal regions in which a \Zboson-veto is applied (SR0noZa/b, SR1noZa/b and SR2noZa/b).
For each mass point in every model considered, the signal region providing the best
expected sensitivity for that model is chosen from each pair. The three selected \Zboson-veto signal regions are combined with each other
and with the remaining three signal regions (SR0Z, SR1Z and SR2Z), taking into account possible correlations of systematic uncertainties between signal regions.
Asymptotic formulas for the test statistic distribution~\cite{cowan} are used when setting model-dependent limits, and signal contamination in the control regions is accounted for.

\subsection{RPV simplified models  \label{sec:limitsRPV}}

\begin{figure}[ht]
\begin{center}
\subfigure[~Chargino NLSP]  { \includegraphics[width=0.5\textwidth]{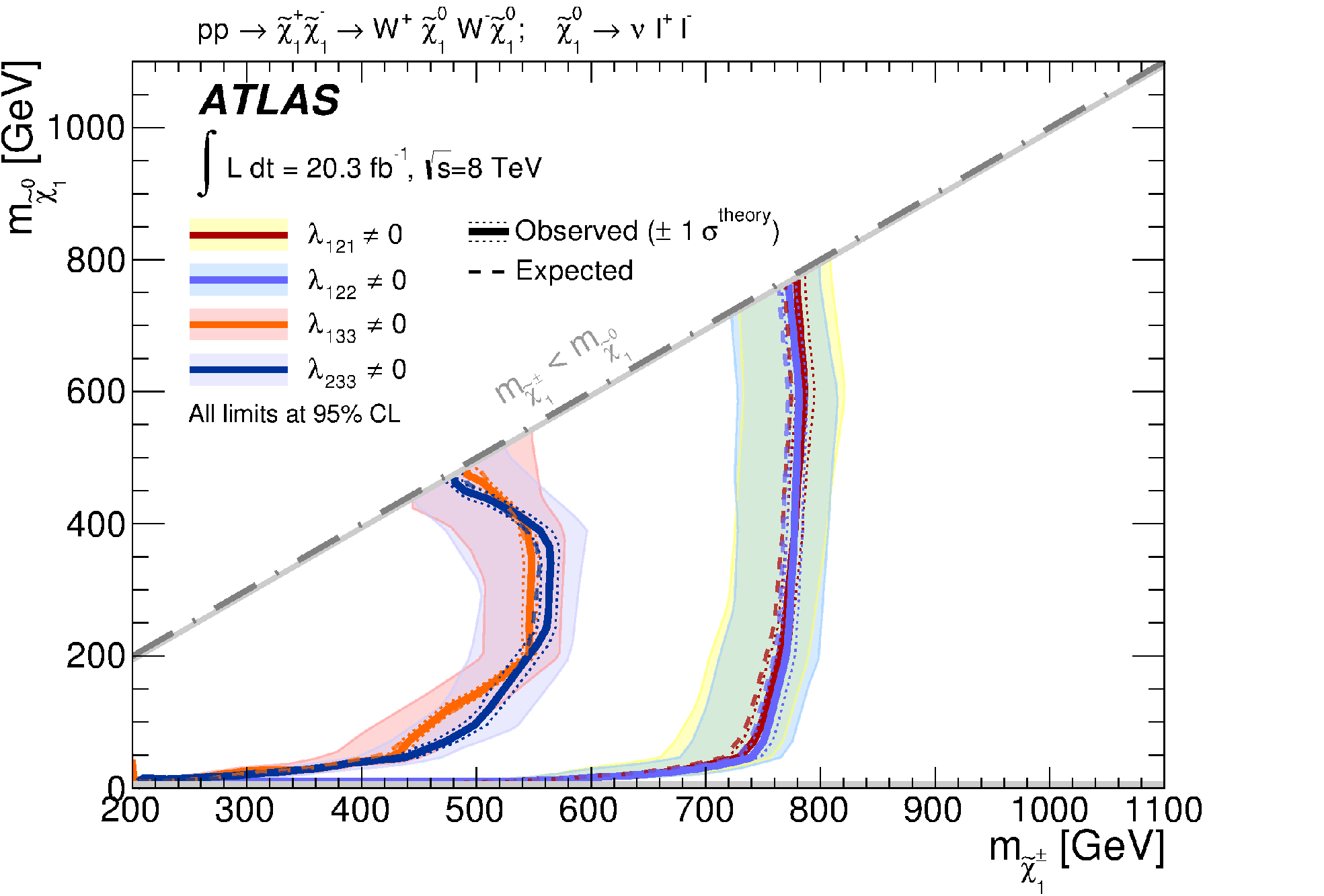} }
\subfigure[~Gluino NLSP]{ \includegraphics[width=0.5\textwidth]{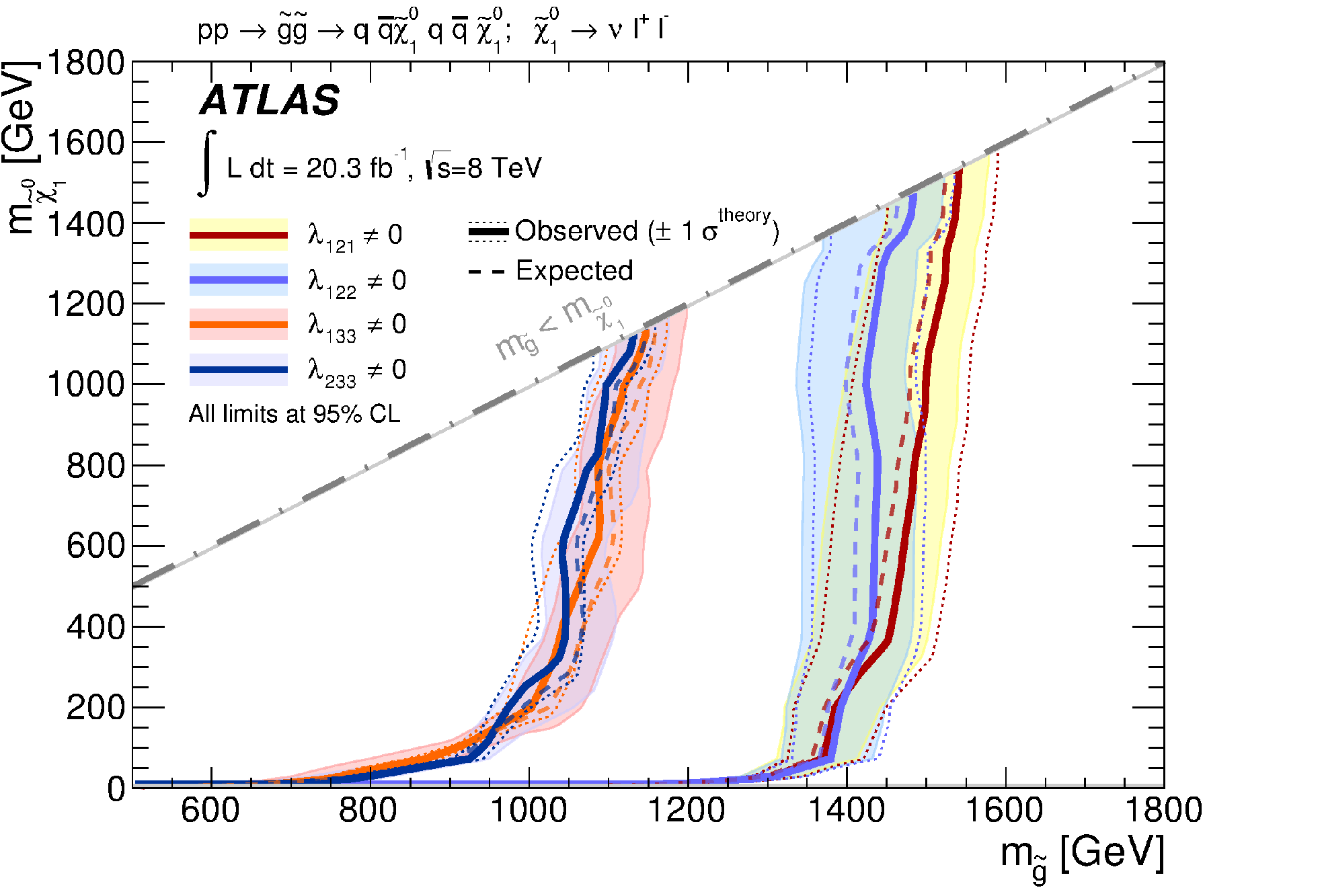} }
 \caption{The observed (solid) and expected (dashed) 95\% CL exclusion limit contours for the RPV (a) chargino NLSP and (b) gluino NLSP simplified models, assuming a promptly decaying LSP.
The exclusion limits include all uncertainties except the theoretical cross section uncertainty for the signal, the effect of which is indicated by the dotted lines either side of the observed exclusion limit contours.
The shaded bands around each expected exclusion limit curve show the $\pm1\sigma$ results.
No events above the diagonal dashed line were generated.
 \label{fig:ExpLimitWinoGluino}}
\end{center}
\end{figure}

\begin{figure}[ht]
\begin{center}
\subfigure[~L-slepton NLSP]{ \includegraphics[width=0.5\textwidth]{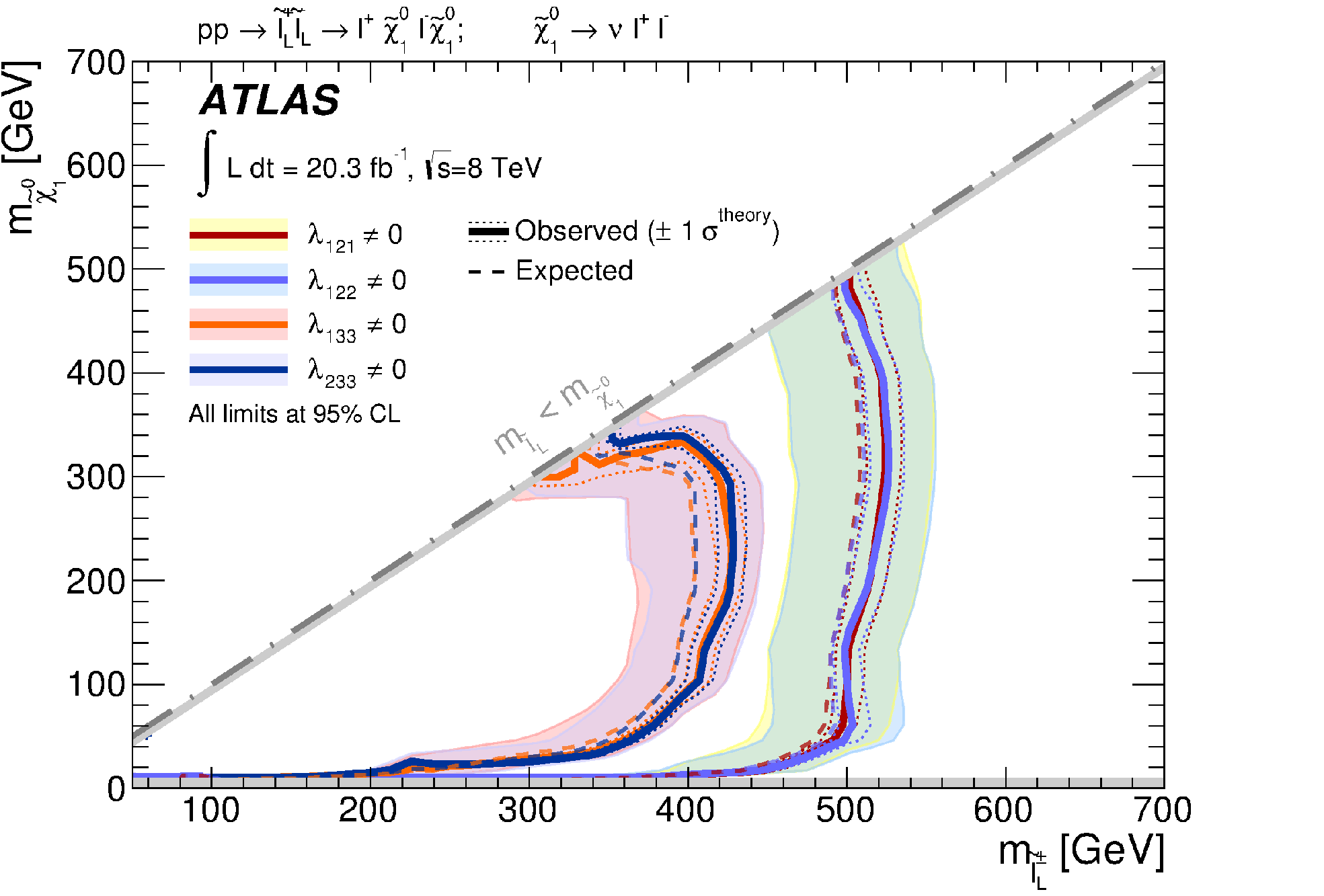} }
\subfigure[~R-slepton NLSP]{ \includegraphics[width=0.5\textwidth]{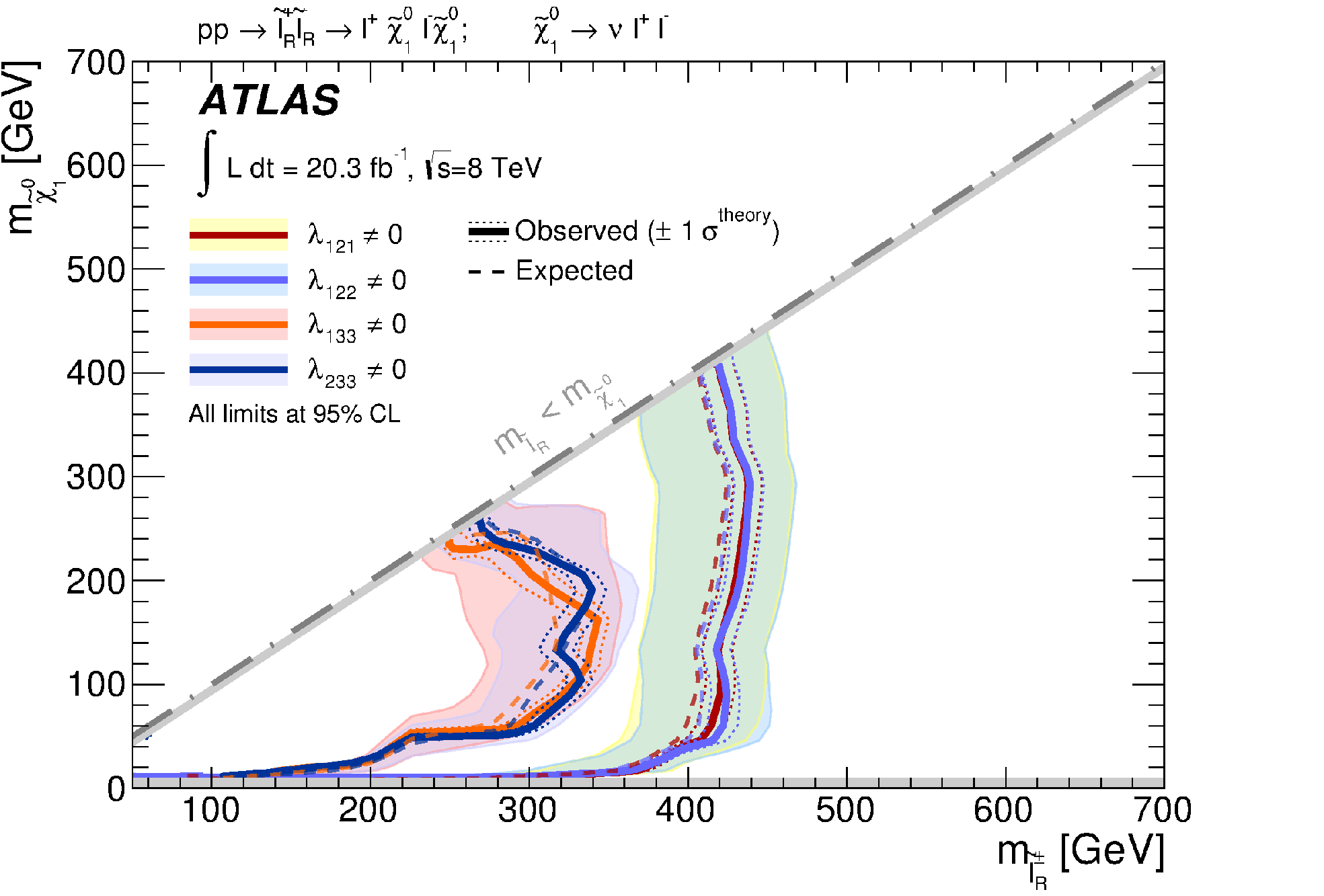} }
\subfigure[~Sneutrino NLSP]{ \includegraphics[width=0.5\textwidth]{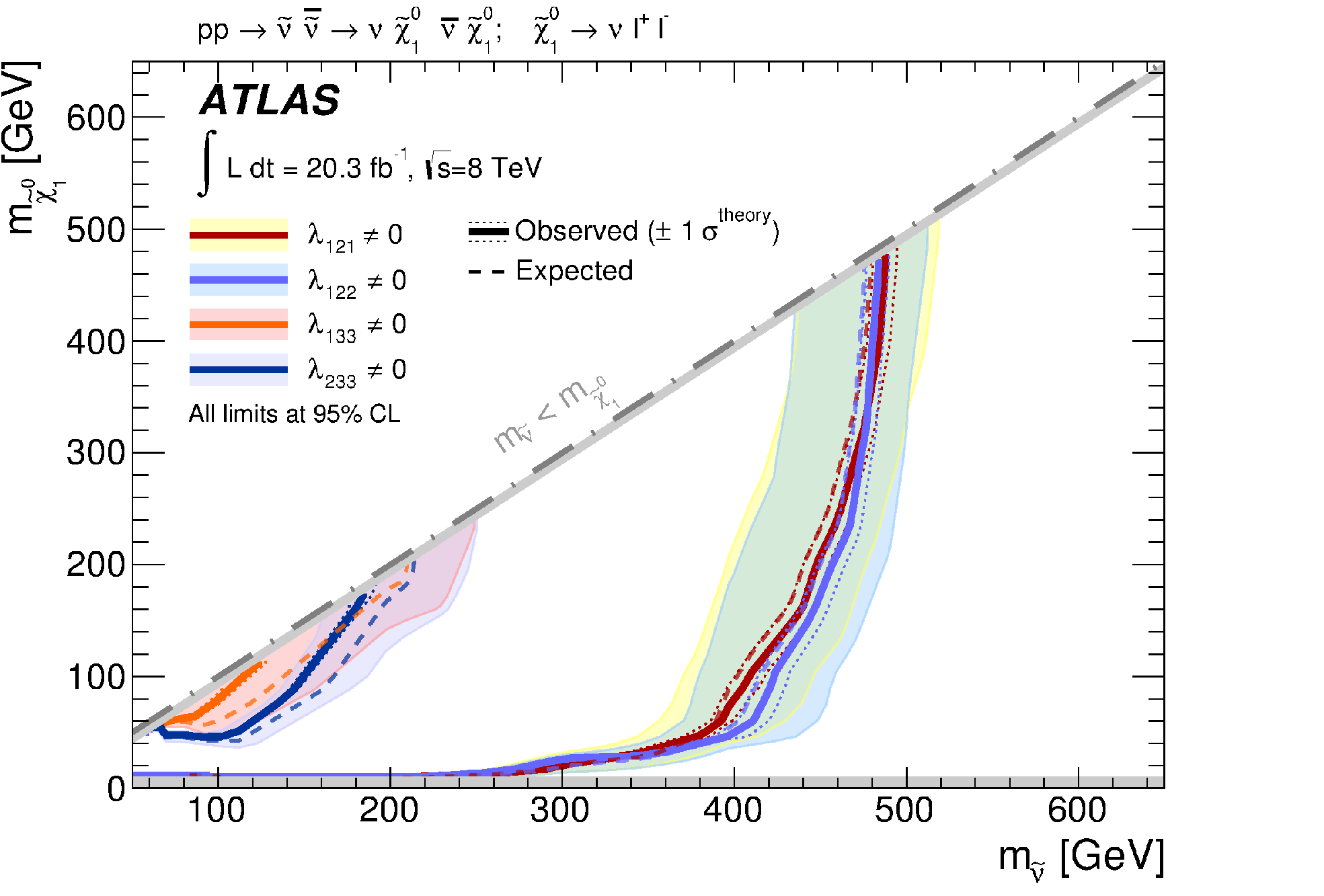} }
 \caption{The 95\% CL exclusion limit contours for the RPV (a) L-slepton NLSP, (b) R-slepton NLSP and (c) sneutrino NLSP simplified models, assuming a promptly decaying LSP.
For further details see \protect\figref{fig:ExpLimitWinoGluino}.
 \label{fig:ExpLimitSlepAndSnu}}
\end{center}
\end{figure}

The observed and expected 95\% CL exclusion limit contours for the RPV chargino NLSP and gluino NLSP simplified models discussed in \Secref{sec:scenarios} are shown in 
\figref{fig:ExpLimitWinoGluino}. 
The colored band around the median expected limit shows the $\pm1\sigma$ variations on the limit, including all uncertainties except the theoretical uncertainty on the
signal cross section. Different choices of \rpvlambda{} parameters correspond to differently colored bands, as per labels in the legend. 
The dotted lines indicate changes in the corresponding observed limit due to $\pm1\sigma$ variations of the signal cross section by the theoretical uncertainty.
The conservative $-1\sigma$ variation is used to quote limits.
Similar conventions are adopted for all exclusion contours and corresponding limits.
Figure~\ref{fig:ExpLimitSlepAndSnu} shows the observed and expected 
95\% CL limit contours for the RPV L-slepton NLSP, R-slepton NLSP and sneutrino NLSP simplified models. 

In all cases, the observed limit is determined primarily by the production cross section 
of the signal process, with stronger constraints on models where \rpvlambda[121] or \rpvlambda[122] dominate, and less stringent limits for tau-rich decays via \rpvlambda[133] or \rpvlambda[233].
Limits on models with different combinations of \rpvlambda{} parameters can generically be expected to lie between these extremes.
The limits are in many cases nearly insensitive to the \ninoone{} mass, except where the \ninoone{} is significantly less massive than the NLSP.
When this is the case [for example, $m_{\ninoone}\lesssim 50\gev$ in \figref{fig:ExpLimitSlepAndSnu}(a)], the \ninoone{} produced in the NLSP decay has substantial momentum in the laboratory frame of reference, 
and its decay products either tend to travel close to the \ninoone{} direction, becoming collimated, or one of the leptons becomes soft.
These effects reduce the analysis acceptance and efficiency, especially if the LSP decays to one or more hadronically decaying taus.
Where the NLSP$\to$LSP cascade may also produce leptons (specifically, the chargino and slepton models), the observed limit may also become weaker as $m_{\ninoone}$ approaches the NLSP mass, and the cascade product momenta fall below threshold.

When the mass of the \ninoone{} LSP is at least as large as 20\% of the NLSP mass, and assuming tau-rich LSP decays,
 lower limits can be placed on sparticle masses, excluding
gluinos with masses less than $950\gev$; winolike charginos with masses less than $450\gev$; and L(R)-sleptons with masses less than 
300 $(240)\gev$.
If instead the LSP decays only to electrons and muons, 
the equivalent limits are approximately $1350\gev$ for gluinos, $750\gev$ for charginos, 490 $(410)\gev$ for L(R)-sleptons, and a  
lower limit of $400\gev$ can also be placed on sneutrino masses.
These results significantly improve upon previous searches at the LHC, where gluino masses of up to $1\tev$~\cite{Chatrchyan:2012mea} and chargino masses of up to $540\gev$~\cite{ATLAS:2012kr} were excluded.

\subsection{RPC simplified models \label{sec:limitsRPC}}

\begin{figure}[ht]
 \begin{center}
 \subfigure[~R-slepton RPC]{\includegraphics[width=0.45\textwidth]{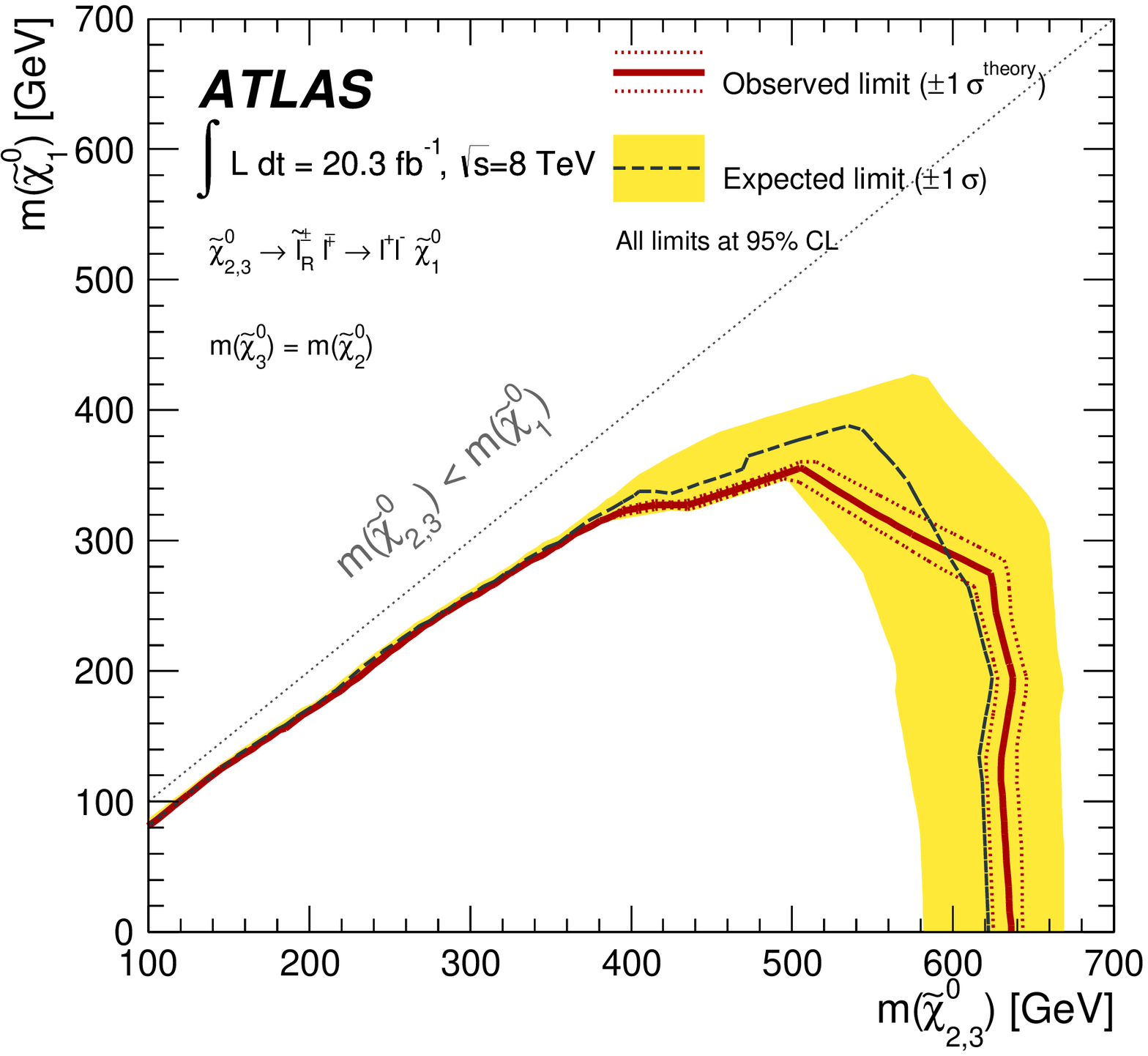}}
 \subfigure[~Stau RPC]{\includegraphics[width=0.45\textwidth]{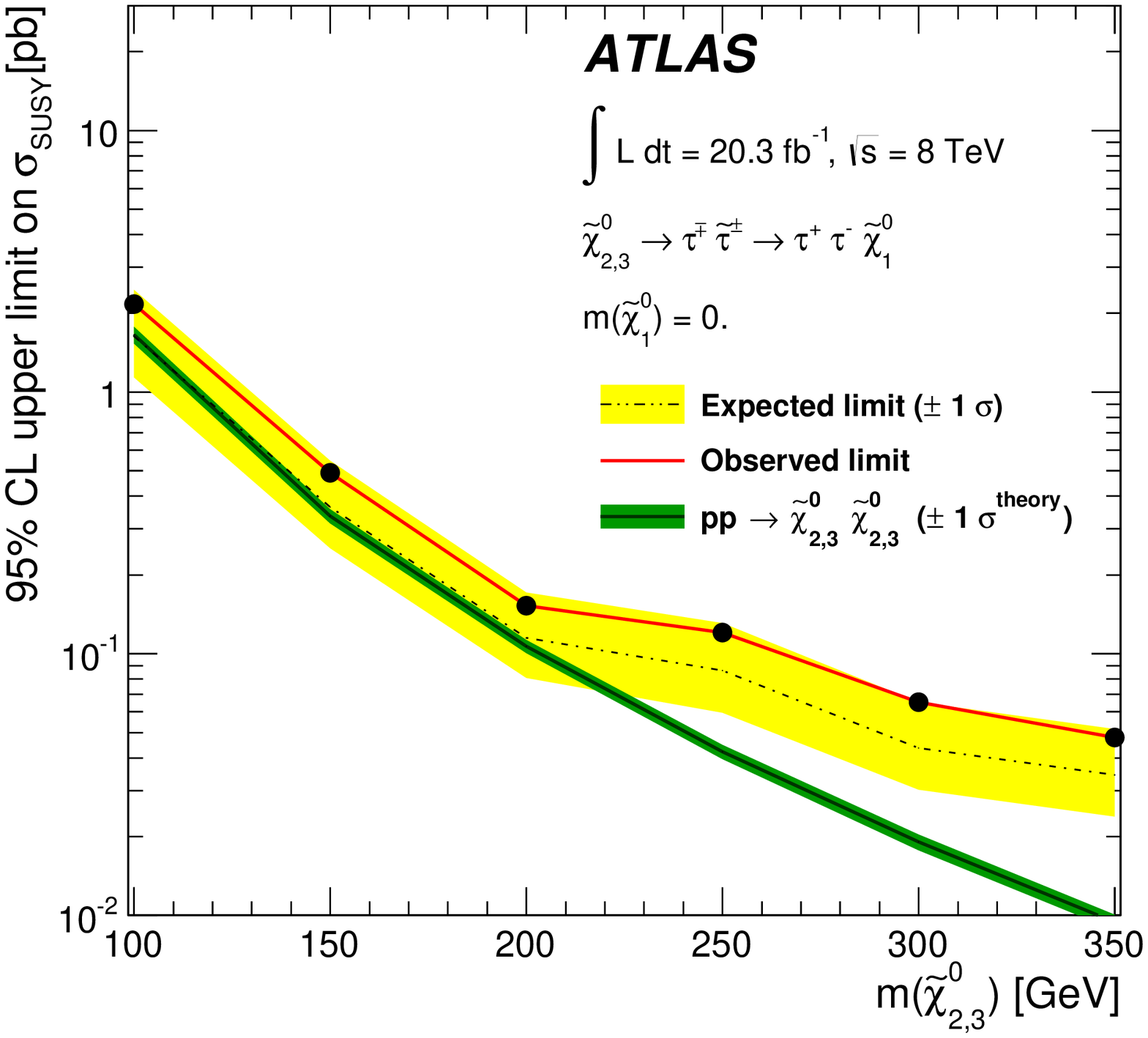}}
 \subfigure[~\Zboson{} RPC]{\includegraphics[width=0.45\textwidth]{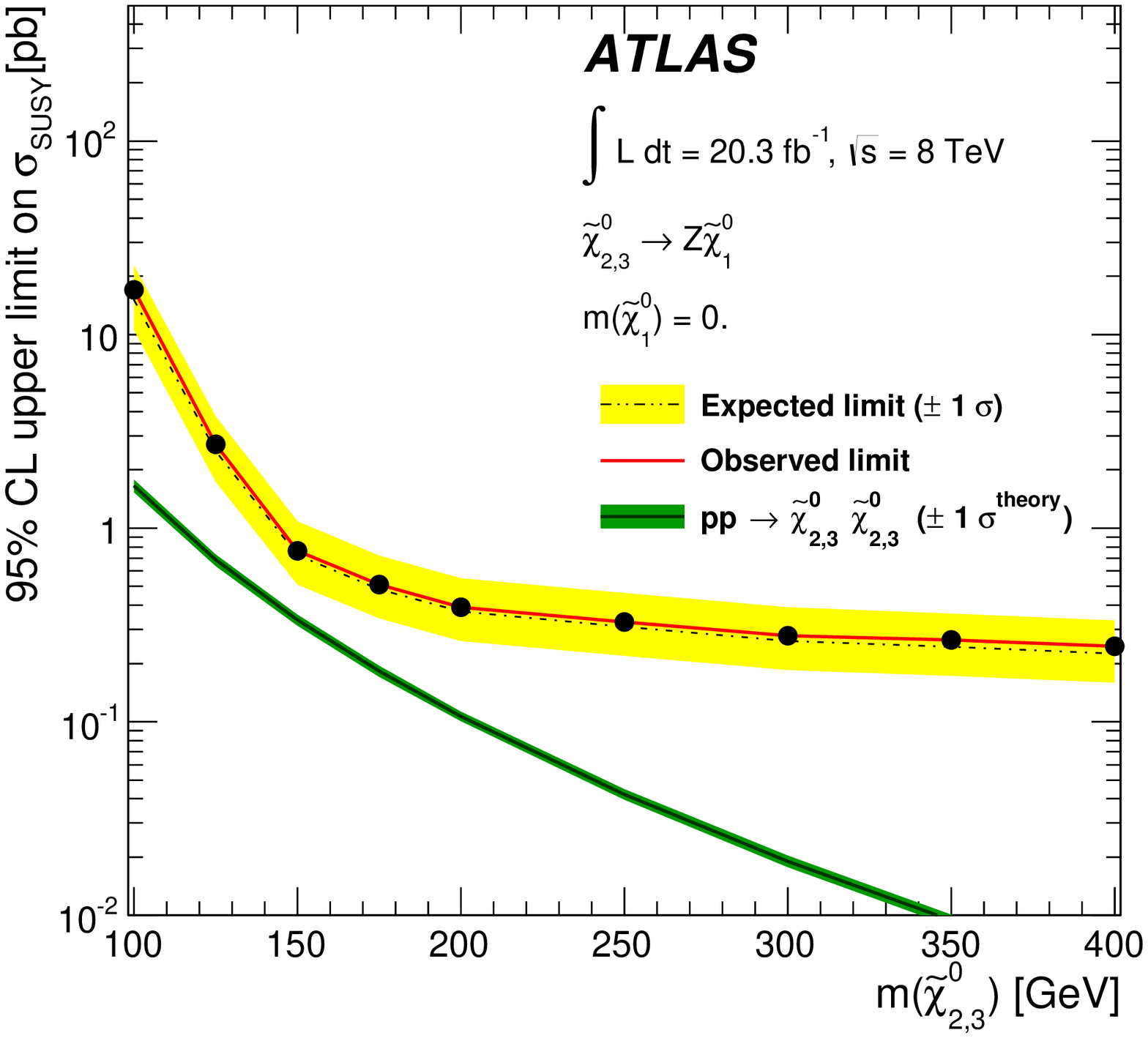}}
   \caption{The 95\% CL exclusion limits for the RPC models:
(a) R-slepton mass exclusion limit; (b) stau model and (c) \Zboson{} model upper limits on the production cross section for a massless LSP.
For further details see \protect\figref{fig:ExpLimitWinoGluino}.
\label{fig:LimitsN2N3}
}
 \end{center}
\end{figure}

The observed and expected  
95\% CL limit contours for the R-slepton RPC simplified models considered in this paper are shown in \figref{fig:LimitsN2N3}(a), 
while \figsref{fig:LimitsN2N3}(b) and \ref{fig:LimitsN2N3}(c) present the observed and expected 
95\% CL limits on the production cross section for the stau and \Zboson{} RPC simplified models, respectively, assuming
zero mass for the \ninoone{}.

The strongest constraints for RPC models are obtained in the R-slepton model.
In this case, \ninotwothree{} with masses of up to $620\gev$ are excluded if the LSP is massless.
As the LSP mass increases, the leptons from the cascade become less energetic, decreasing the analysis acceptance.
The maximum \ninoone{} mass that can be excluded by this analysis is $340\gev$.
In the region allowed by the LEP ($m_{\ninotwo,\ninothree}\gtrsim 100\gev$~\cite{Abdallah:2003xe,Abbiendi:2003sc,Heister:2002mn,Acciarri:1999km}), 
no limits are set on the stau or \Zboson{} models.

\subsection{RPC GGM models} \label{sec:limitsGGM}

\begin{figure}[ht]
 \begin{center}
 \subfigure[~GGM $\tan\beta$=1.5]{\includegraphics[width=0.45\textwidth]{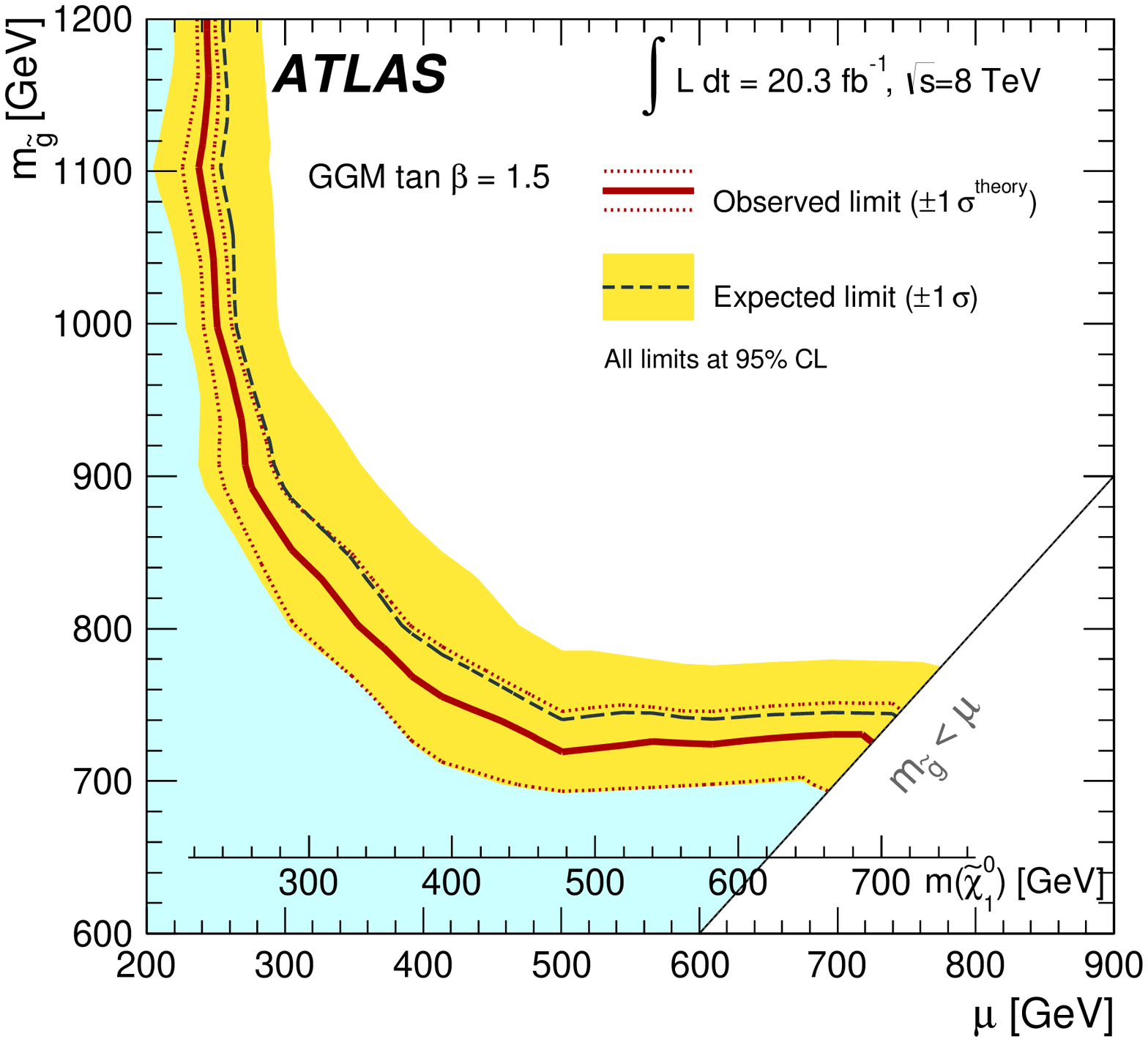}}
 \subfigure[~GGM $\tan\beta$=30]{\includegraphics[width=0.45\textwidth]{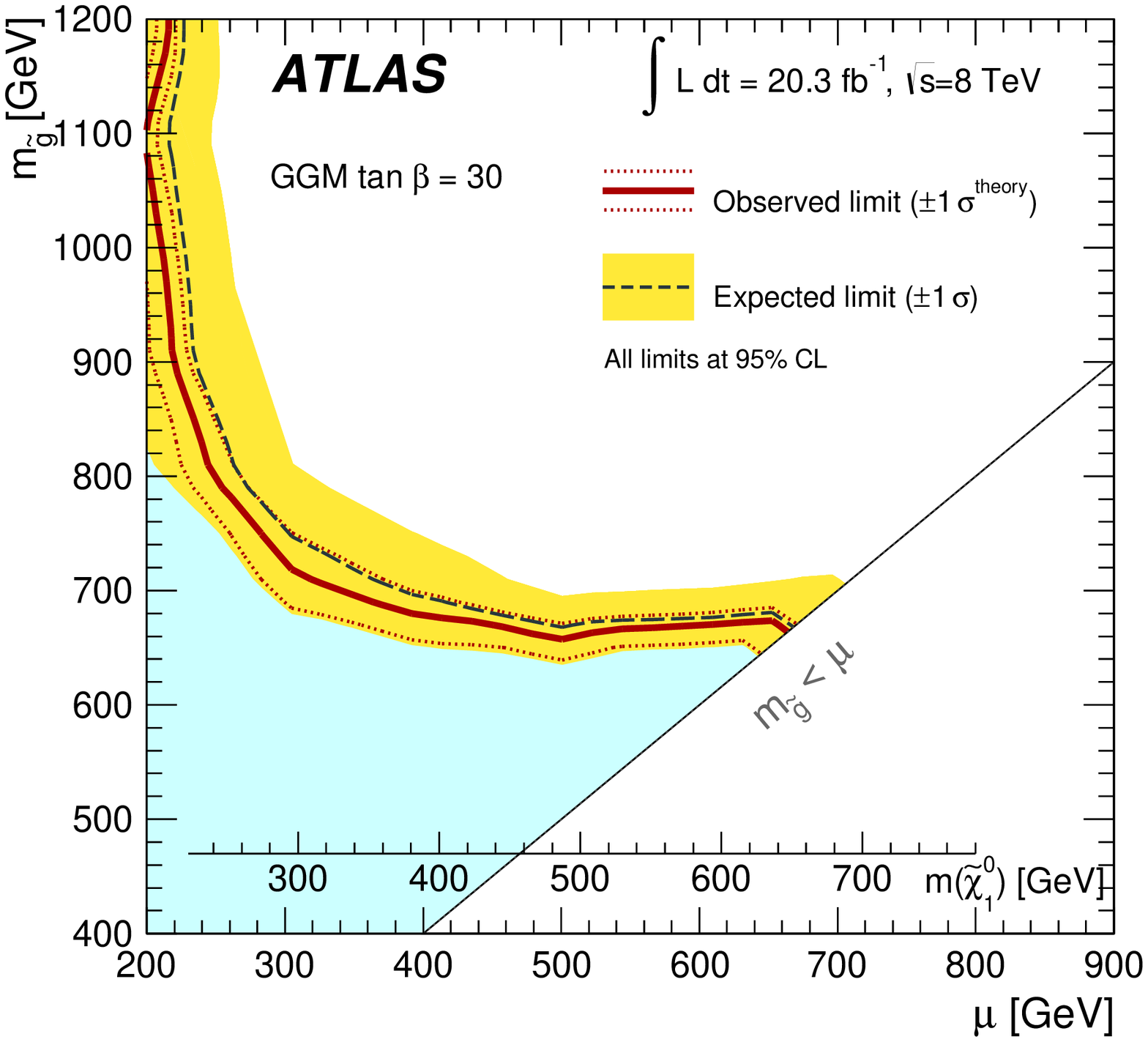}}
   \caption{The 95\% CL exclusion limit contours for the (a) $\tan\beta = 1.5$ and (b) $\tan\beta = 30$ GGM models.
The lower shaded area shows the excluded region.  
For further details see \protect\figref{fig:ExpLimitWinoGluino}.
\label{fig:LimitsGGM}
}
 \end{center}
\end{figure}

The observed and expected 
95\% CL limit contours for the two GGM models considered in this paper are shown in \figref{fig:LimitsGGM}.  
Only regions with a \Zboson{} boson requirement are statistically combined to extract these limits. 

Independently of the value of $\mu$, gluinos with $m_{\gluino}\,$$<700\gev$ are excluded for $\tan\beta=1.5$.
For very large gluino masses, the direct production of \ninoone, \chinoonepm{} and \ninotwo{} becomes dominant, and values of $\mu$ between $200$ and about $230\gev$ are excluded for any gluino mass.
For the larger value of $\tan\beta=30$, the limits are weaker: gluinos with masses less than about $640\gev$ are excluded at 95\% CL.

\section{Summary}

A search has been performed for SUSY signals in final states with four or more leptons using the ATLAS detector, based on 
a data sample corresponding to 20.3~\ifb{} of 
proton-proton collisions delivered by the LHC at $\rts=8\tev$ in 2012. The analysis targets lepton-rich
RPV and RPC SUSY signals, including those from GGM SUSY, which can be either enriched in or depleted of 
\Zboson-boson production.
No significant deviation is observed from SM expectations, within statistical and systematic uncertainties.
The null result is interpreted by providing 95\% CL upper limits on the visible cross section of new processes within each signal region, which lie between 0.17 and 0.45~fb, depending on the final state.

Limits are also placed on sparticle masses in specific SUSY models.
In RPV models where the LSP decays only to electrons and muons, the 95\% CL lower mass limits are the following:
$1350\gev$ for the gluino, $750\gev$ for winolike charginos and 490 $(410)\gev$ for L(R)-sleptons. 
Slightly less stringent limits are placed on the same parameters for RPV models with tau-rich decays.
In both cases the mass of the LSP is assumed to be at least as large as 20\% of the NLSP mass.
A limit of $400\gev$ can be placed on sneutrino masses for RPV models with electron and muon decays of the LSP.

The strongest constraints for RPC models are obtained in the R-slepton model, where \ninotwothree{} with masses of up to $620\gev$ are excluded 
if the LSP is massless.

For the GGM model with $\tan\beta=1.5$, values of $\mu$ between $200$ and about $230\gev$ are excluded for any gluino mass,
and gluinos with $m_{\gluino}\,$$<700\gev$ are excluded independently of the value of $\mu$. 
For $\tan\beta=30$, gluinos with masses less than about $640\gev$ are excluded at 95\% CL.

\FloatBarrier


\acknowledgments

We thank CERN for the very successful operation of the LHC, as well as the
support staff from our institutions without whom ATLAS could not be
operated efficiently.

We acknowledge the support of ANPCyT, Argentina; YerPhI, Armenia; ARC,
Australia; BMWF and FWF, Austria; ANAS, Azerbaijan; SSTC, Belarus; CNPq and FAPESP,
Brazil; NSERC, NRC and CFI, Canada; CERN; CONICYT, Chile; CAS, MOST and NSFC,
China; COLCIENCIAS, Colombia; MSMT CR, MPO CR and VSC CR, Czech Republic;
DNRF, DNSRC and Lundbeck Foundation, Denmark; EPLANET, ERC and NSRF, European Union;
IN2P3-CNRS, CEA-DSM/IRFU, France; GNSF, Georgia; BMBF, DFG, HGF, MPG and AvH
Foundation, Germany; GSRT and NSRF, Greece; ISF, MINERVA, GIF, I-CORE and Benoziyo Center,
Israel; INFN, Italy; MEXT and JSPS, Japan; CNRST, Morocco; FOM and NWO,
Netherlands; BRF and RCN, Norway; MNiSW and NCN, Poland; GRICES and FCT, Portugal; MNE/IFA, Romania; MES of Russia and ROSATOM, Russian Federation; JINR; MSTD,
Serbia; MSSR, Slovakia; ARRS and MIZ\v{S}, Slovenia; DST/NRF, South Africa;
MINECO, Spain; SRC and Wallenberg Foundation, Sweden; SER, SNSF and Cantons of
Bern and Geneva, Switzerland; NSC, Taiwan; TAEK, Turkey; STFC, the Royal
Society and Leverhulme Trust, United Kingdom; DOE and NSF, United States of
America.

The crucial computing support from all WLCG partners is acknowledged
gratefully, in particular from CERN and the ATLAS Tier-1 facilities at
TRIUMF (Canada), NDGF (Denmark, Norway, Sweden), CC-IN2P3 (France),
KIT/GridKA (Germany), INFN-CNAF (Italy), NL-T1 (Netherlands), PIC (Spain),
ASGC (Taiwan), RAL (UK) and BNL (USA) and in the Tier-2 facilities

\bibliography{paper}

\clearpage
\onecolumngrid
\clearpage
\input{atlas_authlist}

\end{document}

%% file: atlas_authlist.tex
\begin{flushleft}
{\Large The ATLAS Collaboration}

\bigskip

G.~Aad$^{\rm 84}$,
B.~Abbott$^{\rm 112}$,
J.~Abdallah$^{\rm 152}$,
S.~Abdel~Khalek$^{\rm 116}$,
O.~Abdinov$^{\rm 11}$,
R.~Aben$^{\rm 106}$,
B.~Abi$^{\rm 113}$,
M.~Abolins$^{\rm 89}$,
O.S.~AbouZeid$^{\rm 159}$,
H.~Abramowicz$^{\rm 154}$
% ...
\bigskip
\\
% ...
$^{11}$ Institute of Physics, Azerbaijan Academy of Sciences, Baku, Azerbaijan\\
% ...
$^{84}$ CPPM, Aix-Marseille Universit{\'e} and CNRS/IN2P3, Marseille, France\\
% ...
$^{89}$ Department of Physics and Astronomy, Michigan State University, East Lansing MI, United States of America\\
% ...
$^{106}$ Nikhef National Institute for Subatomic Physics and University of Amsterdam, Amsterdam, Netherlands\\
% ...
$^{112}$ Homer L. Dodge Department of Physics and Astronomy, University of Oklahoma, Norman OK, United States of America\\
$^{113}$ Department of Physics, Oklahoma State University, Stillwater OK, United States of America\\
% ...
$^{116}$ LAL, Universit{\'e} Paris-Sud and CNRS/IN2P3, Orsay, France\\
% ...
$^{152}$ Institute of Physics, Academia Sinica, Taipei, Taiwan\\
% ...
$^{154}$ Raymond and Beverly Sackler School of Physics and Astronomy, Tel Aviv University, Tel Aviv, Israel\\
% ...
$^{159}$ Department of Physics, University of Toronto, Toronto ON, Canada\\
% ...
\end{flushleft}
